\documentclass[]{aa}
\usepackage{amsmath}
\usepackage{txfonts}
\usepackage{natbib}
\bibpunct{(}{)}{;}{a}{}{,}

\usepackage[dvips]{graphicx}
\usepackage{url}
\usepackage{aalongtable}
\usepackage{lscape}

\usepackage[morefloats=40]{morefloats}

\newcommand{\Ks}[1]{K_{s #1}}
\newcommand{\rEW}{\langle W'\rangle}

\newcommand{\FeH}[1]{\mathrm{[Fe/H]_{#1}}}

\defcitealias{Saviane2012}{S12}
\defcitealias{Gullieuszik2009}{G09}
\defcitealias{Rutledge1997}{R97}
\defcitealias{Carretta2009}{C09}
\defcitealias{CG1997}{CG97}
\defcitealias{Harris2010}{H10}

\newcommand{\Sav}{\citetalias{Saviane2012}}
\newcommand{\Rut}{\citetalias{Rutledge1997}}
\newcommand{\Har}{\citetalias{Harris2010}}
\newcommand{\Carr}{\citetalias{Carretta2009}}
\newcommand{\CarGr}{\citetalias{CG1997}}

\begin{document}

\title{Deriving Metallicities from Calcium Triplet Spectroscopy in combination with Near Infrared Photometry\thanks{Based on observations gathered with ESO-VISTA telescope (proposal ID 172.B-2002).}}
\titlerunning{Deriving Metallicities from CaT in combination with nIR Photometry}
\author{F. Mauro\inst{1} 
\and C. Moni Bidin\inst{1,2} 
\and D. Geisler\inst{1}
\and I. Saviane\inst{3}
\and G.~S. Da Costa\inst{4}
\and A.~C. Gormaz-Matamala\inst{1}
\and S. Vasquez\inst{3,5}
\and A.-N. Chen\'e\inst{1,6,7}
\and R. Cohen\inst{1}
\and B. Dias\inst{3,8}
}
\institute{Departamento de Astronom\'ia, Universidad de Concepci\'on, Casilla 160-C, Concepci\'on, Chile
\and
Instituto de Astronom\'ia, Universidad Cat\'olica del Norte, Av. Angamos 0610, Antofagasta, Chile
\and
European Southern Observatory, Ave. Alonso de Cordova 3107, Casilla 19, 19001 Santiago, Chile
\and
Research School of Astronomy \& Astrophyiscs, Australian National University, Canberra ACT 0200 Australia.
\and
Instituto de Astrof\'isica, Facultad de Fis\'ica, Pontificia Universidad Cat\'olica de Chile, Av Vicu\~na MacKenna 4860, Santiago, Chile
\and
Gemini Observatory, Hawaii, USA
\and
Departamento de de Fis\'ica y Astronom\'ia, Universidad de Valpara\'iso, Av. Gran Breta\~na 1111, Playa Ancha, Casilla 5030, Chile
\and
Instituto de Astronomia, Geof\'\i sica e Ci\^encias Atmosf\'ericas, Universidade de S\~ao Paulo, Rua do Mat\~ao 1226,  Cidade Universit\'aria, S\~ao Paulo, 05508-900, SP, Brazil
}

\abstract{{When established with sufficient precision the ages, metallicities and kinematics of the Galactic globular clusters (GGCs) can shed much light on the dynamical and chemical evolution of the Galactic halo and bulge.}
While the most fundamental way to determine GC abundances is via high resolution spectroscopy, in practice this method is limited to only the brighter stars in the nearest and less reddened objects.
This restriction has, over the years, led to the development of a large number of techniques{ that measure the overall abundance indirectly, from parameters that correlate with overall metallicity}.  
One of the most efficient methods is the measurement of the equivalent width (EW) of the Calcium II Triplet (CaT) at $\lambda\approx 8500$\AA{} in red giants, corrected for the luminosity and temperature effects {via $V$ magnitude differences from the horizontal branch (HB).}}
{We establish a similar method in the NIR, combining the power of the differential {magnitudes} technique with the advantages of NIR photometry in minimizing differential reddening effects}
{We use the $\Ks{}$ magnitude difference between the star and the reddest part of the HB (RHB) or of the Red Clump (RC) to {generate} reduced equivalent widths (rEW) {from} the datasets presented in Saviane et al. (2012) and Rutledge et al. (1997)
Subsequently we calibrated these rEW against three different metallicity scales: the one presented in Carretta et al. (2009), the metallicity values given in Harris (2010) and a version of the former corrected via high-resolution spectroscopic metallicities.}
{We calculated the calibration relations for the two datasets {and} the three metallicity scales and found that they are approximately equivalent, with differences almost negligible.
We compared our nIR calibrations with the {corresponding} optical ones, and found them to be equivalent, establishing that the luminosity-corrected rEW using the $\Ks{}$ magnitude is compatible with the one obtained from the $V$ magnitude.
We {then} used the metallicities obtained {from} the calibration to {investigate the internal metallicity distributions of the GCs}.}
{We {have} established that the ([Fe/H]:rEW) relation is independent from the magnitude used for the luminosity correction {and find} that the calibration relations {only} change slightly {for} different metallicity scales.
The CaT technique using NIR photometry is {thus} a powerful tool to derive metallicities.
{In particular,} it can be used to study the internal metallicity spread of a GC.
{We confirm} the presence of at least two metallicity populations in NGC\,6656 {and} find that several other GCs present peculiar metallicity distributions.}


\maketitle

\section{Introduction}
\label{s:intro}

Stellar population studies are one of our most powerful tools to study a wide variety of fundamental problems in stellar and galactic astrophysics.
In particular, globular clusters (GCs) are perfect laboratories in this regard.
GCs are testbeds for the understanding not only of stellar evolution and dynamics, but also of the formation of stellar exotica and the processes leading to disruption of massive stellar systems.
GCs are cornerstones of the distance scale and serve as dynamical probes of a galaxy's complex kinematics and interaction history.
They are unexcelled as tracers of the structure, formation and chemical evolution of a galaxy and its distinct components.
Galactic GCs (GGCs) represent one of the fundamental systems that allow a reconstruction of the early evolution of the Milky Way: the knowledge of their ages, metallicities and kinematics has shed much light on the dynamical and chemical evolution of the Galactic halo and bulge \citep{Zinn1985,Minniti1995,Ferraro2009}.

While the most fundamental way to determine GC abundances is via high resolution (HR) spectroscopy, in practice this method is limited to only the brighter stars in the nearest GCs.
This restriction has led over the years to the development of a large number of techniques aimed {at} indirectly measure the overall metal abundance, using parameters such as line strength, blanketing, or giant branch effective temperature, all of which correlate with overall metallicity.
Even if these indices strictly provide only criteria for ranking clusters by abundance, actual metallicities can be determined with appropriate calibration.

One of the most efficient methods is the measurement of the equivalent width (EW) of the Calcium II Triplet (CaT) at $\lambda\approx 8500$\AA{} in red giants \citep{Olszewski1991,Armandroff1991}.
The CaT technique has many advantages.
It is one of the most efficient ways to build up a large sample of accurate metallicity and velocity measurements  even in distant GGCs.
The brightest stars in the optical and IR in clusters older than $\approx 1$Gyr are the red giants, and are thus the natural targets for precision measurements of cluster abundances and velocities.
The CaT lines are extremely strong and near the peak flux of unreddened RGB stars, and the technique only requires moderate resolution ($R\sim 3000$).
Because there are many giants in a typical GC, the derived mean abundance can be determined much more robustly than that based on only one or a few stars.
A reasonable sample of stars must be observed in order to ensure cluster membership, especially in bulge GCs (BGC) where membership on the bright RGB may be as low as 20-50\% due to strong field contamination \citep{Saviane2012}.
Observing in the near Infrared (NIR) is also very advantageous for reddened BGCs, where optical indices can be
strongly absorbed.
Many authors have confirmed the accuracy and repeatability of CaT abundance measurements in combination with broad-band optical photometry and shown its very high sensitivity to metallicity and insensitivity to age \citep[e.g., ][]{Cole2004}.
Additionally, as reported by \citet{Carrera2013}, the strength of the CaT lines depends mainly on iron abundances, rather than on the Ca abundance, as has been pointed out also by several other investigations \citep[e.g.][]{Idiart1997,Battaglia2008}.

In view of all of these advantages, many GGCs have had a sample of their RGB stars observed using CaT.
A seminal paper in this regard is that of \citet[R97]{Rutledge1997}.
They observed a total of 976 giants in 52 GGCs and showed that the CaT is both a very efficient and accurate technique for deriving GC velocities and metallicities.
Recently, \citet[S12]{Saviane2012} began an attempt to fill in CaT data for the large sample of GGCs remaining without such measurements.
They obtained CaT abundances for 20 new GGCs.
Still, this leaves more than one half of the GGCs without CaT data, including most bulge GCs (BGCs).
The BGC system is one of the most important in our Galaxy and a thorough knowledge of metallicities and velocities can help to constrain bulge formation and evolution models.
Nevertheless, substantial crowding or large and possibly variable reddening have combined to limit attempts to use even the CaT technique on many of these BGCs.
One of the main reasons is because the traditional CaT technique, although it nominally involves only observations in the near IR, also requires optical photometry in order to calibrate the metallicity.
It is well known that the CaT lines, in addition to being very metallicity sensitive, also depend on effective temperature and especially luminosity, and these effects must be removed in order to properly derive the metallicity.
Traditionally, this is done by defining a reduced EW (rEW or $W'$) for the sum of some combination of the 3 lines, which is then corrected for luminosity and temperature effects using the slope of the RGB, in particular the magnitude difference in V between the star and the horizontal branch (HB), $V_{HB}-V$.
This differential method is very powerful as it also removes any dependence on distance or mean reddening.
Unfortunately, it also requires good optical photometry, which is often problematic for BGCs.
Indeed, for many BGCs $V_{HB}$ is only very poorly known.
Clearly, it would be very advantageous to develop a similar technique without these problems.

Here we establish a similar method in the IR, using as the fiducial magnitude the $\Ks{}$ magnitude of the reddest part of the HB (RHB) or of the Red Clump (RC).
This combines the power of the differential technique with the advantages of IR observations in minimizing extinction and reddening effects.

A major advantage of this work is the possibility to exploit databases that are homogeneous both in terms of spectroscopy and photometry.
Our photometric dataset consists of a catalog of GCs observed as part of the Vista Variables in the Via Lactea
(VVV) Survey \citep{Saito2012}, calibrated on the system of the  Two Micron All Sky Survey \citep[2MASS,][]{2MASS}.
This proves to be the ideal catalog for this purpose, since it is integrated with the 2MASS PSC.
We used the spectroscopic dataset presented in \Sav{}, and also in \Rut{}, while we adopted as the metallicity reference the \citet{Carretta2009} scale, based on UVES and GIRAFFE HR spectra.

We note that an initial attempt at involving NIR photometry to calibrate CaT was made by \citet{Olszewski1991}, where they used the absolute $I$ magnitude of the red giant in order to correct for luminosity and temperature effects.
Unfortunately, this requires an accurate distance, which is certainly problematic for BGCs.
Another attempt of calibration of CaT in the nIR was made by \citet{WarrenCole2009} using spectra of 133 red giant stars from 10 Galactic open clusters and two Galactic globular clusters, and  \citet{ZinnWest1984} as metallicity scale.
They found a linear correlation.
However, the \citet{ZinnWest1984} and \citet{Carretta2009} scales are not correlated properly by a simple linear relation, but by at least a quadratic expression.
{Previous works that used NIR photometry to correct CaT are \citet{Lane2010b} and \citet{WarrenCole2009}, but the former used as reference level the $\Ks{}$ magnitude of the Tip of the RGB, while the latter took the value of $\Ks{}$ at the RR Lyrae instability strip for the GCs.}

The CaT method can also be applied to derive the metallicity for red giants in any stellar population for which $V_{HB}$ is known, e.g. dwarf spheroidal galaxies, the Magellanic Clouds, extragalactic Globular Clusters or even M31 \citep{Battaglia2011,Parisi2010,Foster2010,Jones1984,Cenarro2008}.
Adapting the technique to the IR allows us to apply it to any stellar population where reddening is problematic, opening up many additional targets for detailed study.

Here we first present our observations and reductions (Section \ref{s:data}).
Next we present the metallicity calibration and discuss individual clusters (Section \ref{s:res}).
Our conclusions are discussed in Section~\ref{s:concl}.

\section{Observations and reductions}
\label{s:data}

Our main target list consists of the GCs analyzed by \citetalias{Saviane2012}.
The nIR imaging collected in the context of the VVV Survey was used for all clusters included in the survey area.
We also checked if any of the GCs in the dataset not observed by the VVV Survey had useful 2MASS photometry, permitting us to determine the RHB position.
Including also these GCs allowed us to better constrain the calibration over a wider metallicity range with better sampling.
We similarly selected GCs from the \citetalias{Rutledge1997} catalog in order to determine a calibration for this dataset as well.
{All clusters analyzed in the current paper are listed in Tables~\ref{tab:HB}. The complete data for all spectroscopic stars are presented in Table~\ref{tab:data}, and the metallicity values are listed in Tables \ref{tab:SC09}-\ref{tab:RH10}, separated for spectroscopic dataset, only available in electronic form.}

\subsection{Spectroscopy}
\label{ss:spectra}


\subsubsection{\Sav{} data}
The data were obtained in the z-band region of giant stars with FORS2 \citep{Appenzeller1998}, working at the Cassegrain focus of VLT/UT1-Antu.
The approach used to assemble the list of clusters observed is discussed in \Sav{}.
All spectra were extracted using the FORS2 pipeline version 1.2 \citep{Izzo2010}.
The absorption lines of the CaT were used both to measure radial velocities and derive metallicities.
Metal-rich clusters were measured with Gaussian plus Lorentzian function fits, while the equivalent widths for metal-poor clusters were computed with Gaussian fits only and transformed onto the  scale established in \citet[G09]{Gullieuszik2009}.
This decision is justified since we have verified that there is a one-to-one correspondence between widths measured with the two methods.

The final $\Sigma W_{S12}$ for each star results from the sum of the EWs of the two strongest CaT lines (8542\AA{}, 8662\AA{}):
\begin{equation}\label{eq:EW-Sav12}
\Sigma W_{S12}=EW(8542\text{\AA{}})+EW(8662\text{\AA{}}).
\end{equation}
We refer to \citetalias{Saviane2012} for a more complete description of the observations, reduction procedure, and selection for cluster membership.

\subsubsection{\Rut{} data}
We also retrieved the spectroscopic data from \citet[][R97]{Rutledge1997}.
Not all of the clusters in \citetalias{Rutledge1997}  were used because of the difficulty to identify the stars in the scanned finding charts.
Anyway, we included all the more metal-poor and metal-rich GCs, the ones included in the VVV Survey, as well as others to cover properly the full metallicity range of GGCs.

In this case the final $\Sigma W$ for each star is the weighted sum of the three CaT lines:
\begin{equation}\label{eq:EW-Rut97}
\Sigma W_{R97}=0.5\cdot EW(8498\text{\AA{}})+EW(8542\text{\AA{}})+0.6\cdot EW(8662\text{\AA{}}).
\end{equation}
We refer to R97 for a more complete description of the observations, reduction procedure, and selection for cluster membership.

\subsection{Photometry}
\label{ss:photo}

\begin{figure}[hp]\begin{center}
\resizebox{\hsize}{!}{\includegraphics[]{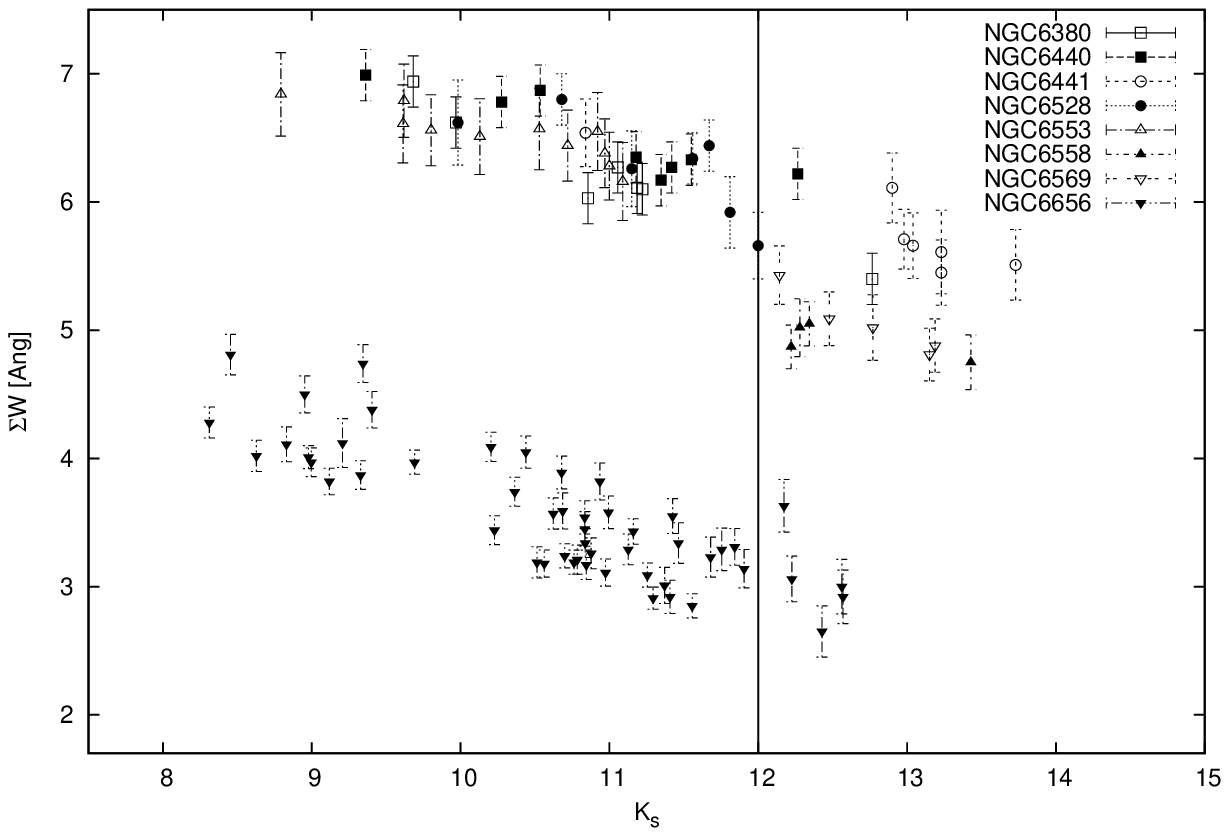}}
\resizebox{\hsize}{!}{\includegraphics[]{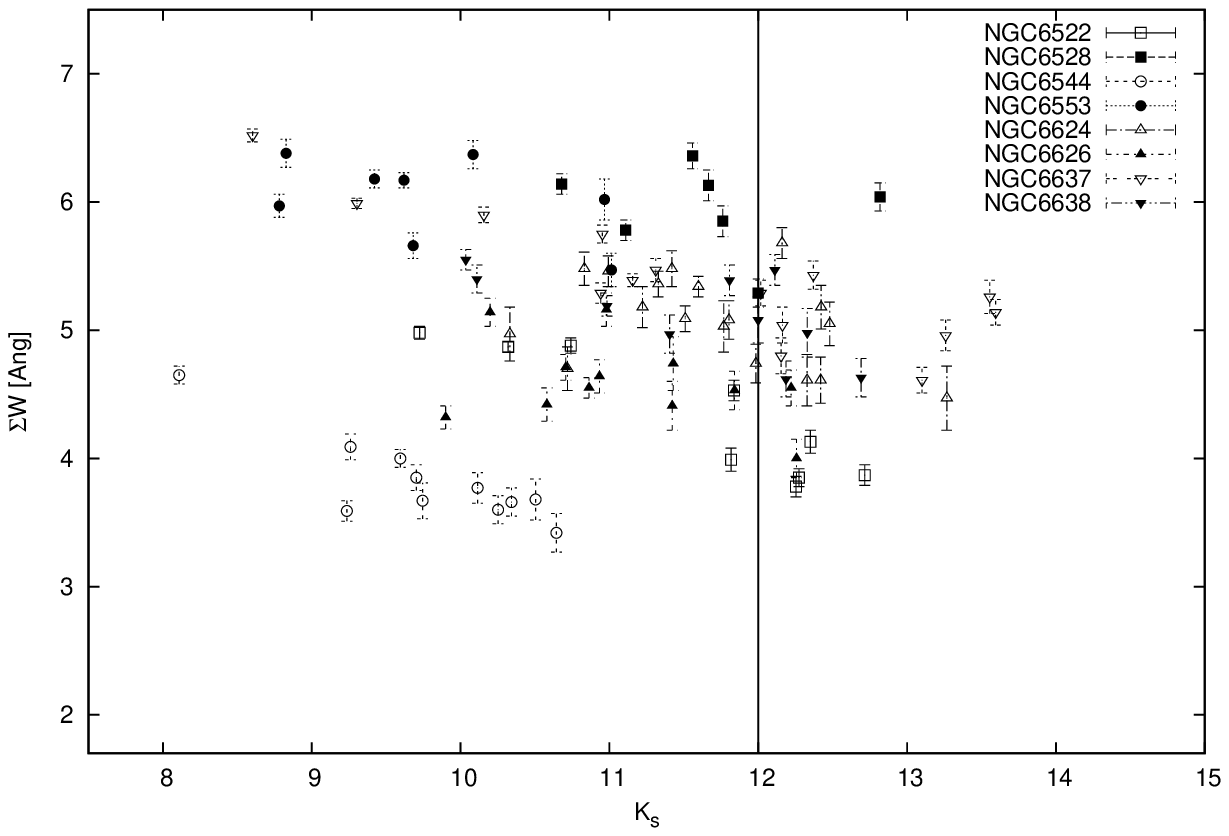}}
\caption{Plot of Ca II line strength against $\Ks{}$ magnitude for VVV clusters in \Sav{} and \Rut{} catalogs.
The vertical bold line at $\Ks{}=12$ is  the saturation limit for the VVV survey.
Vertical bars on each point show the measurement uncertainty in the line strengths.
The measurement uncertainty in magnitude are too small to be noticed.}
\label{fig:W_Ks}
\end{center}\end{figure}

The VVV Survey  \citep{Minniti2010,Catelan2011} is one of six ESO Public Surveys carried out at the 4-meter Visible and Infrared Survey Telescope for Astronomy (VISTA), scanning the Galactic bulge ($-10\leq l\leq+10$, $-10\leq b\leq +5$) and the adjacent part of the southern disk ($-65\leq l\leq-10$, $-2\leq b\leq +2$).
The survey collects data in five NIR bands ($ZYJHK_\mathrm{s}$) with the VIRCAM camera \citep{Emerson2010}, an array of sixteen 2048$\times$2048~pixel detectors with a pixel scale of $0\farcs 341/pix$.
VVV images extend several magnitudes fainter than the 2MASS, and enjoy increased spatial resolution \citep{Saito2010}.
Both of these factors are  particularly important for mitigating contaminated photometry in crowded regions near the Galactic center and the cores of globular clusters.
The VVV survey {provides} precise multi-epoch {$\Ks{}$-}photometry for 39 Galactic globular clusters{, that permits to obtain $\Ks{}$ magnitudes with good accuracy.}
{For this reason, we preferred $\Ks{}$ magnitudes, instead of $J$ or $H$ ones, to correct the equivalent widths.}

\begin{figure}[ht]\begin{center}
\resizebox{\hsize}{!}{\includegraphics[]{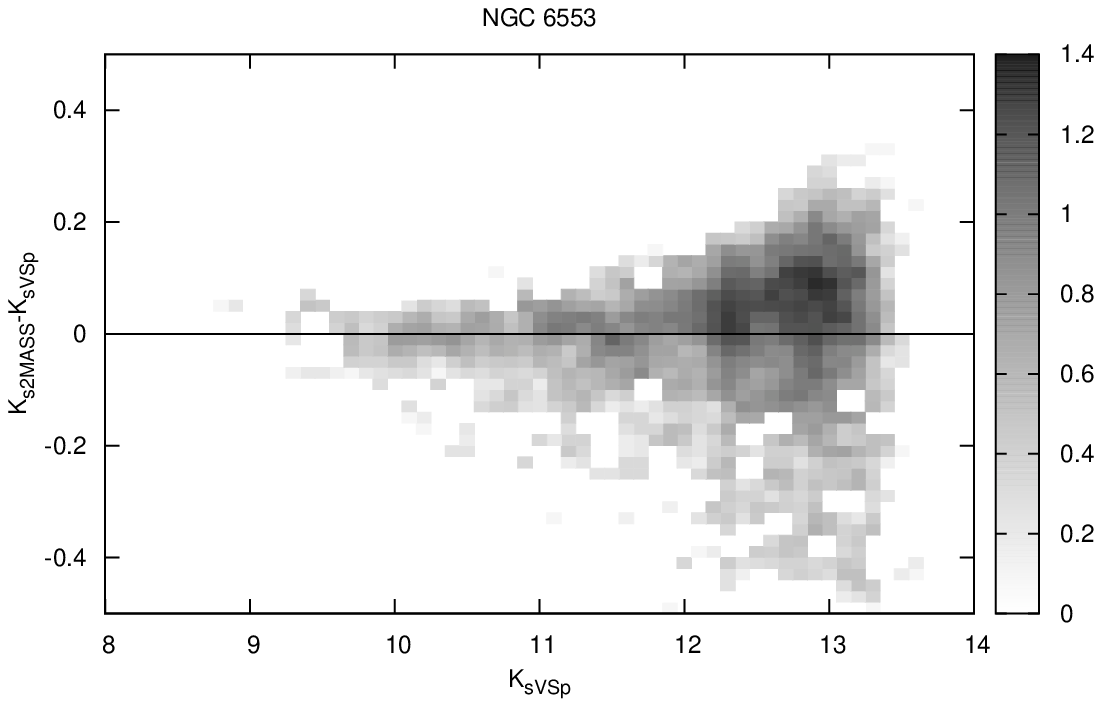}}
\resizebox{\hsize}{!}{\includegraphics[]{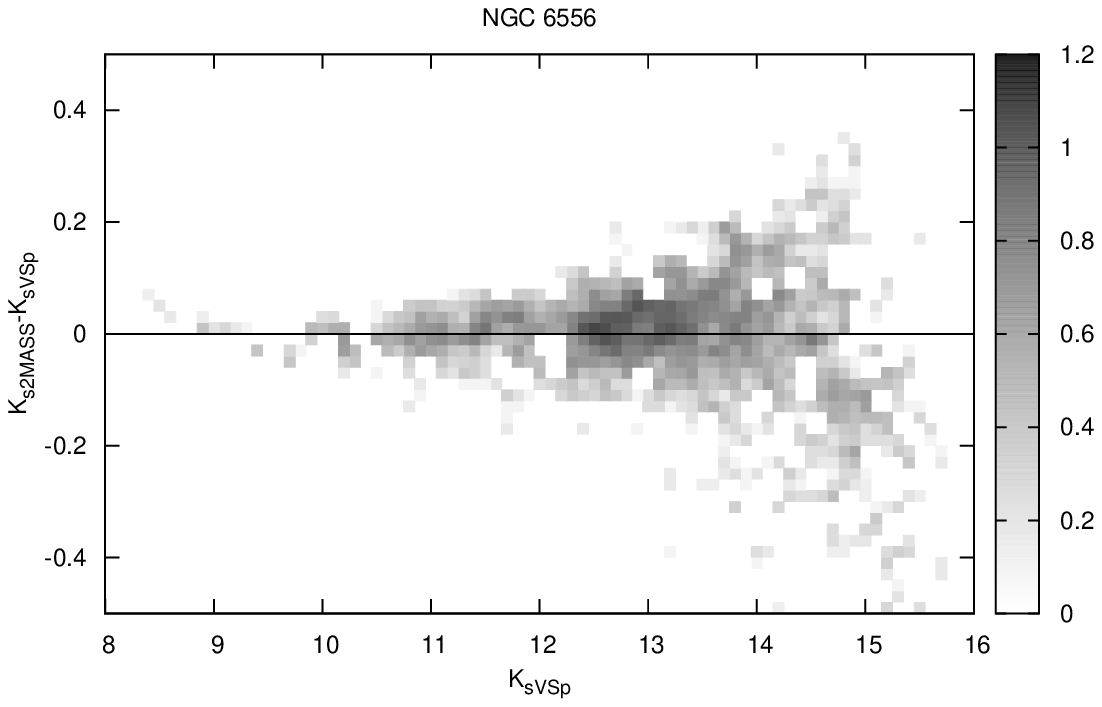}}
\caption{Density map in logarithmic scale of the photometric differences in $\Ks{}$ between 2MASS and VSp catalogs, as a function of $\Ks{}$ magnitude obtained with VVV-SkZ\_pipeline (VSp, upper figure for NGC\,6553, lower figure for NGC\,6656).
No systematic offset exists, especially for the brighter ($\Ks{}<12$) stars, which are saturated in the
VVV data.}
\label{fig:2M-VSp}
\end{center}\end{figure}

We retrieved from the Vista Science Archive website\footnote{\url{http://horus.roe.ac.uk/vsa/}} the VVV images containing the GGCs targeted by \citetalias{Saviane2012} and \citetalias{Rutledge1997}, pre-reduced at the Cambridge Astronomical Survey Unit (CASU)\footnote{\url{http://casu.ast.cam.ac.uk/}} with the VIRCAM pipeline \citep{Irwin2004}.
We performed PSF-fitting photometry using the VVV-SkZ\_pipeline code \citep{VSpaper}, based on DAOPHOT suite\citep{DAOPHOT,ALLFRAME}, on the single 2048$\times$2048~pixel chips extracted from the stacked VVV pawprints \citep{Saito2012}.
The photometry was tied to the 2MASS system, as described in \citet{MoniBidin2011} and \citet{Chene2012}.
The use of the 2MASS system as the standard photometric system permitted us to integrate our photometric database with the 2MASS PSC.
The use of the VVV-SkZ\_pipeline was fundamental for this work, since it is the only photometric procedure for VVV data that provides accurate photometry even for partially saturated stars, which are the bulk of the giants observed spectroscopically.
As can be seen in Figures \ref{fig:W_Ks}, almost 90\% of the observed stars are brighter than the saturation limit for the VVV survey ($\Ks{}=12$), but our photometry is still reliable up to $\Ks{}=9-10$, as the comparison with 2MASS photometry demonstrates in Figure~\ref{fig:2M-VSp}.

The stars with spectroscopic CaT measurements were identified in our VVV photometry.
In some cases it was not possible to find a corresponding star in the VVV catalogs, i.e. very bright stars ($\Ks{}<9$) completely saturated on the VVV images.
It this case, their photometric data were obtain from the 2MASS PSC catalog.
The GC HP1 was excluded due to the difficulty in identifying the cluster RHB.
The data for all spectroscopic stars are presented in Table~\ref{tab:data}, only available in electronic form.

\subsubsection{Determination of the HB-level magnitude}
\label{sss:HBlevel}

The magnitude at the HB level was determined by the position of the peak in the luminosity distribution of the reddest part of the HB.
The HB of some metal-poor GCs (like NGC\,6121, NGC\,6397 or NGC\,6809) is well populated also at the red end, permitting an accurate determination of the RHB magnitude.
For most of the metal-poor GCs, however, the red HB is not easily detectable in the color-magnitude diagram (CMD).
We determined {theoretically a ``first-guess''} position of the RHB in these cases, and compared it with the peaks in the luminosity distribution for the stars located along and next to the RGB, $2-3'$ from the cluster center.
We used the metal-poor GCs with well-defined RHB to calibrate this procedure.
The determination of the {initial} position was performed {considering} the results given by several procedures.
We calculated a ``theoretical'' value based on \citet{Bressan2012} and \citet{Girardi2001}, corrected for distance modulus and reddening of the GC.
An empirical value was also calculated from the $V_{HB}$ value listed in \citet[2010 edition, hereafter H10]{Harris1996}, corrected for distance modulus and reddening of the GC, for a mean $(V-\Ks{})$ color taken from \citet{Bressan2012}.
The accuracy of these two methods depends strongly on the accuracy of the photometric parameters of the cluster.
The CaT datasets include also the $\Delta V=V_{HB}-V$ for each star; we fitted the points in the $(\Delta V; \Ks{})$ plane with a linear {and quadratic\footnote{The locus of the points in the $(V_{HB}-V;\Ks)$ plane is not perfectly linear, but slightly convex.}} relations: the intercepts give an estimate of $\Ks{}$ magnitude of RGB stars that in the optical have the same luminosity of the RHB.
We noted that this method can have an uncertainty as large as 0.5~mag, and depends strongly on the accuracy of $V_{HB}$ and on the magnitude range covered by the data.
We used these estimates, together with the values presented in \citet{Valenti2007,Valenti2010}, {as an initial position} to identify RHB among the peaks in the $\Ks{}$ luminosity distribution.
Additionally, we used the empirical calibration for the RGB bump \citep{Valenti2007} applied to the parameters listed in \Har{} to determine the position of the bump in the luminosity distribution, to avoid confusion between the peaks associated with the two different features.
The results are presented in Table~\ref{tab:HB}.

\begin{table}[!ht]\begin{center}
\caption{List of RHB levels}
\begin{tabular}{l c c c}
ID & Photometric Source & $\Ks{}(HB)$ & $\sigma$ \\
\hline
NGC6380 & VVV & 13.80 & 0.05 \\
NGC6440 & VVV & 13.64 & 0.05 \\
NGC6441 & VVV & 14.38 & 0.05 \\
NGC6522 & VVV & 13.16 & 0.05 \\
NGC6528 & VVV & 13.11 & 0.05 \\
NGC6544 & VVV & 10.50 & 0.05 \\
NGC6553 & VVV & 12.27 & 0.05 \\
NGC6558 & VVV & 13.30 & 0.10 \\
NGC6569 & VVV & 14.30 & 0.05 \\
NGC6624 & VVV & 13.28 & 0.05 \\
NGC6626 & VVV & 13.00 & 0.05 \\
NGC6637 & VVV & 13.51 & 0.05 \\
NGC6638 & VVV & 13.70 & 0.05 \\
NGC6656 & VVV & 11.70 & 0.10 \\
NGC2808 & 2MASS & 14.02 & 0.05 \\
NGC3201 & 2MASS & 12.05 & 0.10 \\
NGC4372 & 2MASS & 12.05 & 0.20 \\
NGC4590 & 2MASS & 13.30 & 0.20 \\
NGC6121 & 2MASS & $\;\;9.97$ & 0.05 \\
NGC6139 & 2MASS & 13.75 & 0.15 \\
NGC6254 & 2MASS & 11.87 & 0.05 \\
NGC6325 & 2MASS & 13.27 & 0.05 \\
NGC6356 & 2MASS & 14.60 & 0.05 \\
NGC6397 & 2MASS & 10.45 & 0.05 \\
NGC6541 & 2MASS & 13.00 & 0.20 \\
NGC6809 & 2MASS & 12.00 & 0.20 \\
NGC6838 & 2MASS & 11.78 & 0.05 \\
NGC7078 & 2MASS & 13.50 & 0.10 \\
NGC7099 & 2MASS & 13.40 & 0.15 \\
Pal7    & 2MASS & 12.55 & 0.05 \\
\hline
\end{tabular}  
\label{tab:HB}                   
\end{center}
\end{table}

The comparison between our $\Ks{}(HB)$ values and the ones deduced from the $(V(HB)-V; \Ks{})$ fit suggests that the value of $V_{HB}$ is  not very accurate for most of the faintest GCs.
In fact, 73\% of GCs with $\Ks{}(HB)<13$ present an agreement between the two values within 0.2~mag, but this fraction reduces to  37\% among fainter GCs.

\subsection{Metallicities}
\label{ss:met}

The metallicities were taken from \citet[][hereafter C09]{Carretta2009}, who measured [Fe/H] for 19 GCs from the analysis of spectra of about 2000 RGB stars using FLAMES@VLT (about 100 stars with GIRAFFE and about 10 with UVES, respectively, in each GC).
With these data, they recalibrated their previous metallicity scale \citep[e.g.][CG97]{CG1997}  to the UVES scale, giving the calibration relations, and assembled a table of [Fe/H] values for 133 clusters present in the \citet{Harris1996} catalog.
The metallicities were computed based on the weighted average of indices published in four different studies, putting them on a single UVES scale.
These values are reliable, but they were not obtained from spectroscopic analysis for all the clusters.
Nevertheless, \Carr{} has the advantage of being  a homogeneous metallicity scale, and it is the main metallicity source for \Har{}.
We considered for the calibration only the values not  derived from \citet{Harris1996} (clusters with a ``1'' in the Notes column of Appendix 1).
The \Sav{} dataset includes 8 of the GCs used in \Carr{} as calibrators (NGC\,2808, NGC\,3201, NGC\,6121, NGC\,6254, NGC\,6397, NGC\,6441, NGC\,6838, and NGC\,7078), while 7 calibrators (NGC\,2808, NGC\,3201, NGC\,4590, NGC\,6121, NGC\,6397, NGC\,6809, and NGC7099) are among the \Rut{} sample used in this work.

We observed that, while the majority of GCs in the $(\FeH{C09};\rEW)$ plane lie within $\sim 1\sigma$ to the fit, others (namely NGC\,6528, NGC\,6558 and NGC\,6569) present larger differences.
These clusters are not part of the 19 \Carr{} calibrators, and their metallicities were derived through the weighted average of literature values.
Checking the sources of the metallicities for these three objects, we found that NGC\,6528 and NGC\,6558 have recent HR spectroscopic metallicities  in poor agreement with the C09 metallicity.
H10 reports four sources of HR spectroscopic metallicity for NGC\,6528.
\citet{Carretta2001} based their estimate  on observations carried out with the High Resolution Echelle Spectrometer (HIRES) at Keck I (final $R\approx 15,000$) of four HB stars.
The \citet{Origlia2005} estimate is based on IR echelle spectra of four bright core giants, acquired with the IR spectrograph NIRSPEC  ($R\sim 25,000$) mounted at the Keck II telescope.
The other two are based on the same spectra acquired by \citet{Zoccali2004} with UVES ($R\sim 45,000-55,000$) of three stars (one HB star and two red giants). 
While the metallicity derived by \citet{Carretta2001} (and used in C09) is $+0.07\pm0.08$~dex, the \citet{Origlia2005} NIRSPEC metallicity is $-0.17\pm0.01$~dex, and the UVES value is $-0.1\pm0.2$~dex and $-0.24\pm0.19$~dex, from \citet{Zoccali2004} and \citet{Sobeck2006} respectively.
The photometric estimates obtained by \citet{Momany2003} are centered on the mean of field stars in Baade's Window, $\FeH{}= -0.25$, as derived by \citet{McWilliam1994}.
Anyway, we notice that the metallicity distributions of the stars in the \Sav{} and \Rut{} catalogs of this GC, according to our calibrations, are both almost flat, covering a metallicity range of $\approx 1$\,dex.
For NGC\,6558, \citet{Barbuy2007} estimate a metallicity of $-0.97\pm0.15$ dex, based on the analysis of HR spectra of five giant stars (two are in common with our sample) acquired at the VLT with the multifiber spectrograph FLAMES in GIRAFFE mode ($R\sim 22,000$).
C09 demonstrates that the metallicity scales obtained with UVES and GIRAFFE do not present systematic differences\footnote{mean difference ``UVES minus GIRAFFE'' of $-0.015\pm 0.008\; dex$ with a rms scatter of 0.037 dex from 19 of their GCs}.
For NGC\,6569, \citet{Valenti2011} find a metallicity of $-0.79\pm0.02$ dex, based on six HR IR echelle spectra acquired with NIRSPEC ($R\sim 25,000$).
The \citetalias{CG1997} metallicity on the UVES/C09 scale is $-0.90$~dex, similar to the averaged photometric value of $\FeH=-0.88$ dex estimated in \citet{Valenti2005}.
For the \Rut{} dataset, NGC\,6624 presents recent HR spectroscopic metallicities  in poor agreement with the \Carr{} metallicity and a difference with our best fits of more than $2\sigma$.
NGC\,6624 was analyzed with HR spectra in \citet{Valenti2011}, where they found  $\FeH{}=-0.69$~dex.
Our calibration yields values of metallicity on the  \Carr{} scale in better agreement with the metallicities estimated with HR spectroscopy than with C09 metallicities, suggesting that the C09 estimates for these GCs may have problems.
Based on these considerations, we decided to construct a second metallicity scale called ``corrected C09'' (C09c) in order to check if these more recent values provide a better calibration relation.
For NGC\,6626, that was not used as a calibration cluster, we used for C09c the value $\FeH{}=-1.28$~dex calculated by \citet{DaCostaArmandroff1995} through CaT method, as comparison.

We additionally calculated the calibration relations also for the metallicities given in \Har{}, since it is a commonly-used source.
All the metallicity values are listed in Tables \ref{tab:SC09}-\ref{tab:RH10}.

\subsection{Dependence of He abundance and contamination from AGB stars}
\label{ss:He}
\citetalias{Saviane2012} highlighted that the CaT method works on the assumption that $V_{HB}$ depends almost exclusively on [Fe/H], the other stellar parameters playing only a secondary role.
While this is true for the age of old stellar systems like GGCs \citep{Salaris2002,Ferraro2006}, cluster-to-cluster differences in helium abundance can instead be significant, and might cause higher residuals in the calibration, supposing these differences are not correlated with metallicity.
In general, halo GGCs share a common He abundance \citep{Buzzoni1983,Zoccali2000,Cassisi2003}, but things might be different for BGCs.
\citet{Nataf2011} recently postulated that bulge stars have a He enhancement $\Delta Y = 0.06$ with respect to halo GCs \citep[see also ][]{Renzini1994} to explain the difference between the luminosity of the RGB bump of the Galactic bulge and that predicted by the luminosity-metallicity relation of GGCs.
\citetalias{Saviane2012} conclude that the cluster-to-cluster scatter in He content should not affect the results of the CaT method, and we refer the reader to their analysis for more details.
We expect similar behavior for  $\Ks{}(HB)$, and the small scatter that the \Carr{} calibrator GCs have around the fitting curves confirms this expectation.

Another potential concern is that, particularly in the differentially reddened clusters, the possibility that AGB stars are included in the selected ``RGB'' samples may lead to additional scatter in the $(\Sigma W , \Ks{} - \Ks{}(HB))$ plane, and thus to increased uncertainty in the derived abundances.
On the other hand this effect is expected to be quite small ($\Delta W=0.04$\AA), as shown by \citet{Cole2000}.

\section{Results}
\label{s:res}

Assuming the relation
\begin{equation}
  \Sigma W=a[\Ks{}(HB)-\Ks{}]+W'\;, \label{eq:W_Ks}
\end{equation}
the slope $a$ is calculated through a least-square fit\footnote{the algorithm is {based} on a series of five lectures presented at ``V Escola Avancada de Astrofisica'' by Peter B. Stetson \url{http://ned.ipac.caltech.edu/level5/Stetson/Stetson_contents.html}}, with the constraint that it must be the same for all the clusters.
To a good approximation, the slope is independent of metallicity within the range spanned by our GCs.

Subsequently, we calculated $W'$ for all stars in each cluster, empirically removing the EW dependence on the star's  gravity and temperature.
We calculated $\rEW$ for each GC, through a weighted average of $W'$ of all stars.
We preferred this approach instead of the intercept value provided by the least-square fit method because, while the two estimates negligibly differ, the uncertainties associated with the fit heavily depend on the number of data, and are thus overestimated.
Finally the [Fe/H] vs. $\rEW$ relation was calculated for each metallicity scale with a polynomial fit to define the best calibration relation.
The unbiased residual mean square (hereafter urms or $\sigma$) was assumed as the uncertainty of our fit.
It consists in the square root of the sum of the square of the residuals, divided by the number of degrees of freedom for error\footnote{The number of data points less the number of the coefficients of the curve used to fit them.}, instead of the number of data points, in order to  remove the bias on the estimate of the variance of the unobserved errors.

\subsection{Calibration of reduced equivalent widths from S12}
\begin{figure}[ht!]\begin{center}
\resizebox{\hsize}{!}{\includegraphics[]{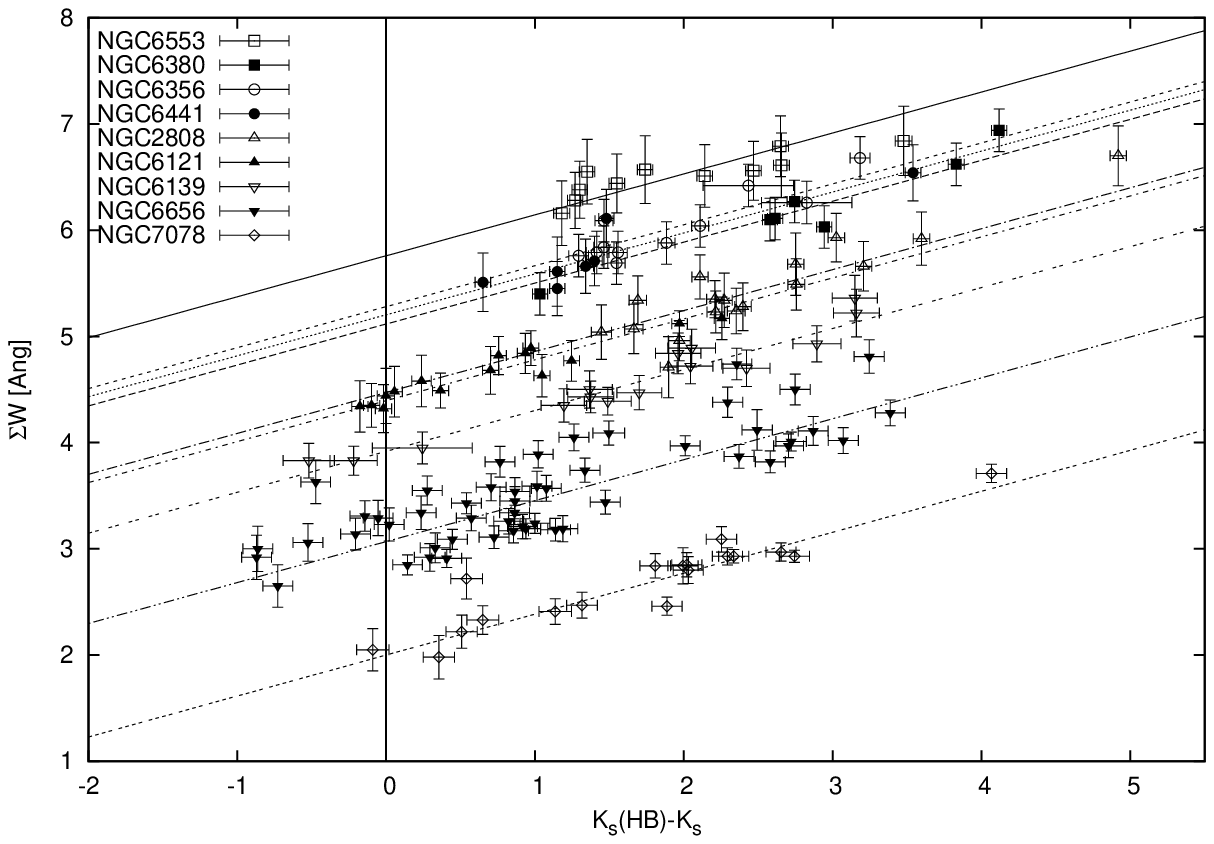}}
\resizebox{\hsize}{!}{\includegraphics[]{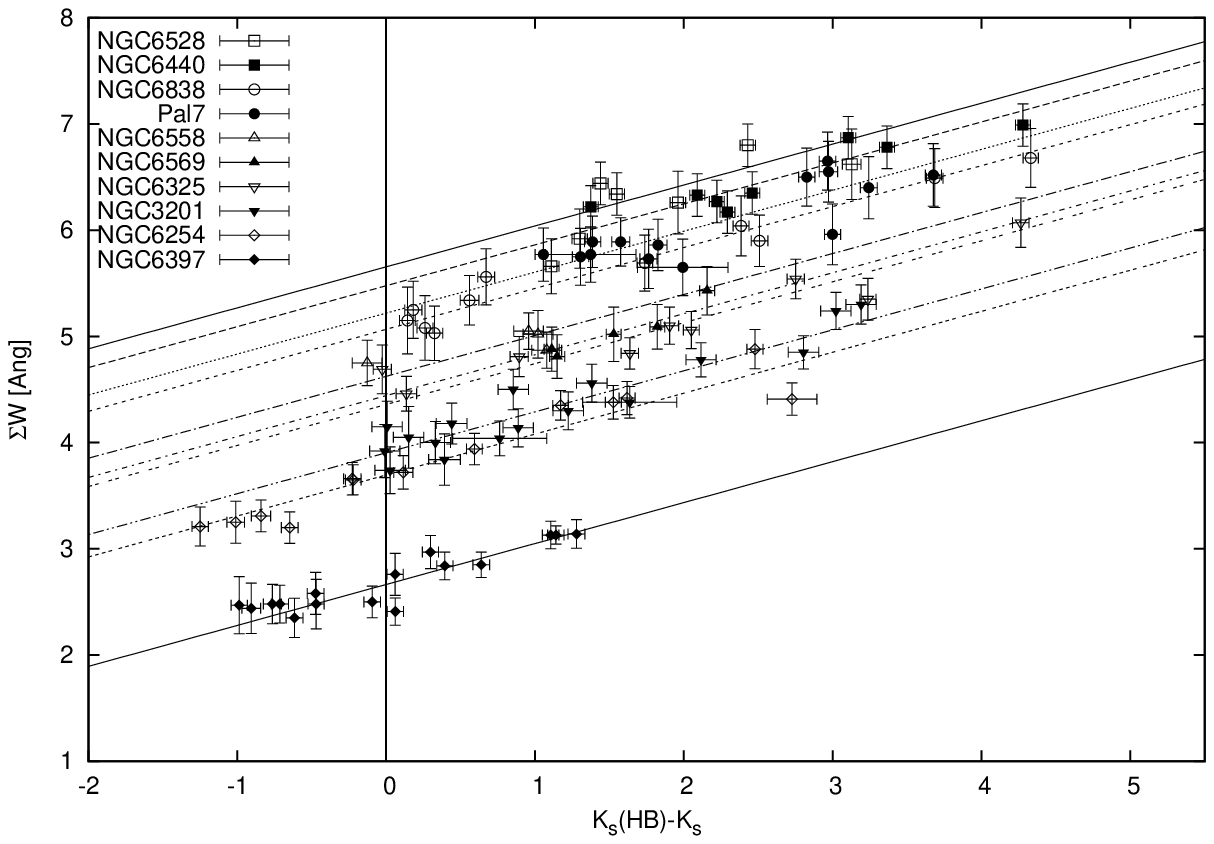}}
\caption{Plot of summed Ca II line strength $\sum W_{S12}$ against magnitude difference from the red clump $\Ks{}(HB)-\Ks{}$ for the selected clusters.
The data for each cluster has been fit with a line of slope $-0.385$ \AA{}/mag with a urms of 0.013 \AA{}/mag. The data are split to avoid too much overlap.}
\label{fig:W_DKs_S}
\end{center}\end{figure}

The $\Sigma W_{S12}$ values are plotted vs. $\Ks{}(RHB)-\Ks{}$ in Figure~\ref{fig:W_DKs_S}. 
For the slope we obtain \mbox{$a=-0.385$~\AA{}/mag} with a urms of 0.013~\AA{}/mag.
The  $\rEW$ cover a range of 2-5.8~\AA.

\begin{figure}[ht!]\begin{center}
\resizebox{\hsize}{!}{\includegraphics[]{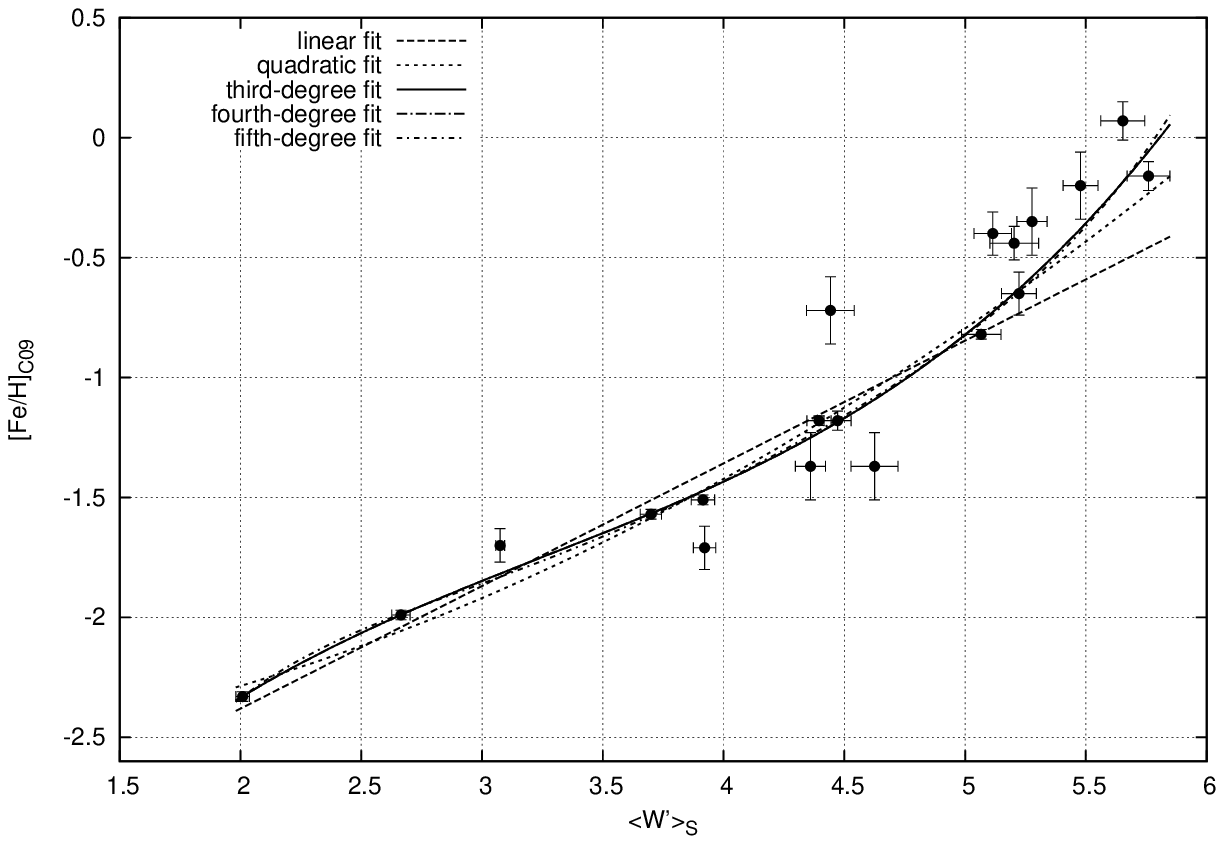}}
\caption{The calibration relations among [Fe/H] on the C09 scale and the nIR $\rEW$ on the G09 scale.}
\label{fig:Fe_W_SC09}
\end{center}\end{figure}

The calibration relations for [Fe/H] on the \citetalias{Carretta2009} scale and the nIR $\rEW$ on the \citetalias{Gullieuszik2009} scale are shown in Figure~\ref{fig:Fe_W_SC09}, defined in a [Fe/H] range from $-2.33$~dex to $+0.07$~dex.
As can be seen in the Figure, both the cubic and the quadratic relations well reproduce the observed trend, while the linear one does not fit properly the more metal-rich GCs.
The fourth- and fifth-order polynomials do not differ noticeably from the third-order one.
The cubic calibration relation is
\begin{equation}\label{eq:Fe_W_3SC09}
\mathrm{[Fe/H]_{C09}}=-4.61 +1.842\rEW -0.4428\rEW^2 +0.04517\rEW^3 ,
\end{equation} 
with a urms of 0.214 dex (0.113 dex considering only the 8 GCs used as calibrators in \Carr{}).
The quadratic calibration relation
\begin{equation}\label{eq:Fe_W_2SC09}
\mathrm{[Fe/H]_{C09}}= -2.63+0.040\rEW +0.0653\rEW^2 ,
\end{equation} 
with a urms of 0.237 dex (0.110 dex considering only the 8 calibrators).
The differences in metallicity between the cubic and the quadratic fit are less than 0.1 dex in absolute value in the range of definition ($2<\rEW<5.5$, $-2.3<\mathrm{[Fe/H]}<-0.3$), increasing up to 0.35 for $\rEW=6$ ($\FeH{}\sim 0.1$).

\begin{figure}[ht!]\begin{center}
\resizebox{\hsize}{!}{\includegraphics[]{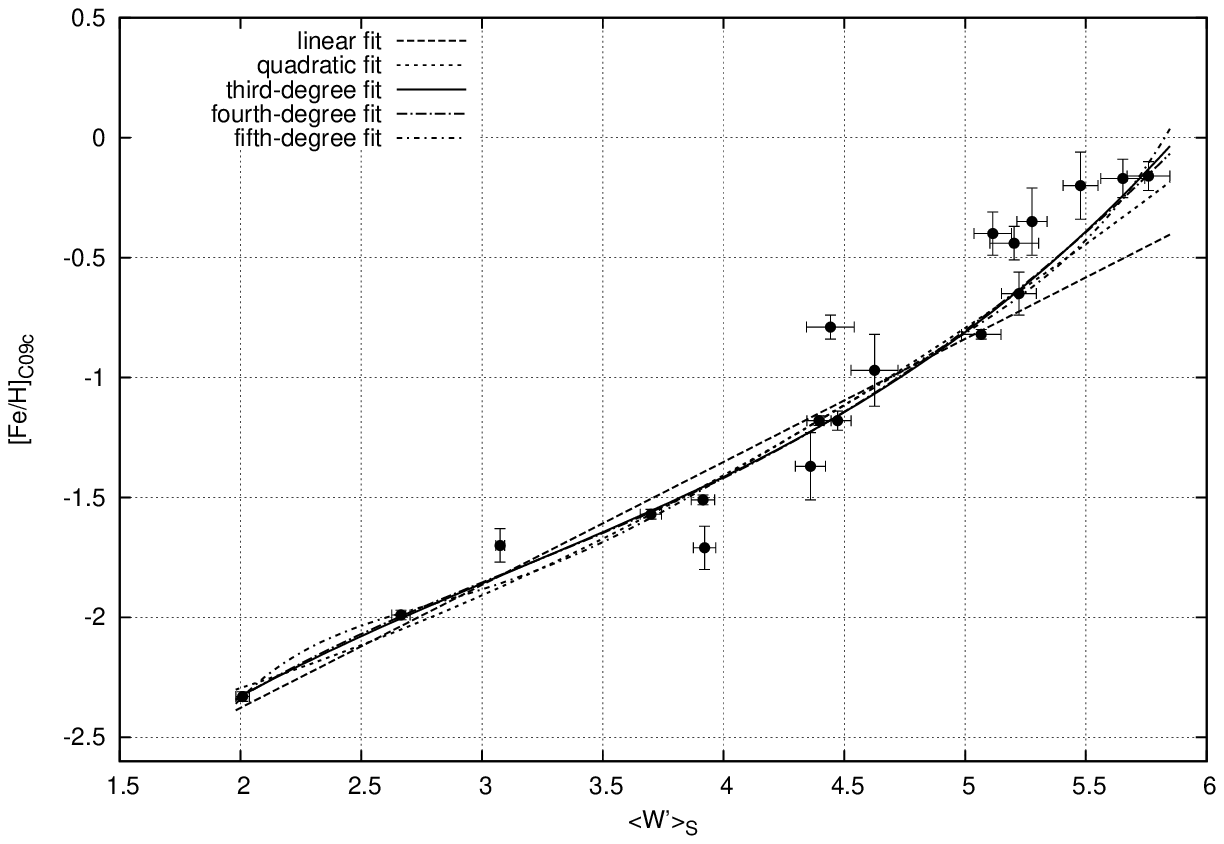}}
\caption{The calibration relations among [Fe/H] on the ``corrected C09'' scale and the nIR $\rEW$ on the G09 scale.}
\label{fig:Fe_W_SC09c}
\end{center}\end{figure}

The calibration relations for [Fe/H] on the C09c scale and the nIR $\rEW$ on the G09 scale are shown in Figure~\ref{fig:Fe_W_SC09c}.
They are defined over an [Fe/H] range from $-2.33$~dex to $-0.17$~dex.
The differences between the cubic and quadratic calibration relations are slightly smaller than for C09 scale.
The cubic calibration relation in this revised system is 
\begin{equation}\label{eq:Fe_W_3SC09c}
\mathrm{[Fe/H]_{C09c}}=-4.09+1.341\rEW -0.2919\rEW^2 +0.03098\rEW^3 ,
\end{equation} 
with a urms of 0.173 dex (0.114 dex considering only the 8 calibrators).
The quadratic calibration relation
\begin{equation}\label{eq:Fe_W_2SC09c}
\mathrm{[Fe/H]_{C09c}}=-2.73+0.103\rEW +0.0568 \rEW^2 ,
\end{equation} 
with a urms of 0.180 dex (0.110 dex considering only the 8 calibrators).

\begin{figure}[ht!]\begin{center}
\resizebox{\hsize}{!}{\includegraphics[]{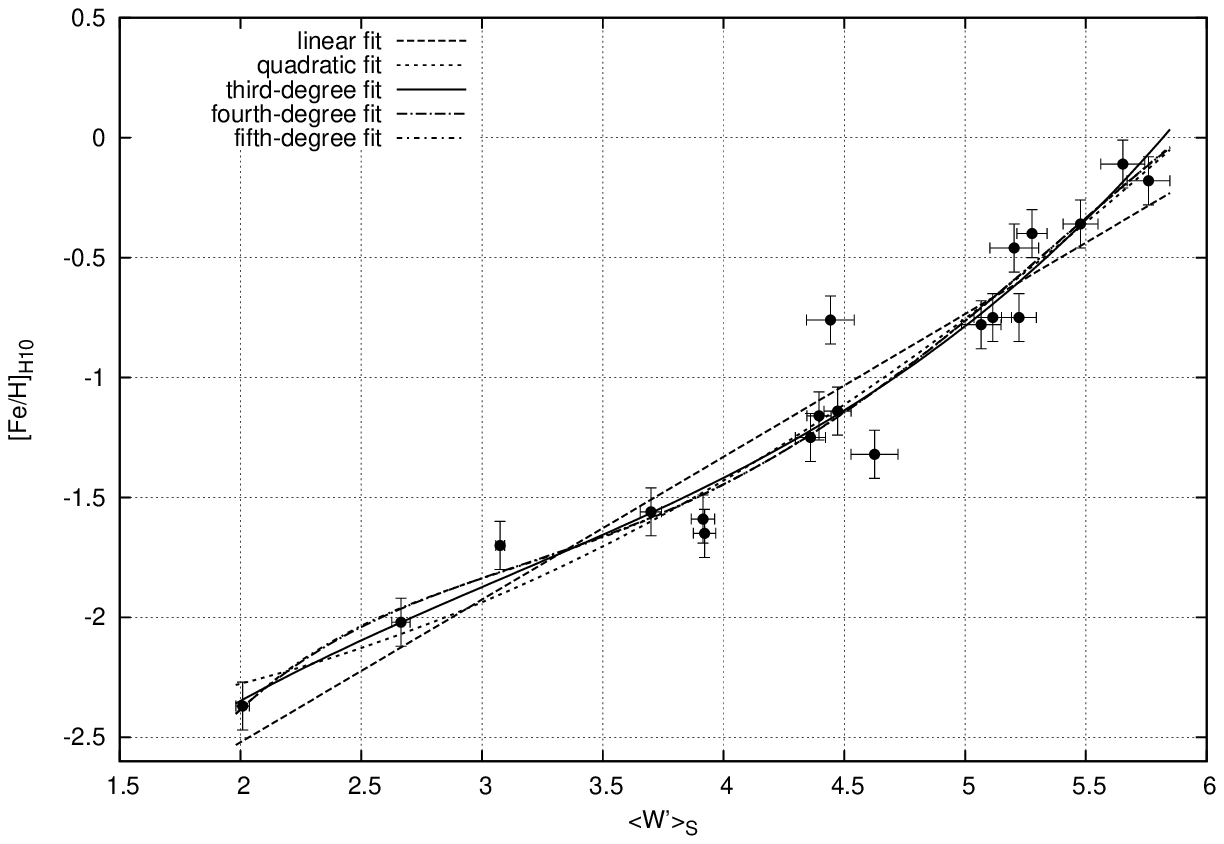}}
\caption{The calibration relations among [Fe/H] on the H10 scale and the nIR $\rEW$ on the G09 scale.}
\label{fig:Fe_W_SH10}
\end{center}\end{figure}

The calibration relations for [Fe/H] on the H10 scale and the nIR $\rEW$ on the G09 scale are shown in Figure~\ref{fig:Fe_W_SH10}.
They are defined over an [Fe/H] range from $-2.37$~dex to $-0.11$~dex.
The differences between the cubic and quadratic calibration relations are slightly smaller than for C09 scales.
The cubic calibration relation is
\begin{equation}\label{eq:Fe_W_3SH10}
\mathrm{[Fe/H]_{H10}}=-4.14 +1.371\rEW -0.3032\rEW^2 +0.03263\rEW^3 ,
\end{equation} 
with a urms of 0.165 dex.
The quadratic calibration relation is
\begin{equation}\label{eq:Fe_W_2SH10}
\mathrm{[Fe/H]_{H10}}=-2.45 -0.080\rEW +0.0839 \rEW^2 ,
\end{equation} 
with a urms of 0.165 dex.

\subsubsection{Comparison between reference scales}

\begin{figure}[ht!]\begin{center}
\resizebox{\hsize}{!}{\includegraphics[]{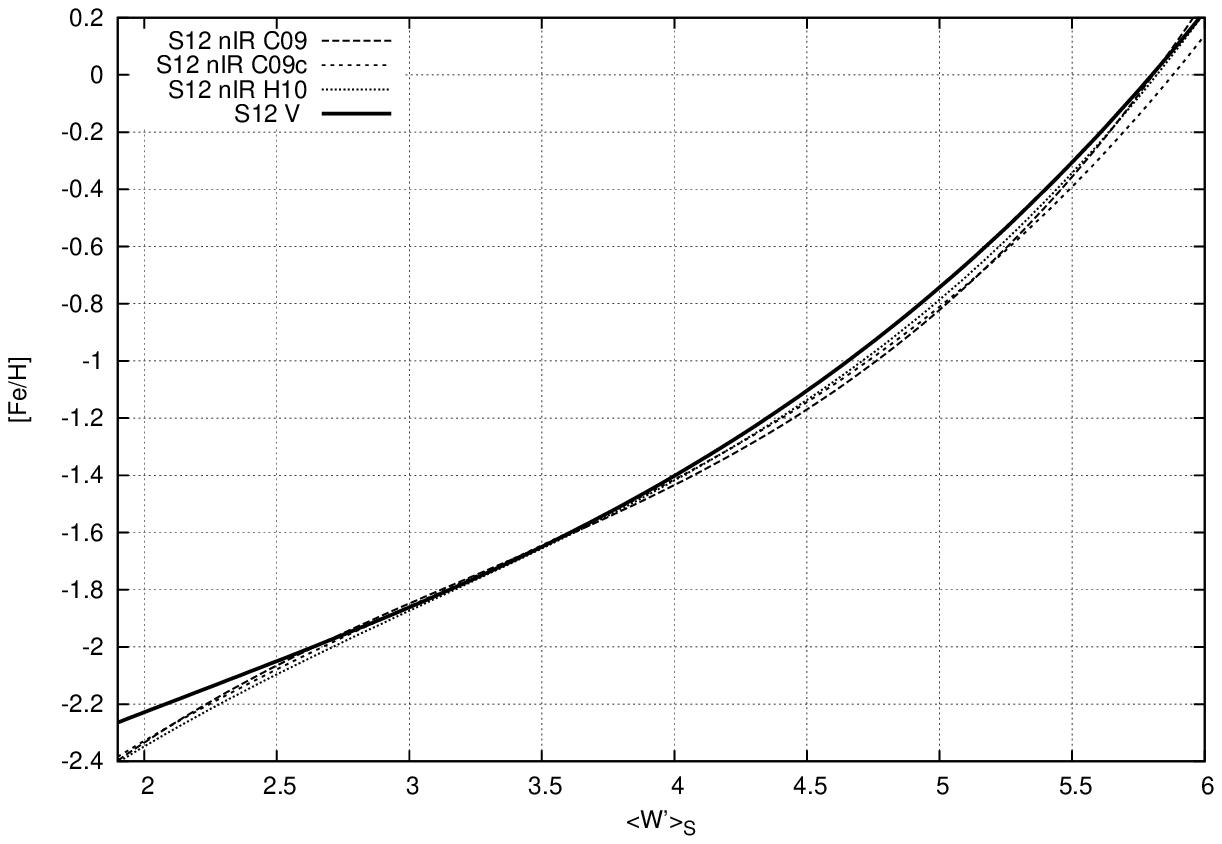}}
\caption{Comparison among the cubic calibration relations [Fe/H] vs nIR $\rEW$ on the G09 scale in the three metallicity scale and the one obtained in \Sav{} for the visible.}
\label{fig:Fe_W_Scomp}
\end{center}\end{figure}

We compared the three calibration relations for the \Sav{} dataset obtained in this study with the one calculated at optical wavelengths in \Sav{}.
The cubic and the quadratic fits do not differ noticeably, and \Sav{} presented a cubic relation, hence the 3rd-order solution was adopted in the comparison.
As can be seen in Figure~\ref{fig:Fe_W_Scomp}, the four relations are approximately the same, with differences less than 0.1 dex over the whole metallicity range.
The three relations from this work differ among each other generally less than $0.02-0.04$~dex, with a maximum value of 0.08~dex between the solutions C09 and C09c at the metal-rich end.
This demonstrates that the three scales are approximately the same, and the smallest values of urms could indicate  a more precise calibration.
This overall comparison demonstrates that the photometric correction to the CaT equivalent widths is independent of the passband used, since the same calibration relation can be adopted when using both optical and NIR magnitudes.
Additionally, the good agreement over the whole metallicity range demonstrates that the HB level was determined with sufficient accuracy in our work.

Comparing the \Carr{} metallicities with those obtained from our equation (see Table~\ref{tab:SC09}), 6 of the 19 GCs present a difference greater than $1\sigma$.
However, the metallicity of three of these is controversial, and is poorly known for the other two.
Only NGC\,6569 among these GCs presents a difference greater than $2\sigma$.
We derived a metallicity of $-1.20$~dex for it, similar to those of NGC\,2808 and NGC\,6121, versus the \Carr{} value of $-0.72$~dex.
We get $\FeH{}=-0.58$~dex (instead of $-0.35$~dex) for NGC\,6356, $-0.19$~dex  (instead of $+0.07$~dex) for NGC\,6528, and $-1.09$~dex (instead of $-1.37$~dex) for NGC\,6558.
For the two GCs with poorly known metallicity, the equation gives $-1.47$~dex for NGC\,6139 (\Carr{} suggests $-1.71$~dex), and $-0.73$~dex for NGC\,6380 ($-0.40$~dex).
The statistic is similar in the comparison with the C09c scale (see Table~\ref{tab:SC09c}): 6 GCs with $|\Delta\FeH{}| >1\sigma$, with NGC\,6569 still the only one with $|\Delta\FeH{}| >2\sigma$.
We obtain now a difference smaller than $1\sigma$ for NGC\,6528 and NGC\,6558, while the discrepancy for NGC\,6440 and NGC\,6441 is now greater than before: the equation gives $-0.41$~dex (against the \Carr{} value of $-0.20$~dex), and $-0.65$~dex ($-0.44$~dex), respectively.
We note that, in these two \Carr{}-based scales, NGC\,6441 is the only one of the 8 \Carr{} calibrators that presents a $\Delta\FeH{} >0.3\sigma$.
For the \Har{} scale (see Table~\ref{tab:SH10}), only four of the 19 GCs present $|\Delta\FeH{}| >1\sigma$, namely NGC\,6139, NGC\,6558, NGC\,6569, and NGC\,6656 ($-1.90$~dex instead of $-1.70$~dex).

\subsection{Calibration of reduced equivalent widths from R97}

\begin{figure}[ht!]\begin{center}
\resizebox{\hsize}{!}{\includegraphics[]{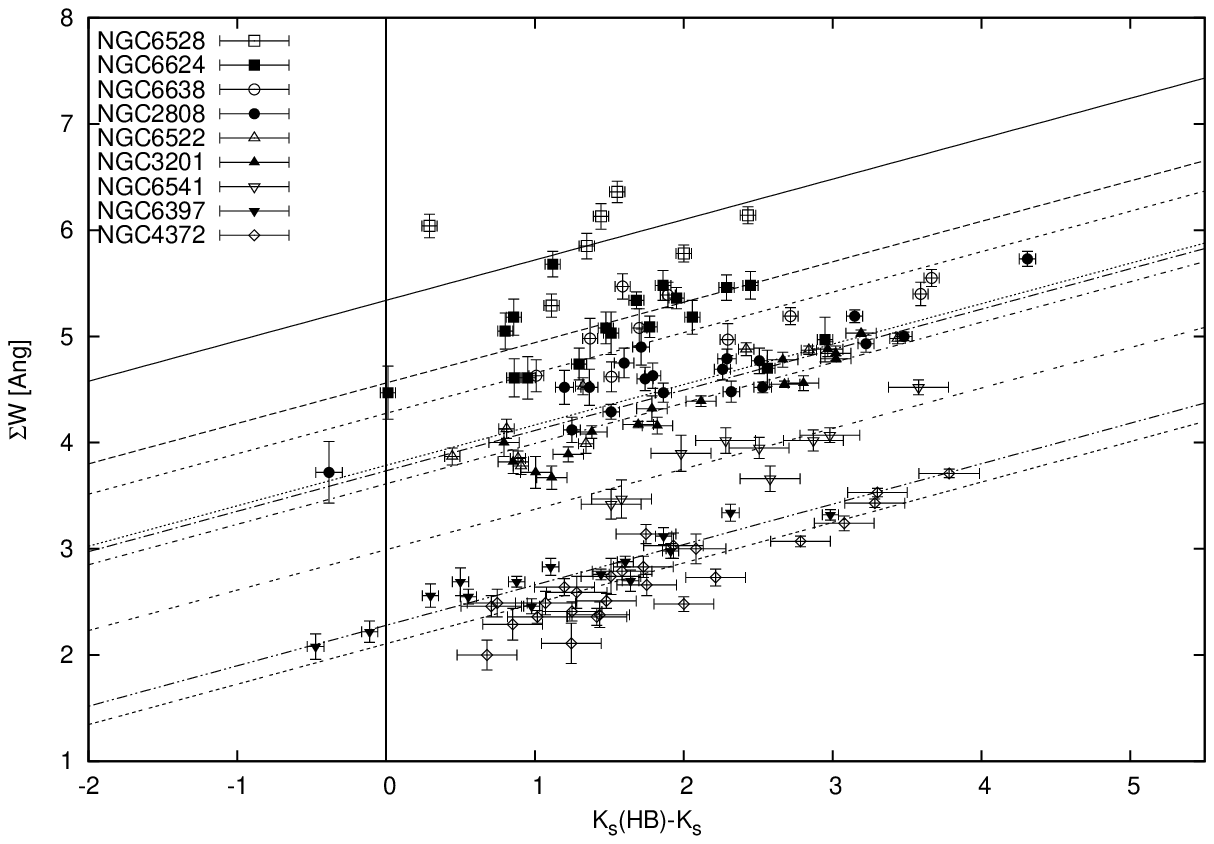}}
\resizebox{\hsize}{!}{\includegraphics[]{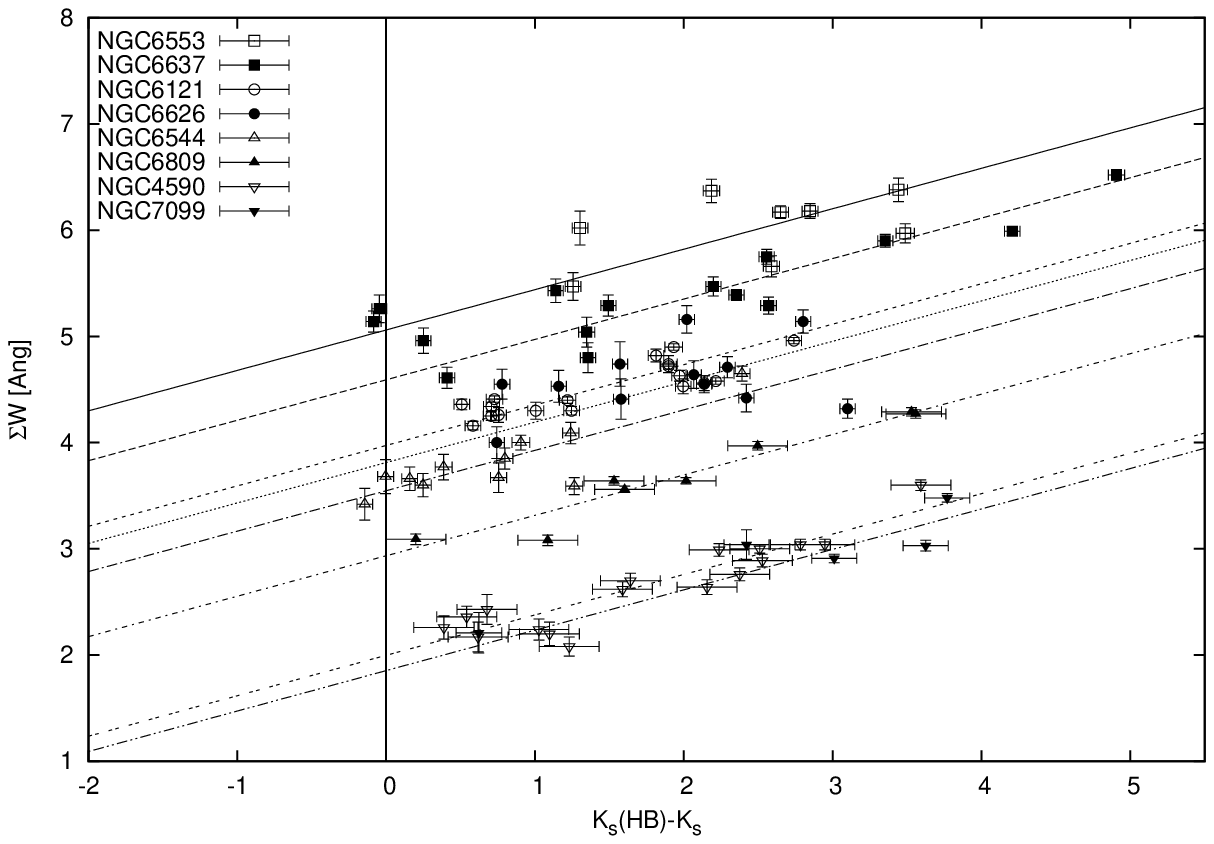}}
\caption{Plot of summed Ca II line strength ($\sum W_{R97}$) against magnitude difference from the red clump $\Ks{}(RHB)-\Ks{}$ for the selected clusters. The data for each cluster has been fit with a line of slope $-0.348$ \AA{}/mag. The data are split to avoid too much overlap.}
\label{fig:W_DKs.R97}
\end{center}\end{figure}

The $\Sigma W_{R97}$ values are plotted vs. $\Ks{} - \Ks{RHB}$ in Figure~\ref{fig:W_DKs.R97}.
For the \Rut{} dataset we obtain  \mbox{$a=-0.380$~\AA{}/mag} with a urms of 0.014~\AA{}/mag.
This value differs from the one for \Sav{} by only 0.005~\AA{}/mag ($\approx 0.4\sigma$), consequently the slope is independent of the way the equivalent width is calculated.
Such variation would cause a change in the mean rEW less than 0.02~\AA{} in our case, which is completely negligible.
The  $\rEW$ cover a range of 1.6-5.1~\AA.

\begin{figure}[ht!]\begin{center}
\resizebox{\hsize}{!}{\includegraphics[]{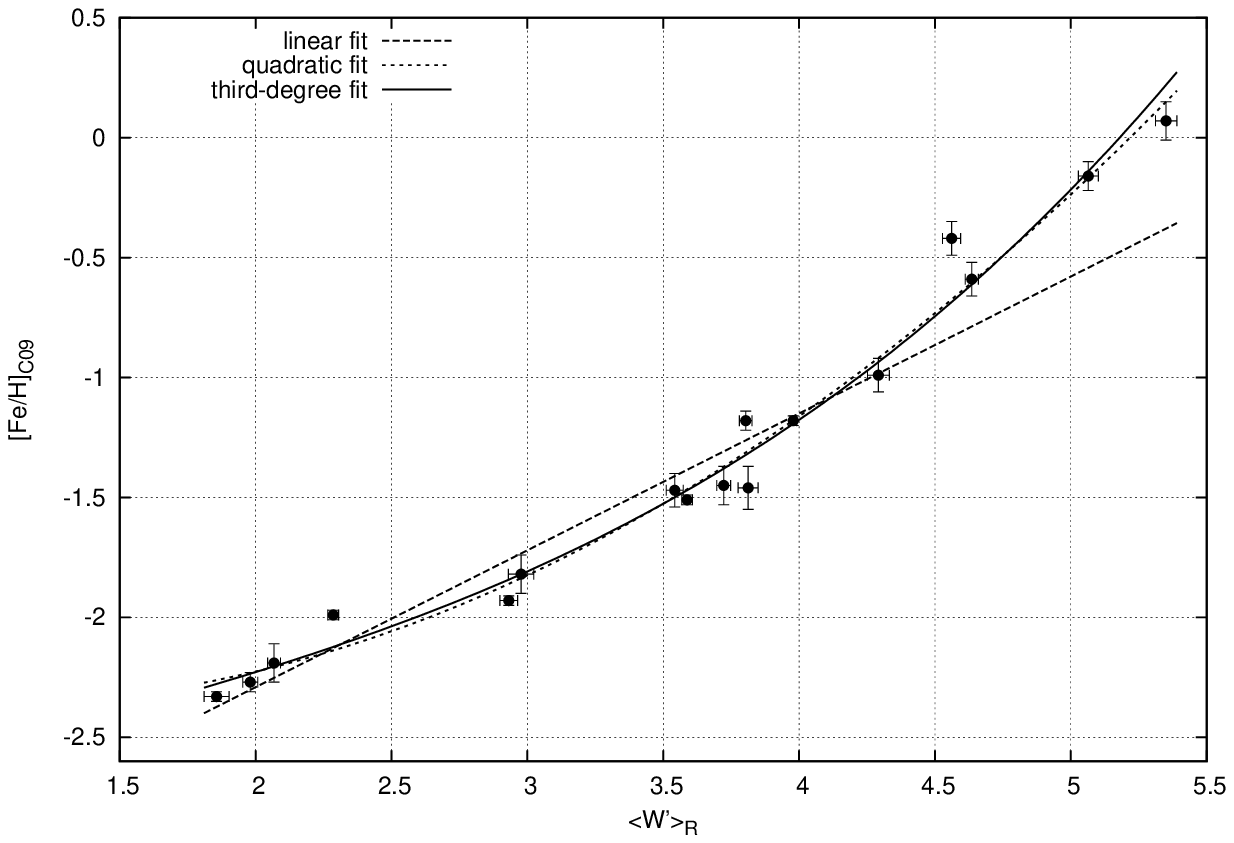}}
\caption{The calibration relations among [Fe/H] on the C09  scale and the nIR $\rEW$ on the R97 scale.}
\label{fig:Fe_W_RC09}
\end{center}\end{figure}

The calibration relations for [Fe/H] on the \citetalias{Carretta2009} scale and the nIR $\rEW$ are shown in Figure~\ref{fig:Fe_W_RC09}, defined over an [Fe/H] range from $-2.33$~dex to $+0.07$~dex.
The calibrations based on a fourth- and fifth-order polynomial are not suitable, and were consequently rejected.
The best fits are obtained with the cubic and quadratic relations, while the linear one does not fit properly the more metal-rich GCs, as observed also for the S12 dataset, or the more metal-poor GCs.
The quadratic calibration relation is
\begin{equation}\label{eq:Fe_W_2RC09}
\mathrm{[Fe/H]_{C09}}= -2.24 -0.254\rEW +0.13094  \rEW^2 ,
\end{equation} 
with a urms of 0.102 dex (0.114 dex considering only the 7 GCs used as calibrators in \Carr{}), while the cubic calibration relation is
\begin{equation}\label{eq:Fe_W_3RC09}
\mathrm{[Fe/H]_{C09}}=-2.90+0.393\rEW -0.0684\rEW^2 +0.01939 \rEW^3 ,
\end{equation} 
with a urms of 0.113 dex (0.130 dex considering only the 7 calibrators).
The differences in metallicity between the cubic and the quadratic fits are less than $0.06$~dex in absolute values in almost all the range of definition ($1.7<\rEW<5.0$, $-2.2<\mathrm{[Fe/H]}<-0.3$), increasing up to 0.2 for $\rEW=5.3$ ($\mathrm{[Fe/H]}\sim 0.0$).

\begin{figure}[ht!]\begin{center}
\resizebox{\hsize}{!}{\includegraphics[]{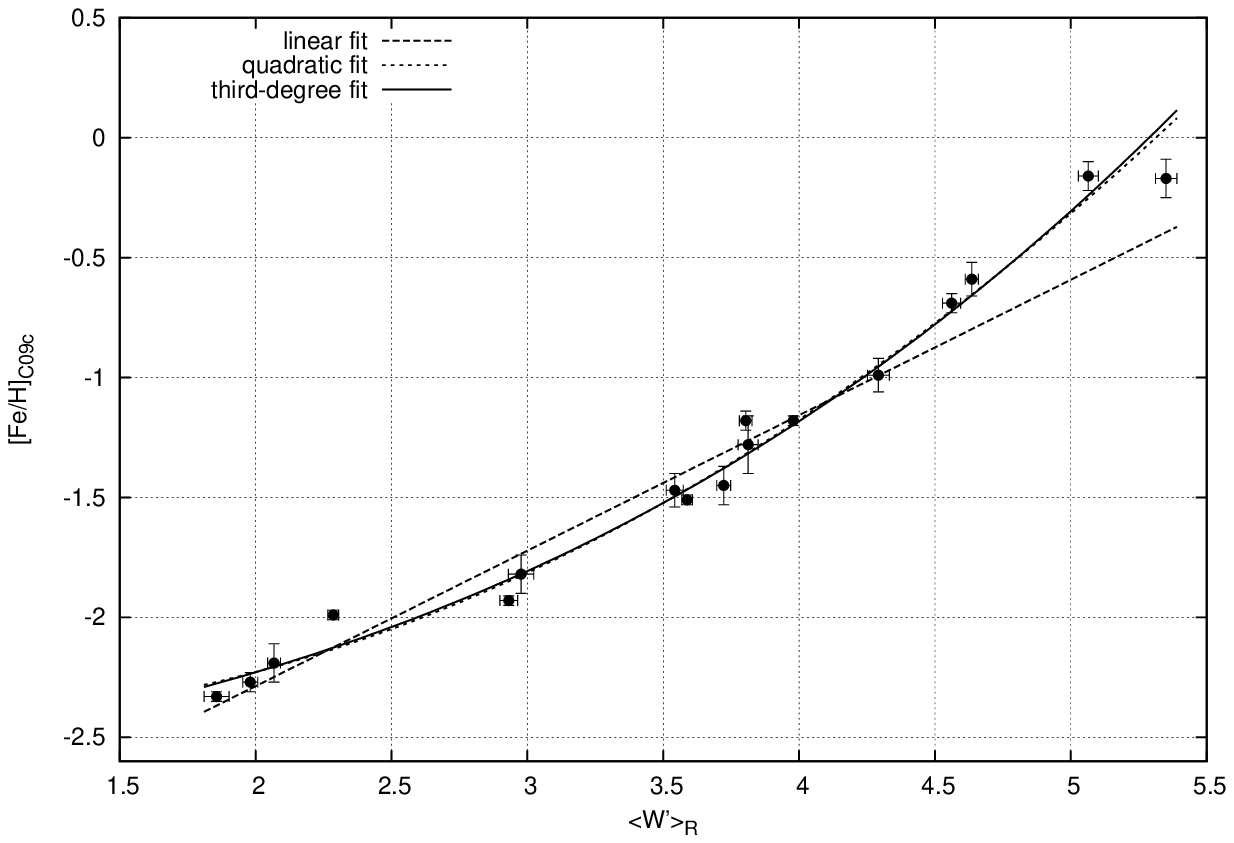}}
\caption{The calibration relations among [Fe/H] on the ``corrected C09'' scale and the nIR $\rEW$ on the R97 scale.}
\label{fig:Fe_W_RC09c}
\end{center}\end{figure}

The calibration relations for [Fe/H] on the C09c scale and the nIR $\rEW$ are shown in Figure~\ref{fig:Fe_W_RC09c}.
They are defined over an [Fe/H] range from $-2.33$~dex to $-0.16$~dex.
The differences between the cubic and quadratic calibration relations are slightly smaller than for the \Carr{} scale.
The quadratic calibration relation is
\begin{equation}\label{eq:Fe_W_2RC09c}
\mathrm{[Fe/H]_{C09c}}=-2.39-0.143\rEW +0.1116\rEW^2 ,
\end{equation} 
with a urms of 0.096 dex (0.115 dex considering only the 7 calibrators).
The cubic calibration relation is
\begin{equation}\label{eq:Fe_W_3RC09c}
\mathrm{[Fe/H]_{C09c}}= -2.66 +0.125\rEW +0.0285\rEW +0.00809\rEW^3 ,
\end{equation} 
with a urms of 0.104 dex (0.133 dex considering only the 7 calibrators).

\begin{figure}[ht!]\begin{center}
\resizebox{\hsize}{!}{\includegraphics[]{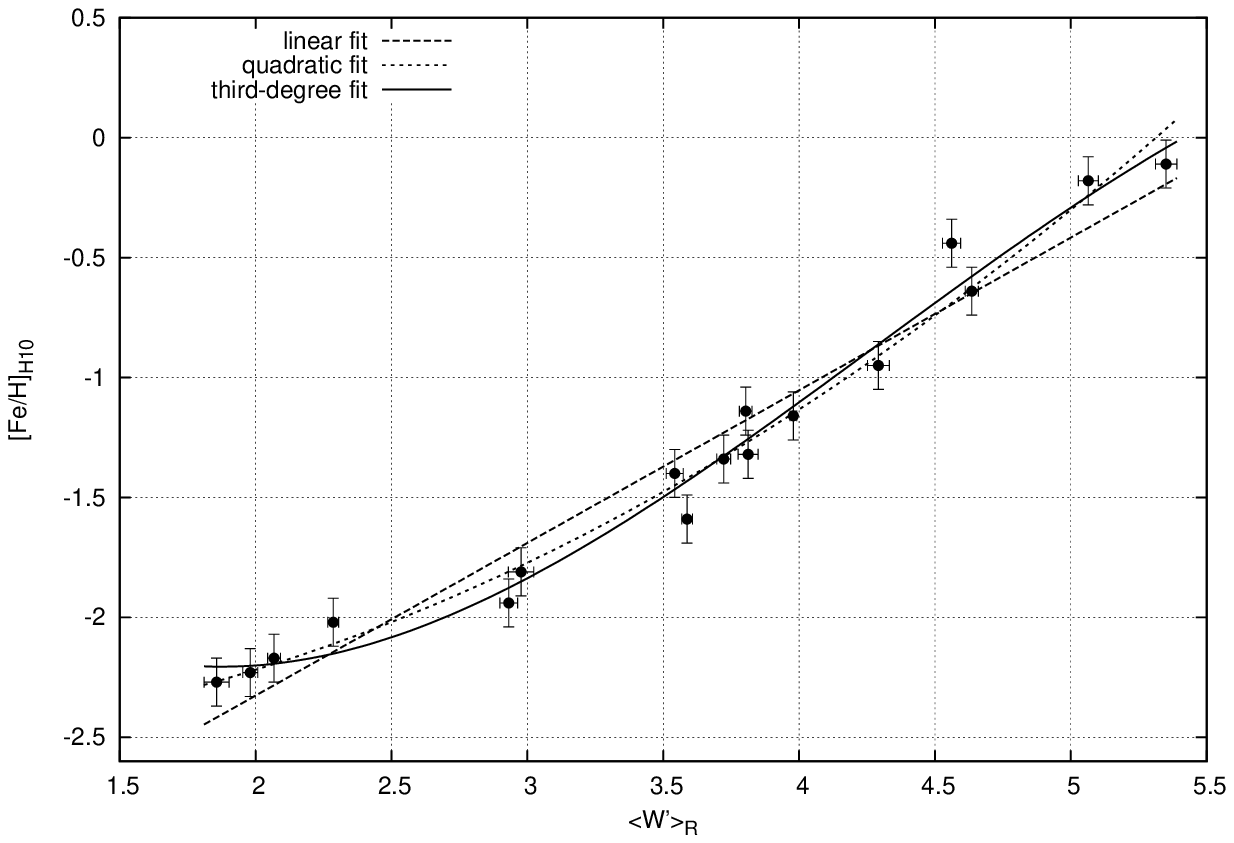}}
\caption{The calibration relations among [Fe/H] on the H10 scale and the nIR $\rEW$ on the R97 scale.}
\label{fig:Fe_W_RH10}
\end{center}\end{figure}

The calibration relations for [Fe/H] on the H10 scale and the nIR $\rEW$ are shown in Figure~\ref{fig:Fe_W_RH10}.
They are defined over an [Fe/H] range from $-2.27$~dex to $-0.11$~dex.
The differences between the cubic and quadratic calibration relations are less than 0.06~dex in the whole range.
The quadratic calibration relation is
\begin{equation}\label{eq:Fe_W_2RH10}
\mathrm{[Fe/H]_{H10}}=-2.53 -0.037\rEW +0.09658 \rEW^2 
\end{equation} 
has a urms dispersion around the fit of 0.110 dex.
The cubic calibration relation is
\begin{equation}\label{eq:Fe_W_3RH10}
\mathrm{[Fe/H]_{H10}}=-0.66-1.821\rEW +0.6211\rEW^2 -0.04848\rEW^3
\end{equation} 
has a urms dispersion around the fit of 0.103 dex.

\subsubsection{Comparison between the reference scales}

\begin{figure}[ht!]\begin{center}
\resizebox{\hsize}{!}{\includegraphics[]{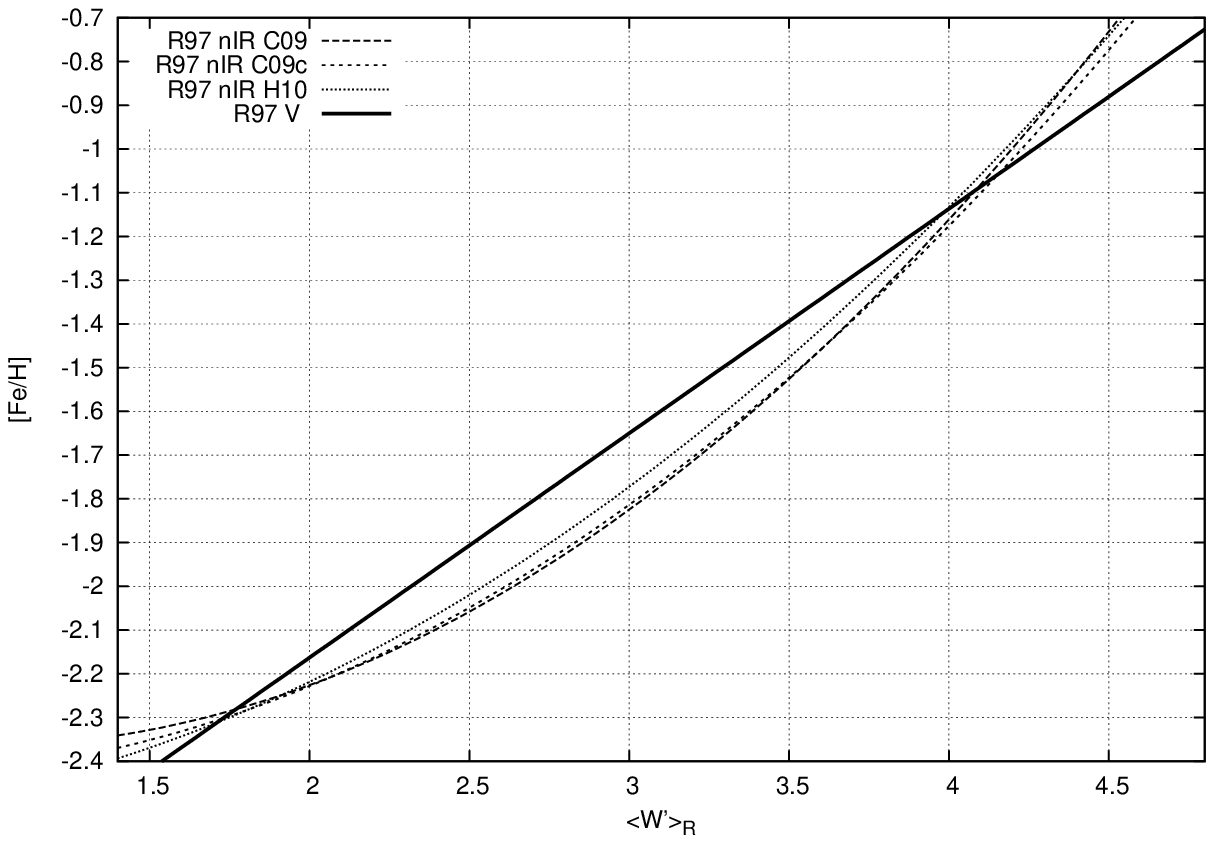}}
\resizebox{\hsize}{!}{\includegraphics[]{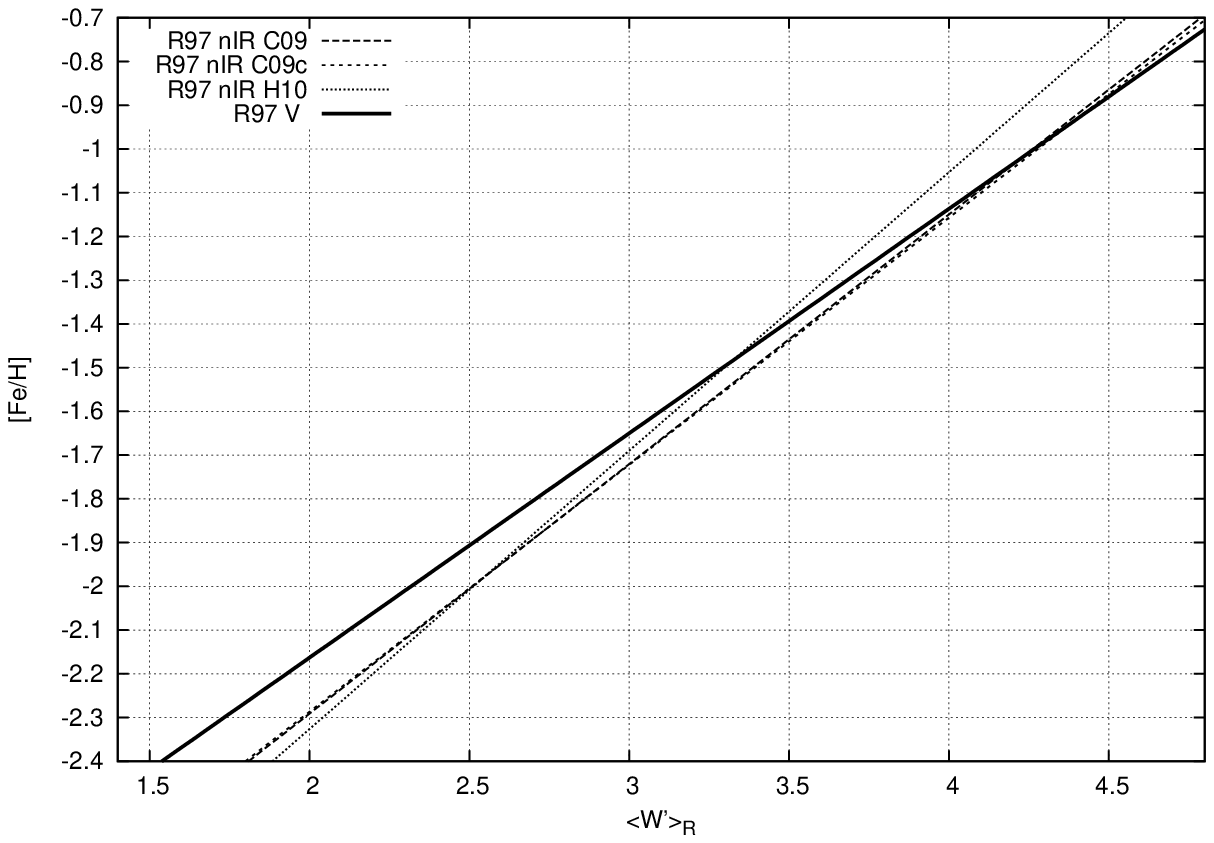}}
\caption{Comparison among the quadratic (upper figure) and linear (lower figure) calibration relations [Fe/H] vs nIR $\rEW$ on the three metallicity scales and the one obtained by \citetalias{Carretta2009} for \Rut{} for the visible.}
\label{fig:Fe_W_Rcomp}
\end{center}\end{figure}

For the comparison between the three metallicity scales, we chose to use the quadratic relations, since they are the ones with the lower urms values.
Our three relations are approximately the same, even for this dataset, as can be seen in Figure~\ref{fig:Fe_W_Rcomp}.
The differences are negligible, being less than $\sim 0.03$~dex over most of the metallicity range, with a maximum value of 0.08~dex at the more metal-rich GCs.
The H10 relation presents a systematic over-estimate of $\sim 0.04$~dex compared to the other two.

The linear relation between the V-band rEW and [Fe/H] presented by \citetalias{Carretta2009} was defined in the ranges $1.5<\rEW<4.7$ and $-2.33<\mathrm{[Fe/H]}<-0.7$, thus excluding the most metal-rich GCs.
Anyway, as can be noted in Figures~\ref{fig:Fe_W_RC09}-\ref{fig:Fe_W_RH10}, in this metallicity range the difference among the cubic and quadratic fits and the linear one is $\lesssim 1\sigma$.
As a matter of fact, if we recalculate the urms for our linear fits in the same metallicity range, we obtain results more similar to the urms of the quadratic equation (0.116, 0.099, and 0.149~dex, respectively for the three scales), showing that a non-linear fit is needed including also the metal-rich end.
We compared both the quadratic and the linear calibration relations for the nIR rEW with the one from \Carr{}.
While the quadratic relations present an obvious difference with the optical one, the comparison among the four linear relations (see lower plot in Fig. \ref{fig:Fe_W_Rcomp}) confirms that the ([Fe/H] vs rEW) relation can be considered almost independent of the passband used for the photometry.

Comparing the \Carr{} metallicities with the ones obtained from our equations (see Table~\ref{tab:RC09}), the quadratic equation yields a metallicity different by more than $1\sigma$ for four GCs, namely  NGC\,2808, NGC\,6397, NGC\,6624, NGC\,6626.
The problems related to the controversial metallicity of the last two GCs were already discussed in Section~\ref{ss:met}.
The cubic solution gives a difference  $\Delta\FeH{} >1\sigma$ even for NGC\,6528.
NGC\,6624 is only cluster with $\Delta\FeH{} > 2\sigma$, with an estimated metallicity of $-0.67$~dex, against the \Carr{} value of $-0.42$~dex.
We obtain $\FeH{}=-1.31$~dex for NGC\,2808, while \Carr{} gives $-1.18$~dex and \Sav{} equation $\sim -1.14$~dex, $-2.14$~dex  for NGC\,6397, versus $-1.99$~dex, for both \Carr{} and \Sav{} equation, and $-1.31$~dex  for NGC\,6626, instead of $-1.46$~dex.
Considering the six GCs in common with \Sav{} dataset, the metallicity obtained for NGC\,3201 and NGC\,6121 are in perfect agreement, while NGC\,6553 presents a difference $\Delta\FeH{} \approx 1\sigma$.
NGC\,6528 shows an even larger discrepancy, but its metallicity is  controversial and, along with NGC\,6553, it is sensitive to the fit uncertainties at high metallicities. 
Considering the C09c scale (see Table~\ref{tab:RC09c}), we have only three GCs with $|\Delta\FeH{}| >1\sigma$, namely NGC\,2808, NGC\,6397 and NGC\,6528.
This last is the only one with $|\Delta\FeH{}| >2\sigma$, with a resulting $\FeH{}=+0.04$~dex, against the adopted value of $-0.17$~dex.
The equation for the \Har{} scale (see Table~\ref{tab:RH10}) presents more discrepancies, since five clusters show $|\Delta\FeH{}| >1\sigma$, namely NGC\,2808, NGC\,3201 ($-1.42$~dex instead of $-1.59$~dex), NGC\,6528 ($+0.04$~dex instead of $-0.11$~dex), NGC\,6624 (again with $\Delta\FeH{} >2\sigma$), and NGC\,6809 ($-1.81$~dex instead of $-1.94$~dex).
The  equations from the \Rut{} dataset presented more discrepancies in the derived metallicities for the \Carr{} calibrators with respect to the \Sav{} data, maybe due to the method used to determine the EWs, or to larger uncertainties in the measured EWs.

\subsection{Metallicity Distributions}
\label{ss:metdistr}

\begin{figure}[ht!]\begin{center}
\resizebox{\hsize}{!}{\includegraphics[]{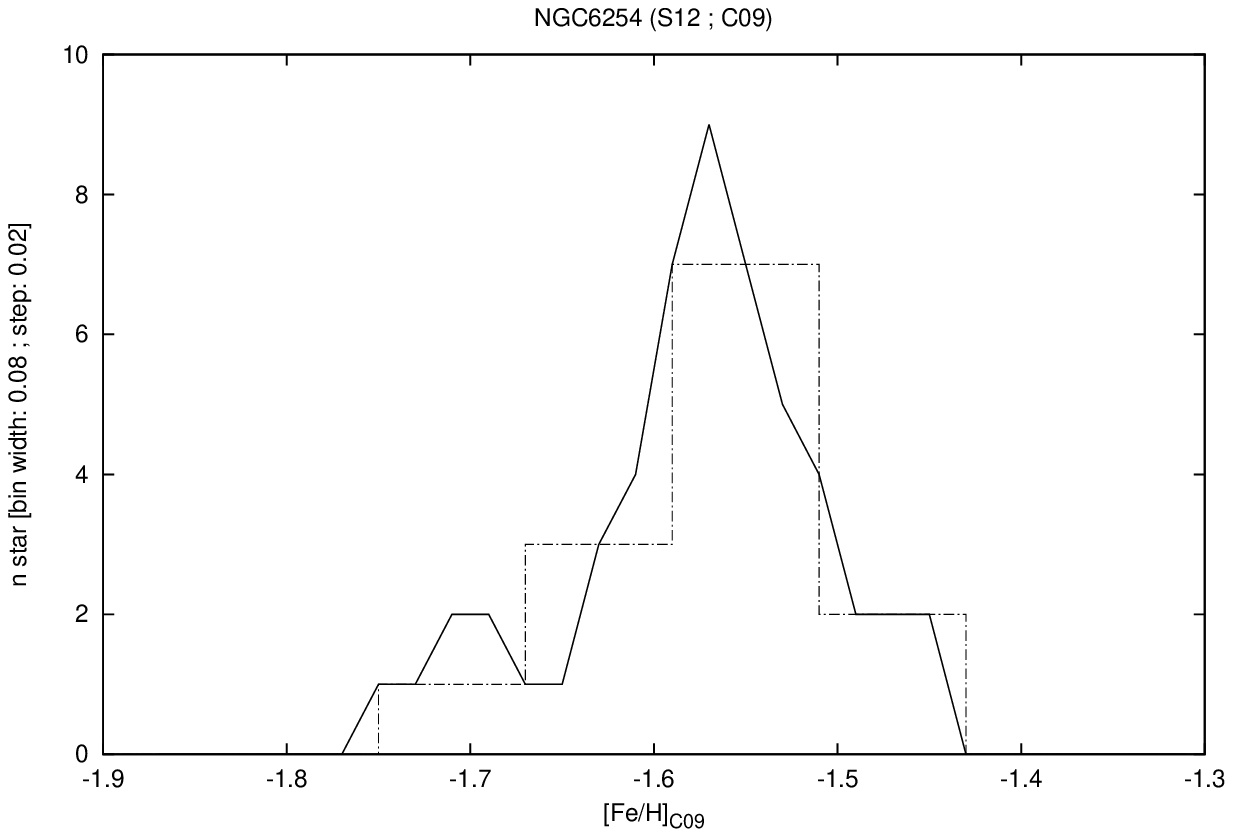}}
\caption{Distribution in metallicity on the C09 scale for NGC\,6254 stars in the S12 sample.}
\label{fig:6254}
\end{center}\end{figure}

\begin{figure}[ht!]\begin{center}
\resizebox{\hsize}{!}{\includegraphics[]{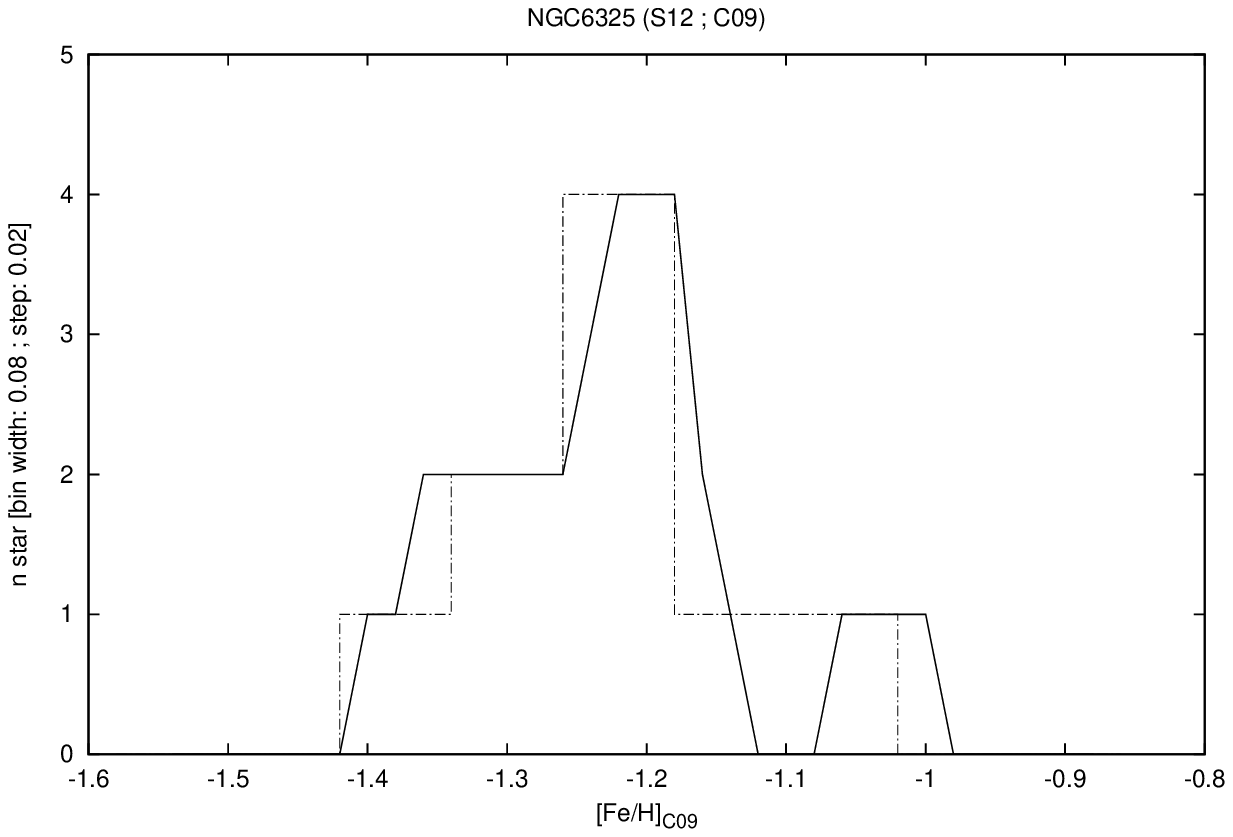}}
\caption{Distribution in metallicity on the C09 scale for NGC\,6325 stars in the S12 sample.}
\label{fig:6325}
\end{center}\end{figure}

\begin{figure}[ht!]\begin{center}
\resizebox{\hsize}{!}{\includegraphics[]{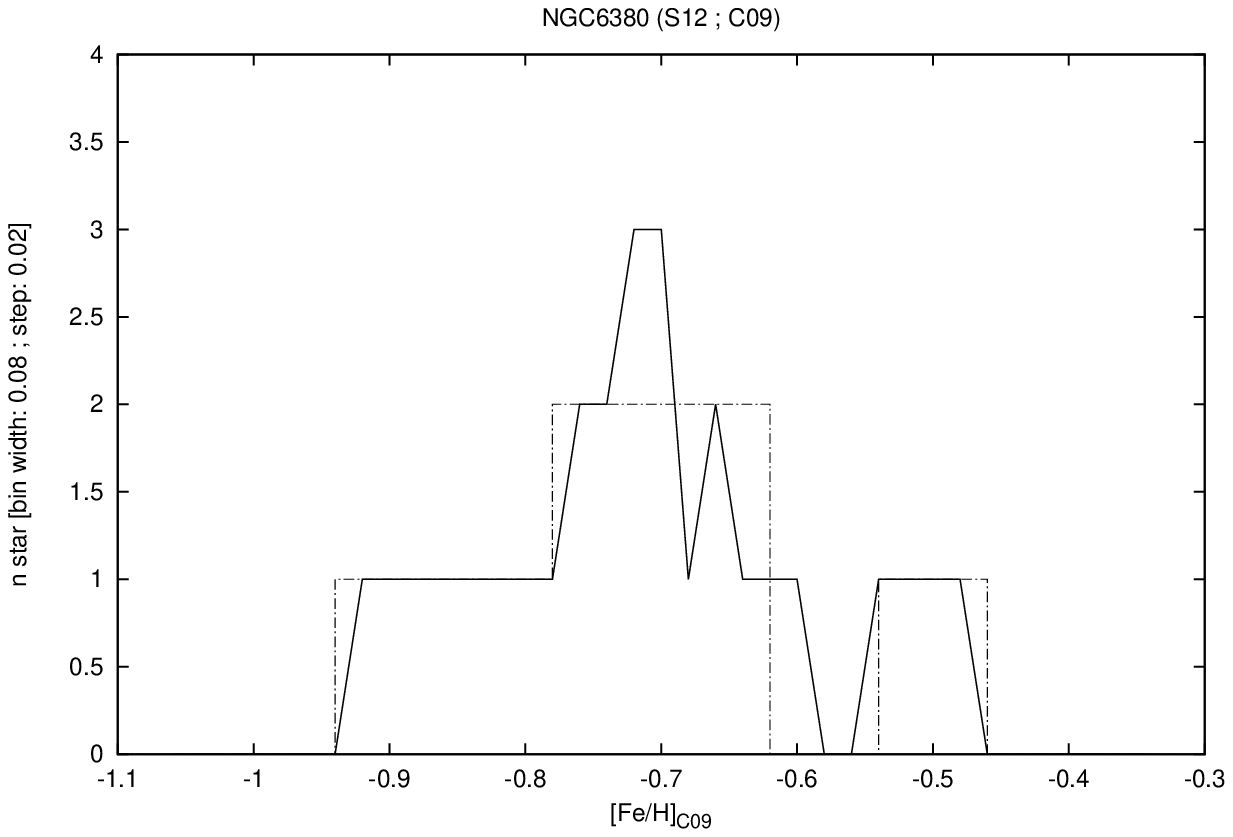}}
\caption{Distribution in metallicity on the C09 scale for NGC\,6380 stars in the S12 sample.}
\label{fig:6380}
\end{center}\end{figure}

\begin{figure}[ht!]\begin{center}
\resizebox{\hsize}{!}{\includegraphics[]{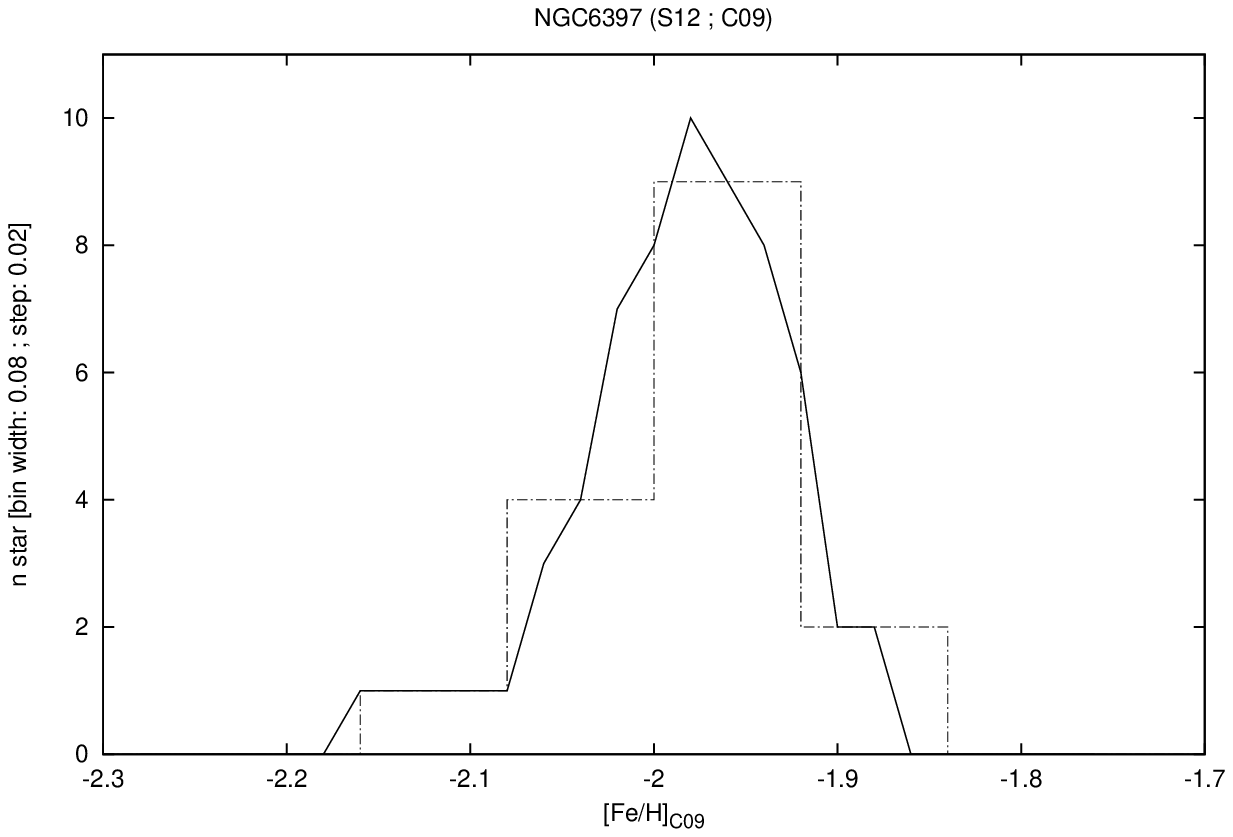}}
\resizebox{\hsize}{!}{\includegraphics[]{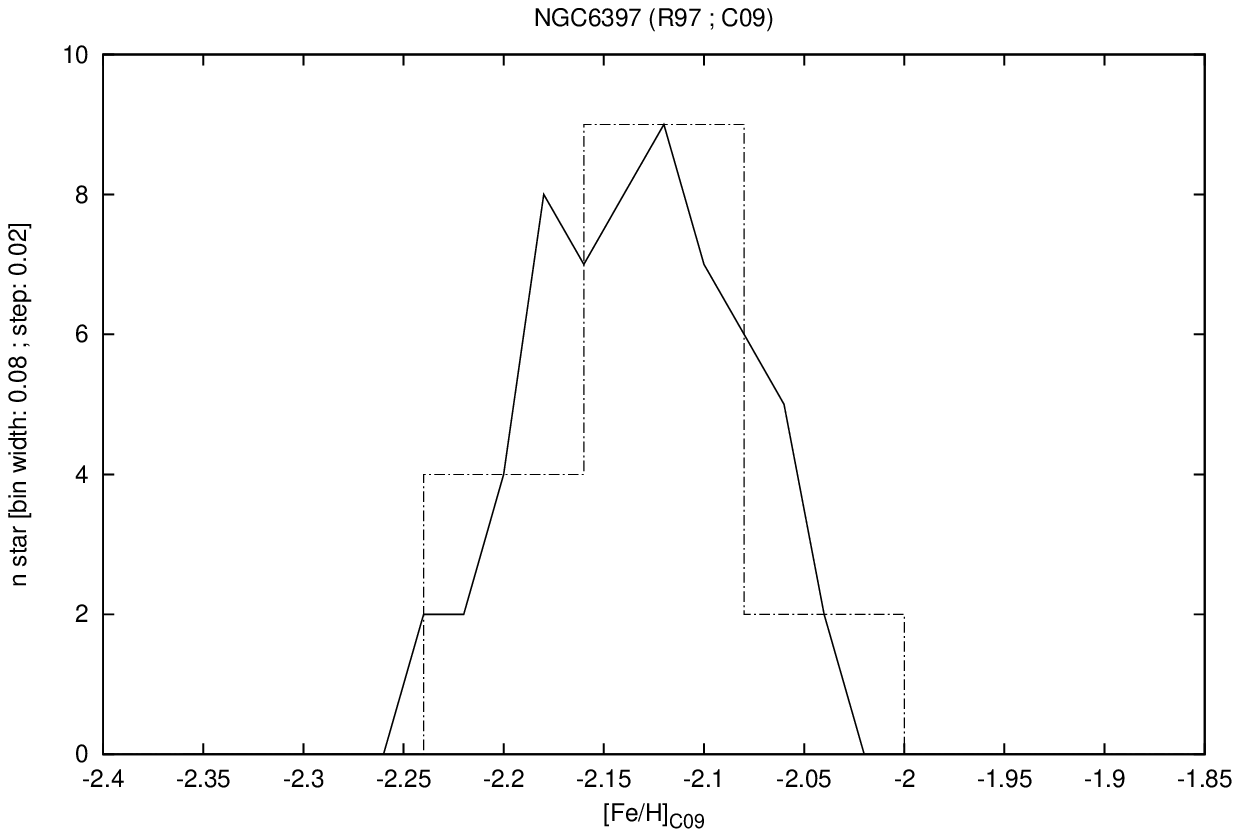}}
\caption{Distribution in metallicity on the C09 scale for NGC\,6397 stars in the S12 (upper plot) and R97 (lower plot) samples.}
\label{fig:6397}
\end{center}\end{figure}

\begin{figure}[ht!]\begin{center}
\resizebox{\hsize}{!}{\includegraphics[]{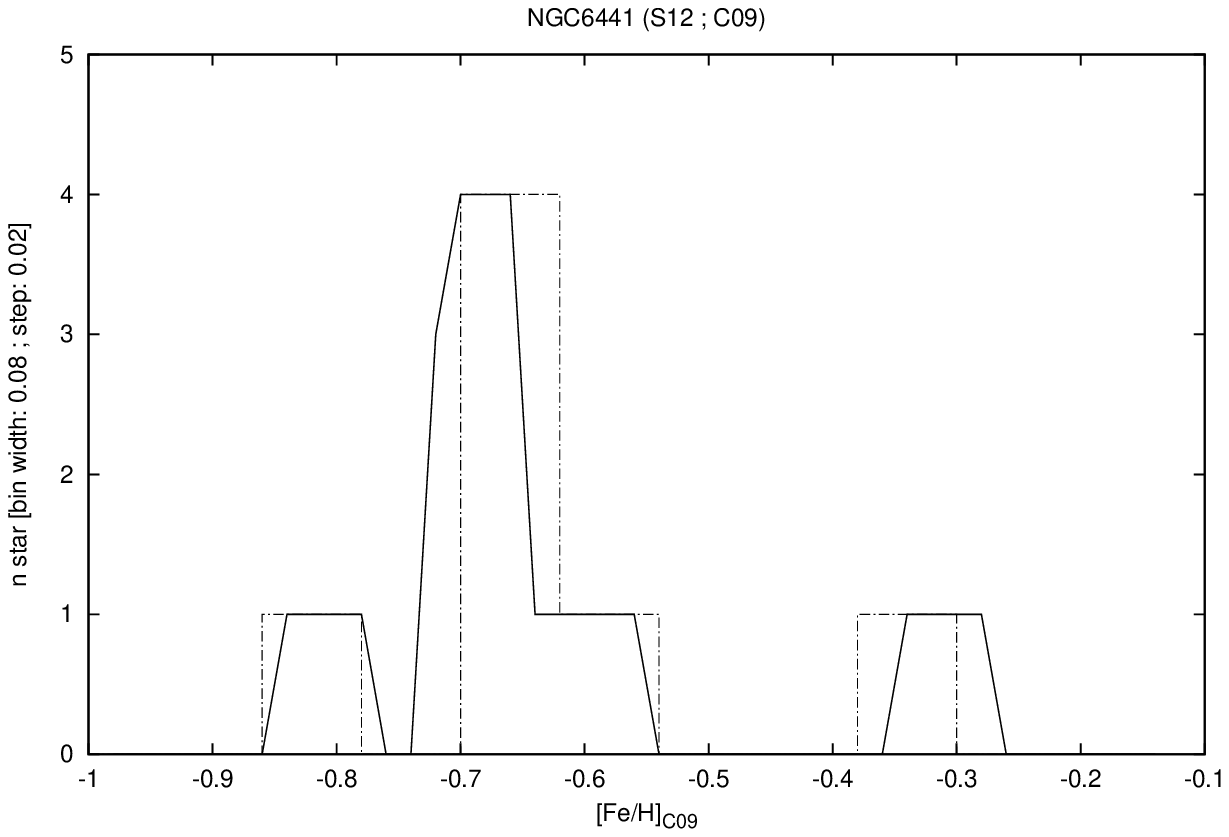}}
\caption{Distribution in metallicity on the C09 scale for NGC\,6441 stars in the S12 sample.}
\label{fig:6441}
\end{center}\end{figure}

\begin{figure}[ht!]\begin{center}
\resizebox{\hsize}{!}{\includegraphics[]{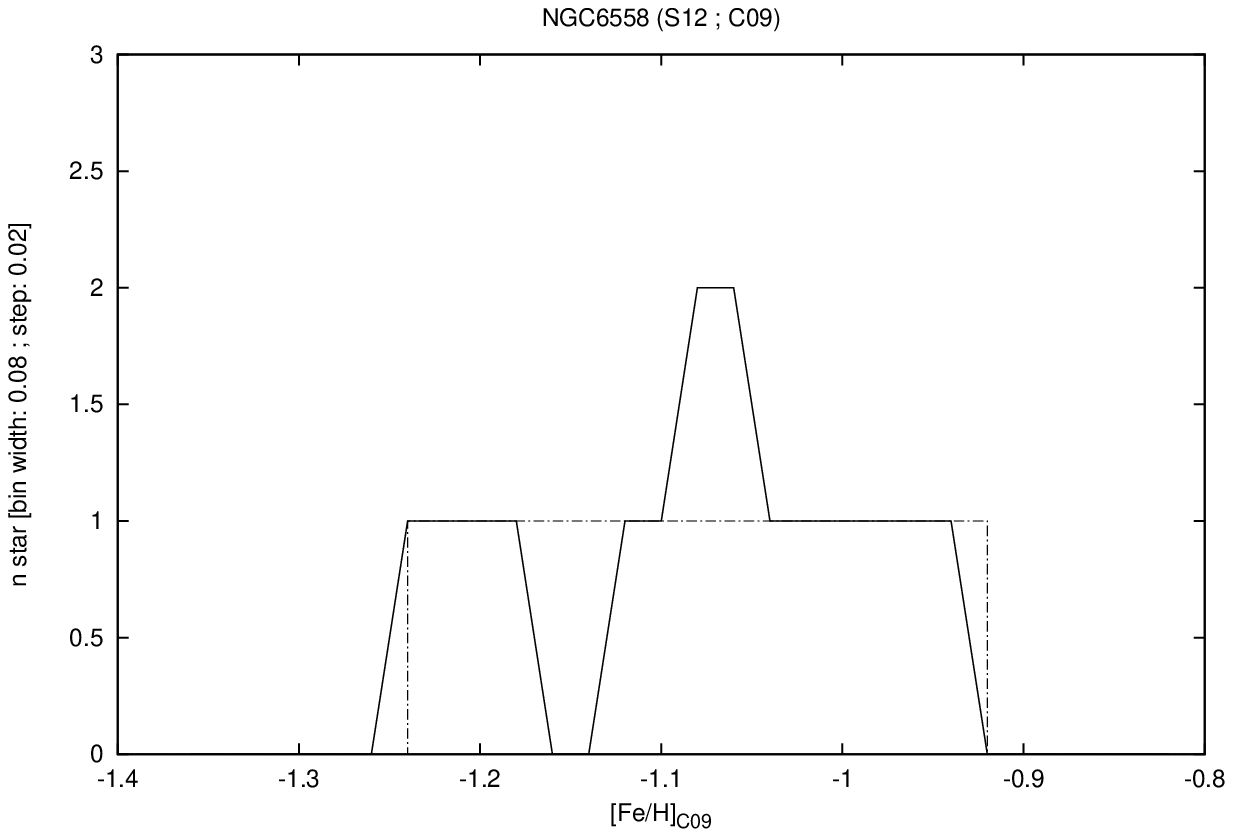}}
\caption{Distribution in metallicity on the C09 scale for NGC\,6558 stars in the S12 sample.}
\label{fig:6558}
\end{center}\end{figure}

\begin{figure}[ht!]\begin{center}
\resizebox{\hsize}{!}{\includegraphics[]{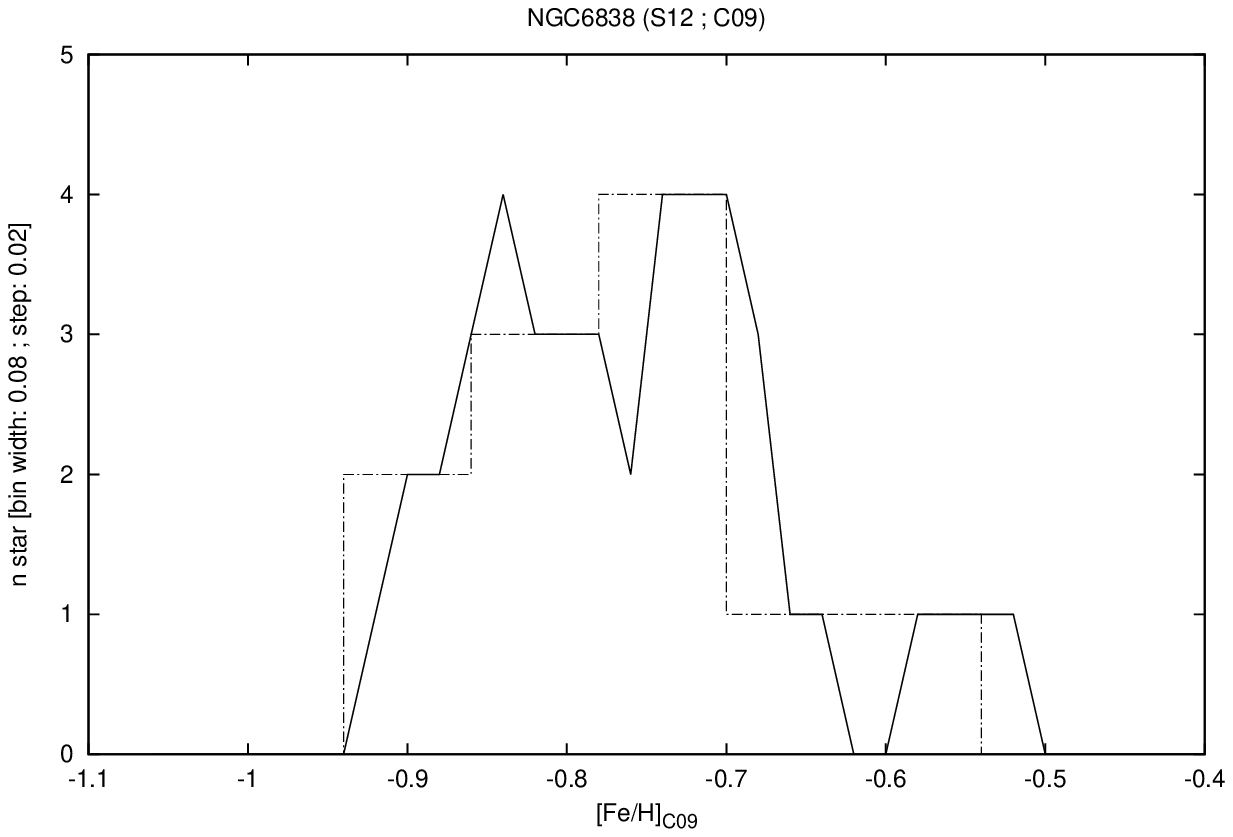}}
\caption{Distribution in metallicity on the C09 scale for NGC\,6838 stars in the S12 sample.}
\label{fig:6838}
\end{center}\end{figure}

\begin{figure}[ht!]\begin{center}
\resizebox{\hsize}{!}{\includegraphics[]{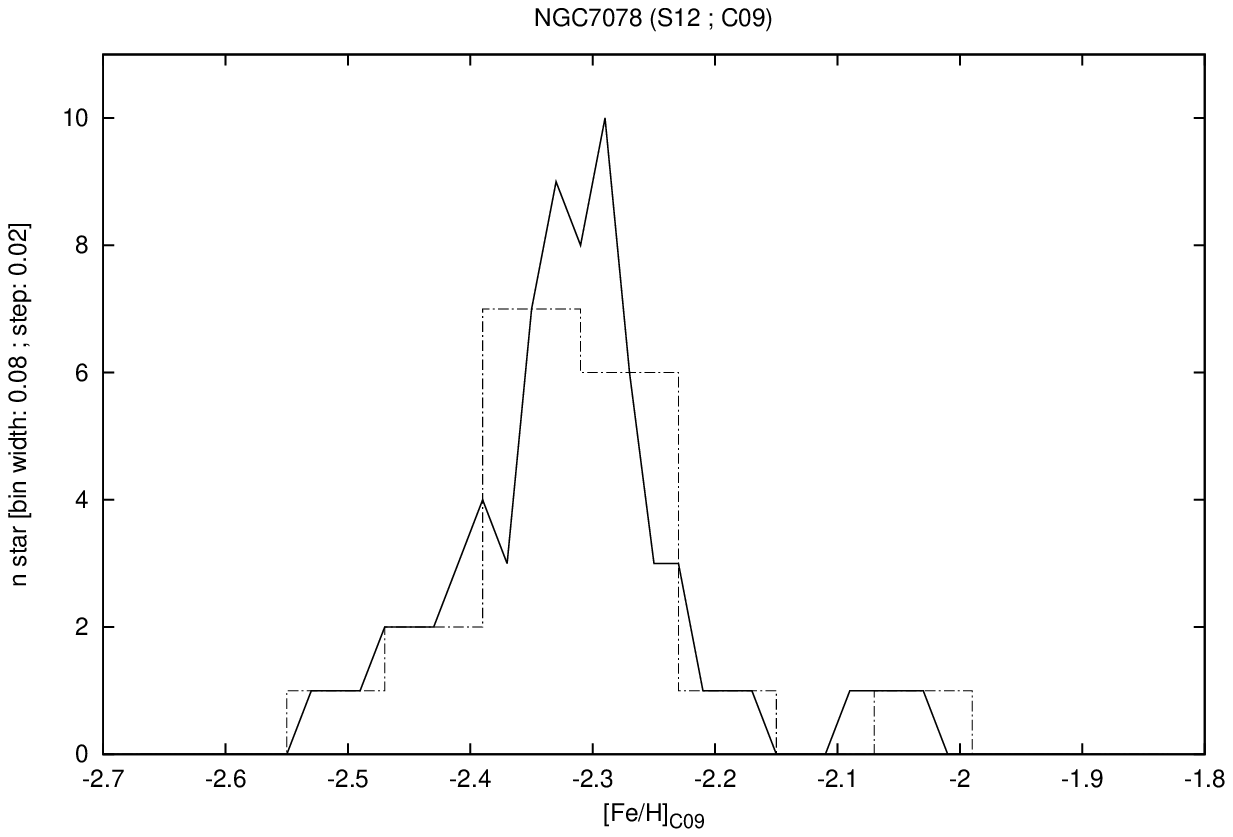}}
\caption{Distribution in metallicity on the C09 scale for NGC\,7078 stars in the S12 sample.}
\label{fig:7078}
\end{center}\end{figure}

\begin{figure}[ht!]\begin{center}
\resizebox{\hsize}{!}{\includegraphics[]{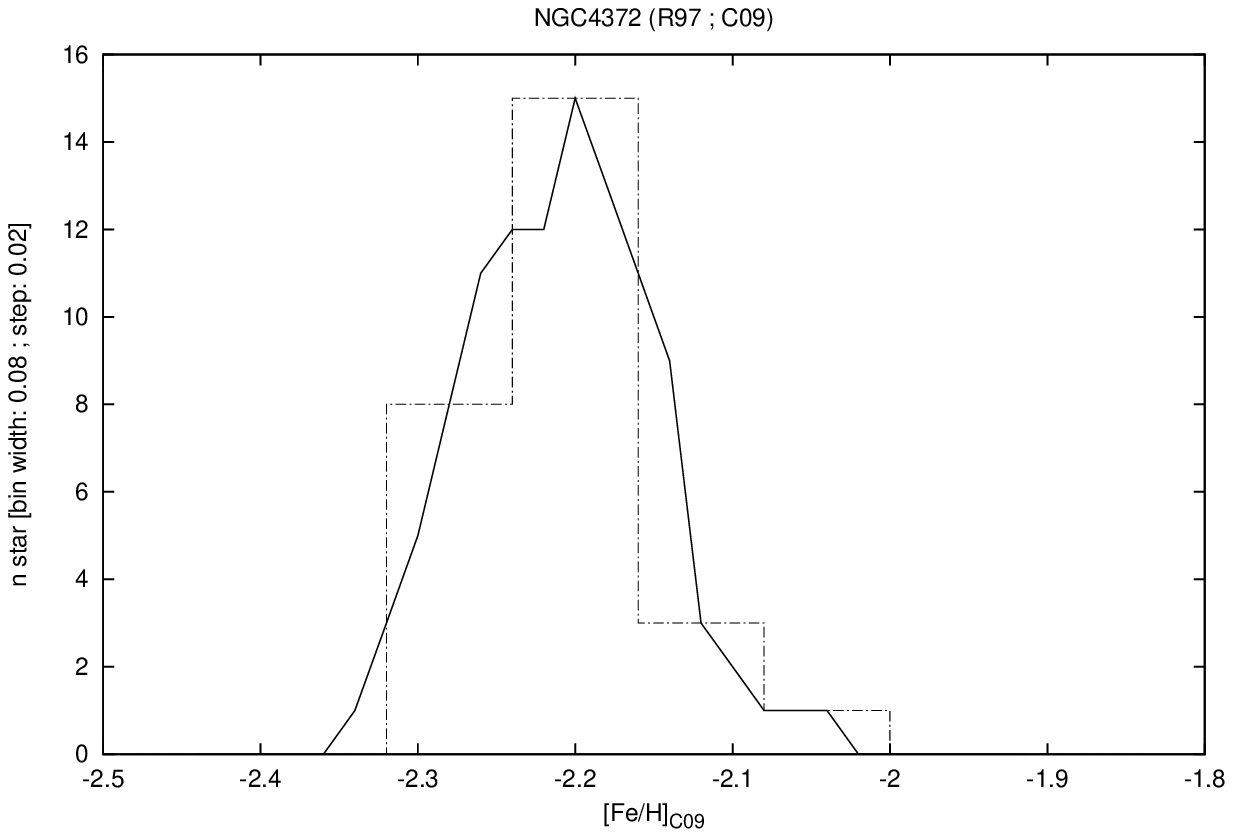}}
\caption{Distribution in metallicity on the C09 scale for NGC\,4372 stars in the R97 sample.}
\label{fig:4372}
\end{center}\end{figure}

\begin{figure}[ht!]\begin{center}
\resizebox{\hsize}{!}{\includegraphics[]{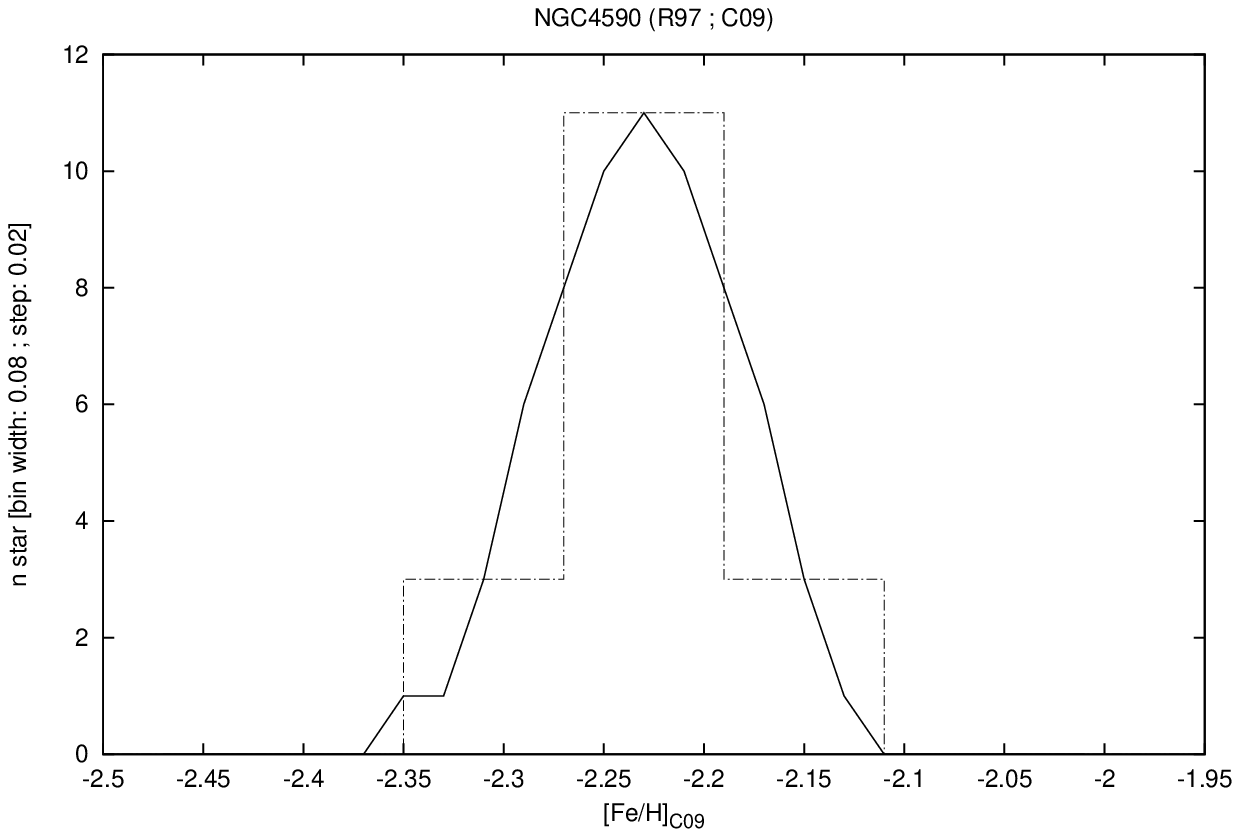}}
\caption{Distribution in metallicity on the C09 scale for NGC\,4590 stars in the R97 sample.}
\label{fig:4590}
\end{center}\end{figure}

\begin{figure}[ht!]\begin{center}
\resizebox{\hsize}{!}{\includegraphics[]{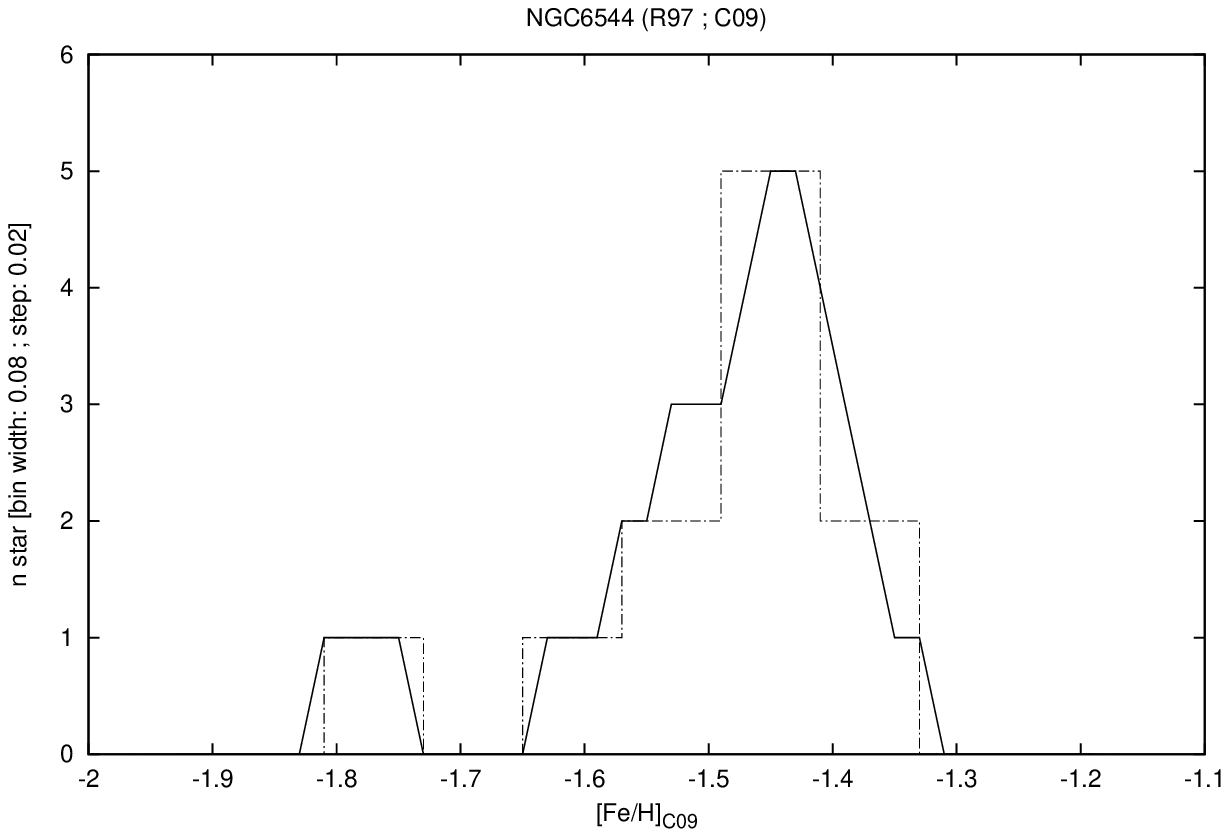}}
\caption{Distribution in metallicity on the C09 scale for NGC\,6544 stars in the R97 sample.}
\label{fig:6544}
\end{center}\end{figure}

\begin{figure}[ht!]\begin{center}
\resizebox{\hsize}{!}{\includegraphics[]{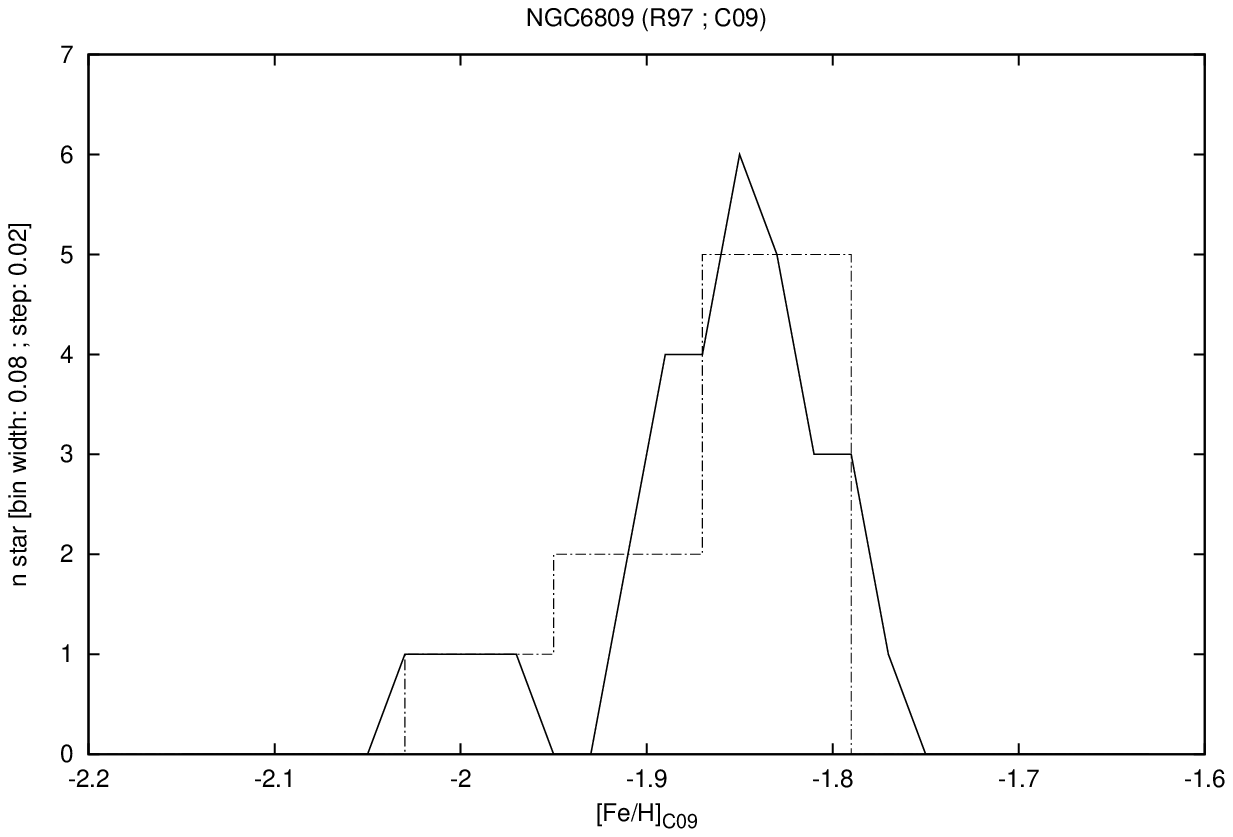}}
\caption{Distribution in metallicity on the C09 scale for NGC\,6809 stars in the R97 sample.}
\label{fig:6809}
\end{center}\end{figure}

\begin{figure}[ht!]\begin{center}
\resizebox{\hsize}{!}{\includegraphics[]{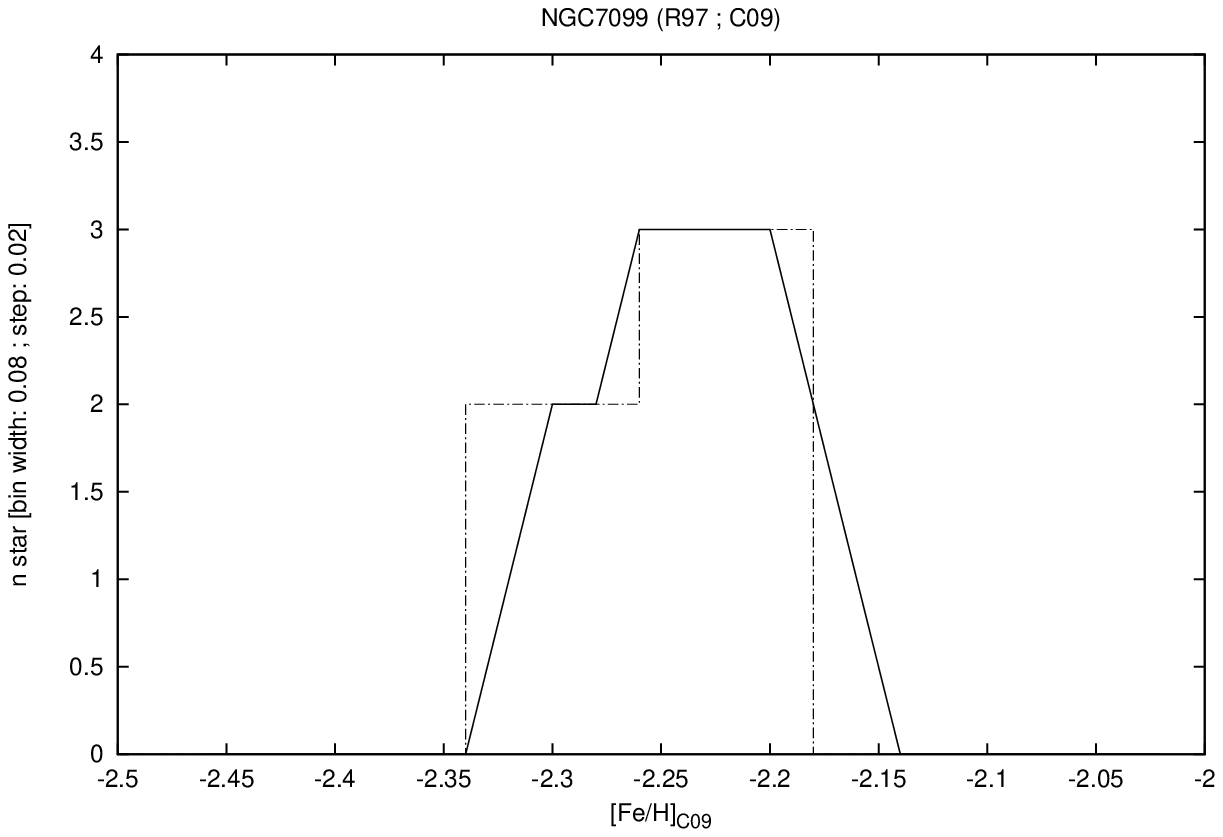}}
\caption{Distribution in metallicity on the C09 scale for NGC\,7099 stars in the R97 sample.}
\label{fig:7099}
\end{center}\end{figure}

We used the cubic calibration relations for the S12 dataset and the quadratic ones for the R97 dataset to calculate the metallicity for each star and obtain the metallicity distribution in each GC.
The results for the clusters discussed in this Section are shown in Figures \ref{fig:2808}-\ref{fig:6638}, while similar histograms for all the clusters in our sample are shown in Figures \ref{fig:6254}-\ref{fig:7099}, only available in electronic form.
The convolved frequency is drawn with a solid line, while the classical frequency is plotted with a dotted histogram.
We chose a bin width of 0.08 dex.
{The convolved frequency was obtained shifting the position of the bin by a step of 0.02 dex, a quarter of the bin width, instead of using bins placed side by side.}
{The analysis of the convolved frequency permits to remove the bias introduced by the choice of the minimum value.}
We show the C09 distribution for the GCs included in both spectroscopic datasets, and both the C09 and H10 distributions for the GCs with only one spectroscopic dataset.

NGC\,6656 (M\,22) presents a clear multimodal distribution (see Figure~\ref{fig:6656}), similar to the one shown in \citet{DaCosta2009}, based on the same spectroscopic data.
In Figure~\ref{fig:6656V} we show the distributions of the metallicities obtained from the optical rEW, and sampled in the same way as our distributions, for both the values calculated in \citet[upper plot]{DaCosta2009}  and \Sav{} (lower plot).
Both studies analyzed 51 stars, but their data reduction differed.
The metallicity bimodality stands out more clearly with our rEW than it did with the optical data from \citet{DaCosta2009} and \Sav{}.
The metallicity distributions based on optical magnitudes have a FWHM of $\approx 0.4$~dex, while it reduces to 0.3~dex when based on nIR photometry.
In the latter case, the peaks are also 30-50\% higher.
This result can be explained by the smaller effect of differential reddening using the $\Ks{}$ data.
The calibration is not a source of these differences, because the ([Fe/H];$\rEW$) relation for optical and nIR data are very similar and approximately linear in the metallicity range under study.
The main peak (located at $\FeH{}\simeq -1.90$~dex in all the distributions) and the overall distribution is quite similar in both plots based on optical photometry, although the secondary peak is more evident in the Da Costa data (with a mean [Fe/H] value of $\sim -1.60$) with respect to the \Sav{} data, where is also located at $\sim -1.70$~dex.
Hence, the differences in data reduction have not influenced the results noticeably.
In our distribution, the two peaks are located at $-1.77$ and $-1.90$~dex, presenting a systematic shift of $\approx 0.08$~dex with respect to the results found by \citet{Marino2009}, that determined a mean [Fe/H] values of $-1.68$ and $-1.82$~dex for  stars rich and poor in s-process element, respectively.


NGC\,6656 is not the only GC in our sample that shows a structured metallicity distribution suggesting a complex mix of stellar populations with different metallicity.
The most peculiar distributions are those of NGC\,6528 and  NGC\,6553, which are nearly flat and cover a $\sim 1$~dex range in metallicity.
Membership cannot be considered certain, since it is based only on position and radial velocity.
They are also the two most metal-rich clusters, and the literature presents controversial values for their metallicity.

\section{Discussion and Conclusions}
\label{s:concl}

\begin{figure}[ht!]\begin{center}
\resizebox{\hsize}{!}{\includegraphics[]{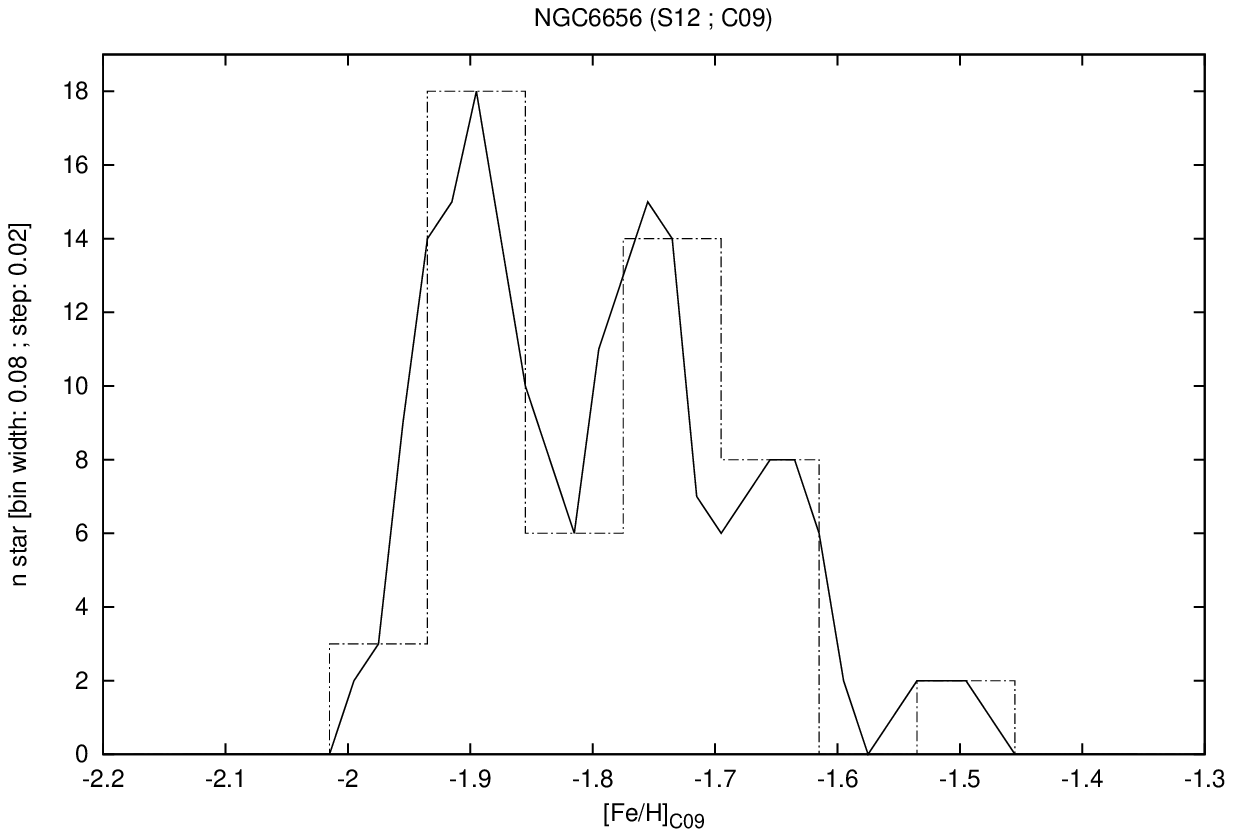}}
\resizebox{\hsize}{!}{\includegraphics[]{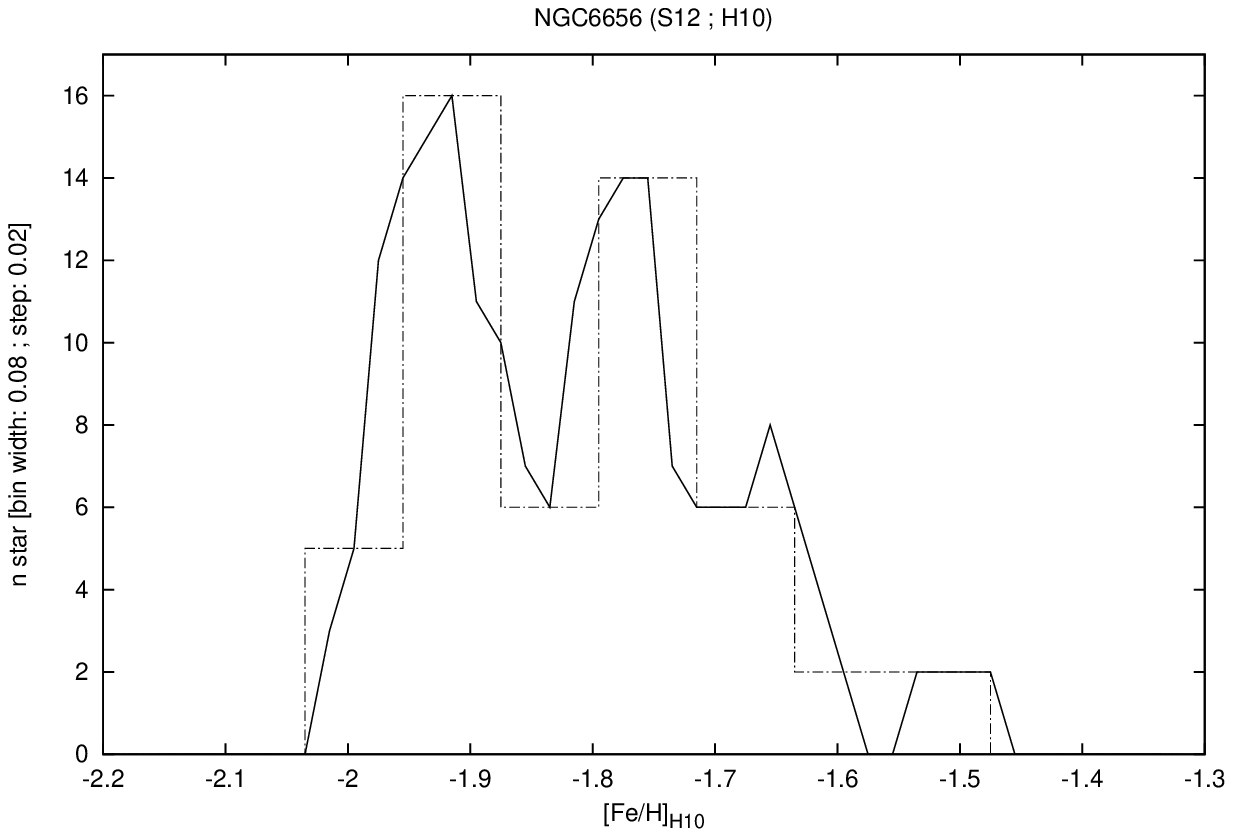}}
\caption{Distribution in metallicity on the C09 (upper plot) and H10 (lower plot) scales for NGC\,6656 stars in the S12 sample.}
\label{fig:6656}
\end{center}\end{figure}

\begin{figure}[ht!]\begin{center}
\resizebox{\hsize}{!}{\includegraphics[]{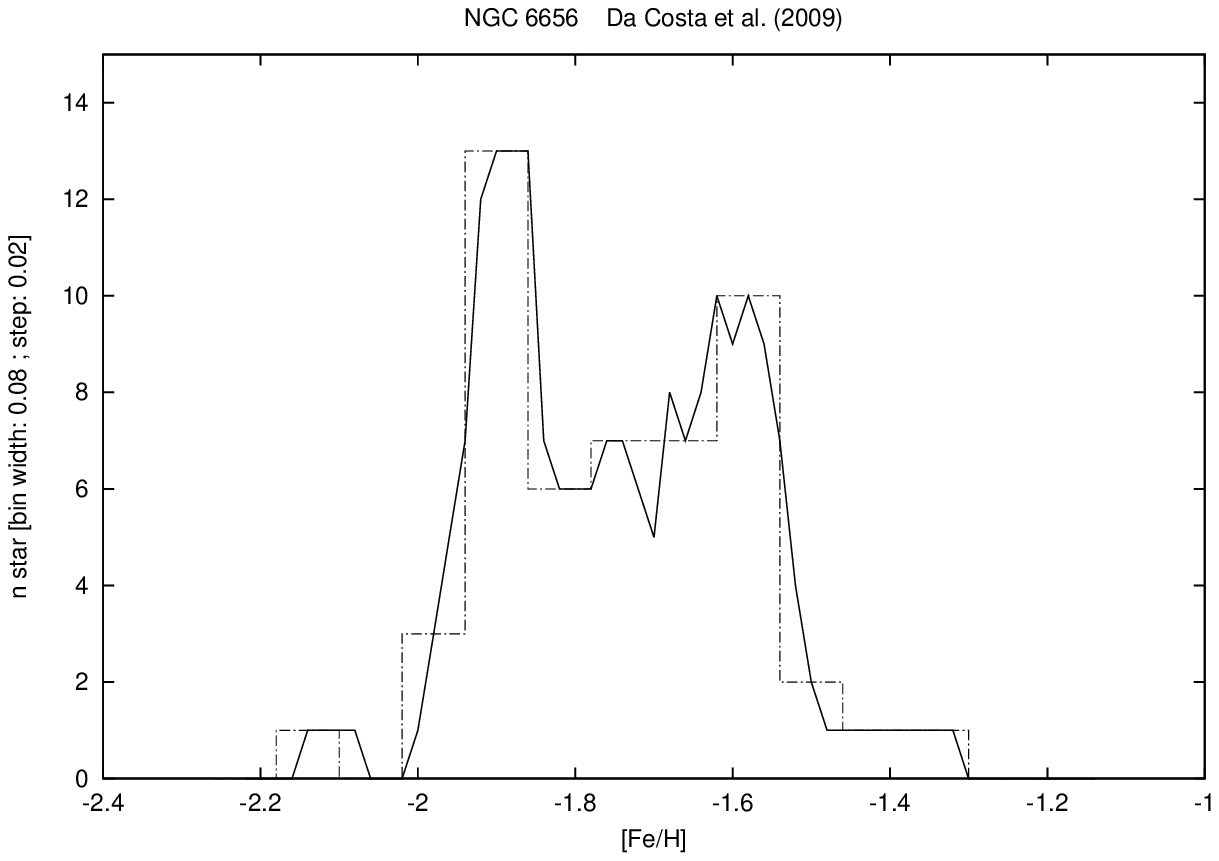}}
\resizebox{\hsize}{!}{\includegraphics[]{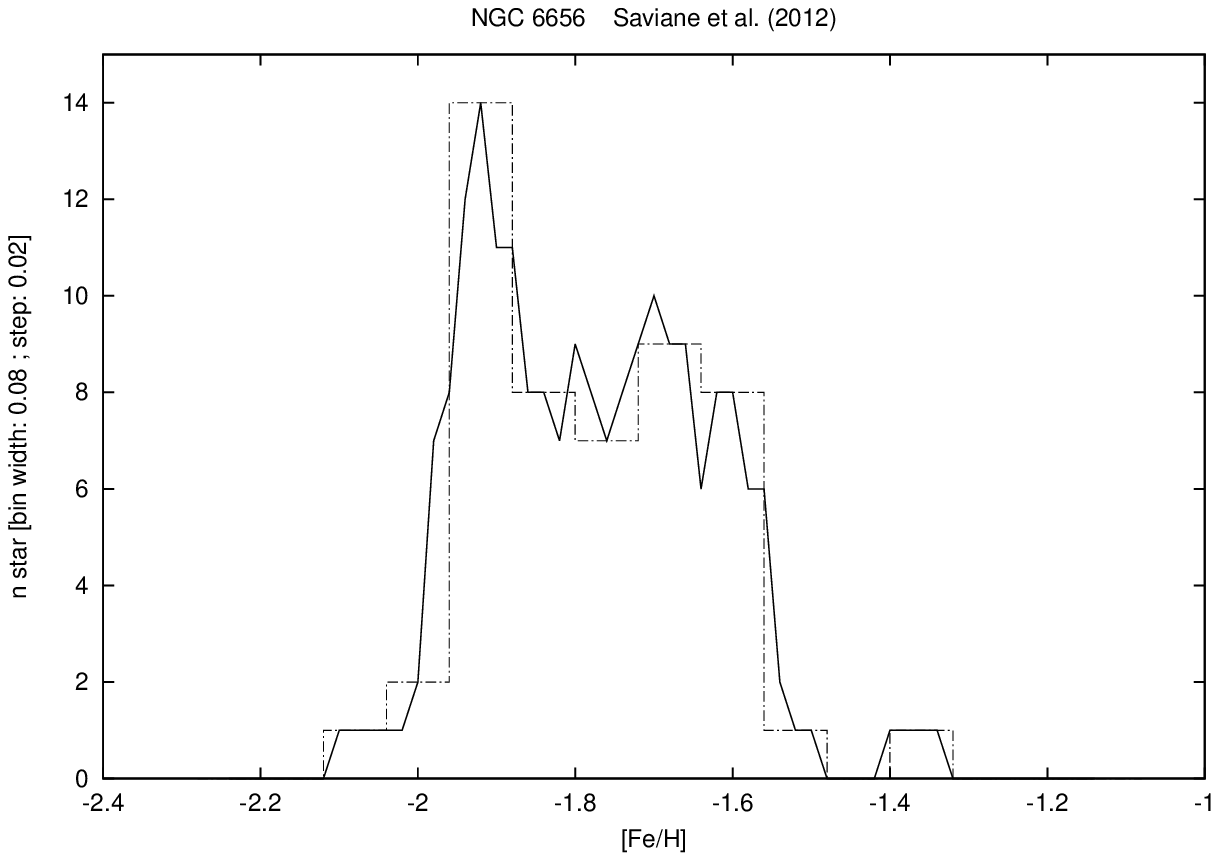}}
\caption{Distribution in metallicity for NGC\,6656 stars from optical rEW.}
\label{fig:6656V}
\end{center}\end{figure}

\begin{figure}[ht!]\begin{center}
\resizebox{\hsize}{!}{\includegraphics[]{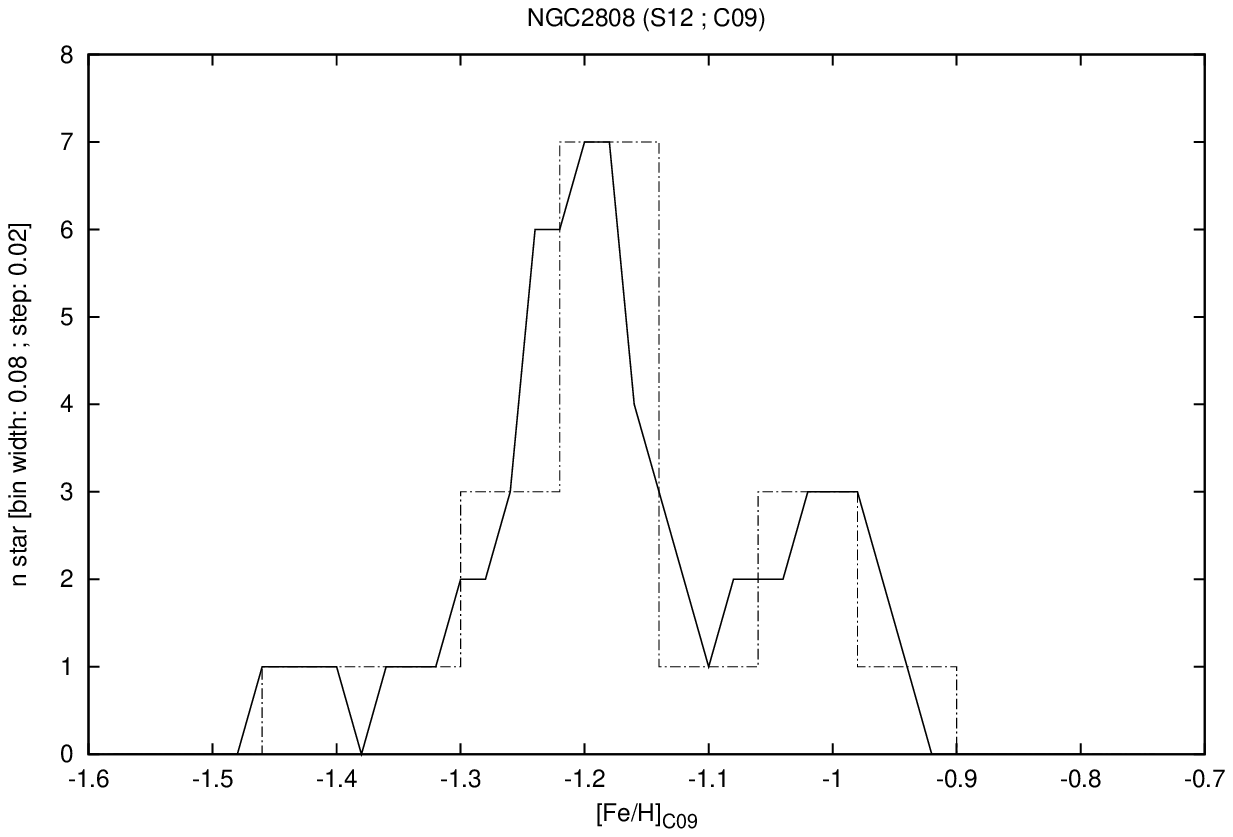}}
\resizebox{\hsize}{!}{\includegraphics[]{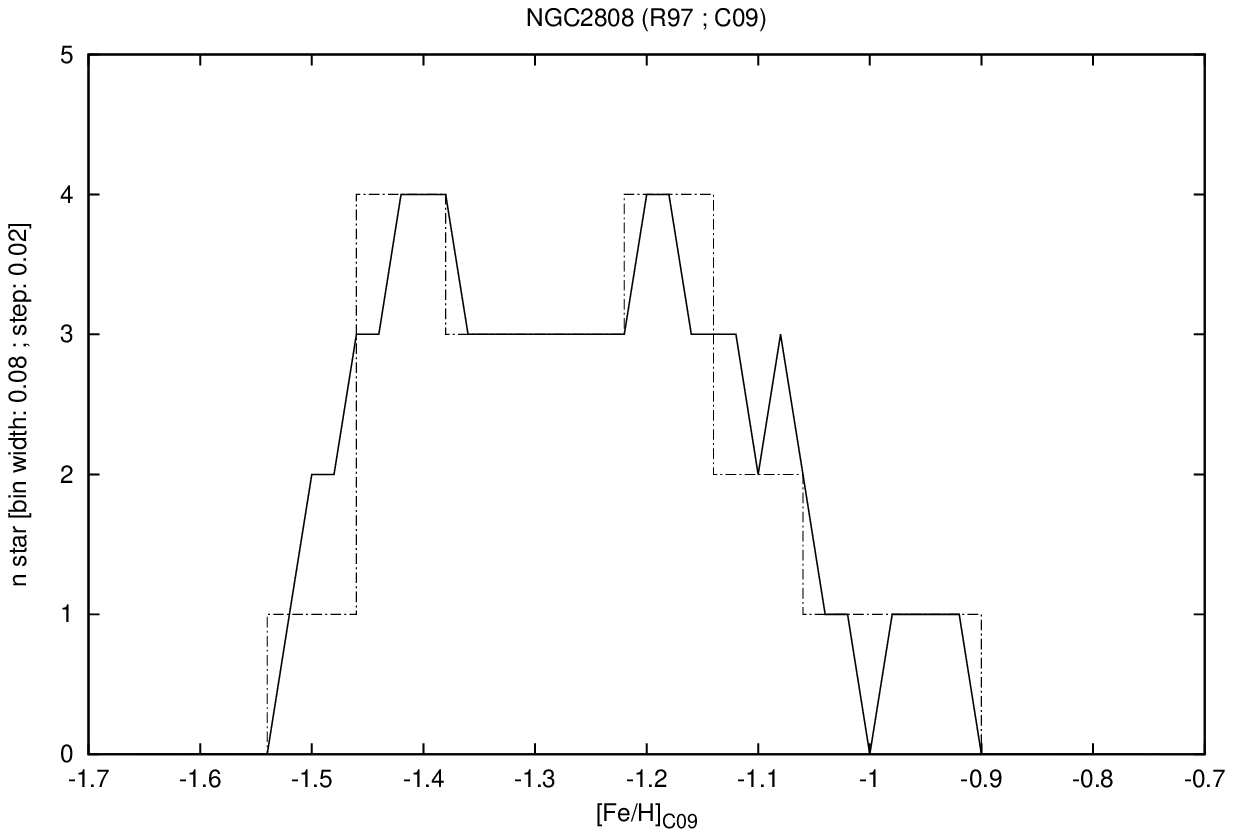}}
\caption{Distribution in metallicity on the C09 scale for NGC\,2808 stars in the S12 (upper plot) and R97 (lower plot) samples.}
\label{fig:2808}
\end{center}\end{figure}

\begin{figure}[ht!]\begin{center}
\resizebox{\hsize}{!}{\includegraphics[]{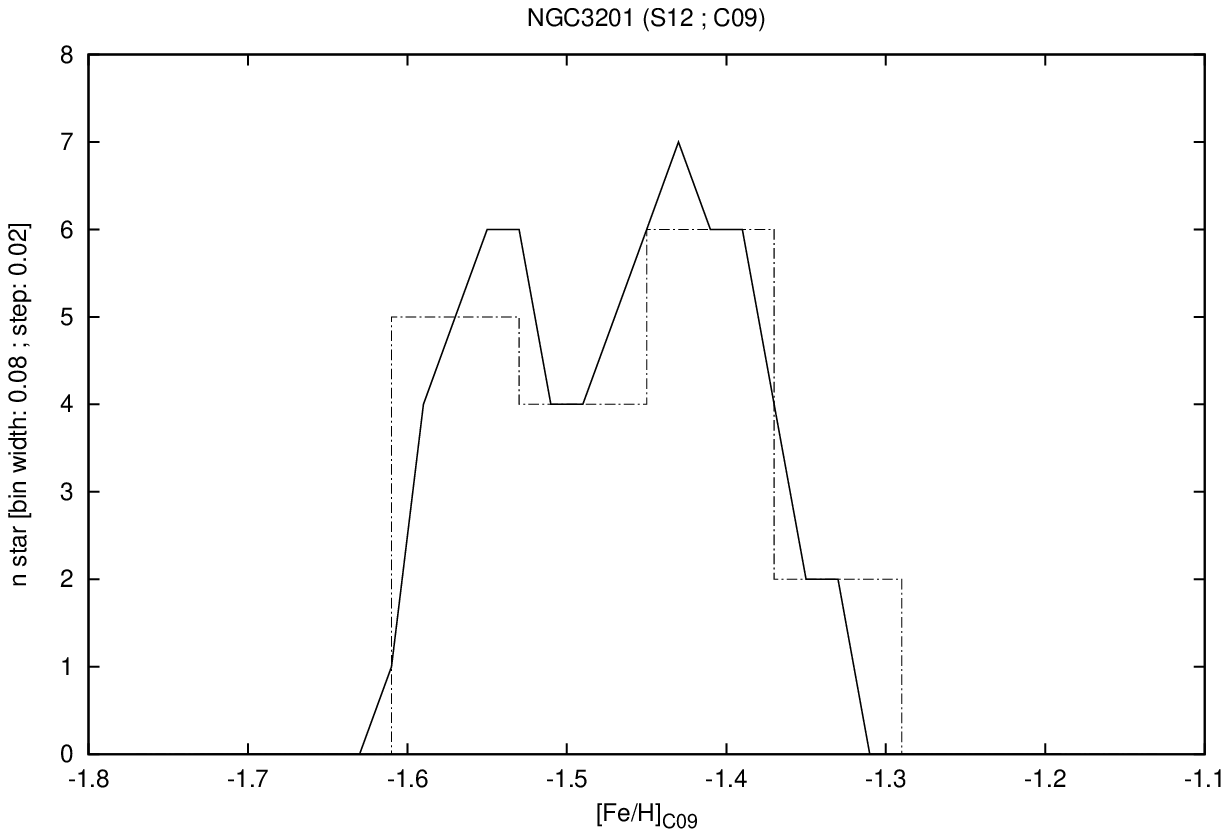}}
\resizebox{\hsize}{!}{\includegraphics[]{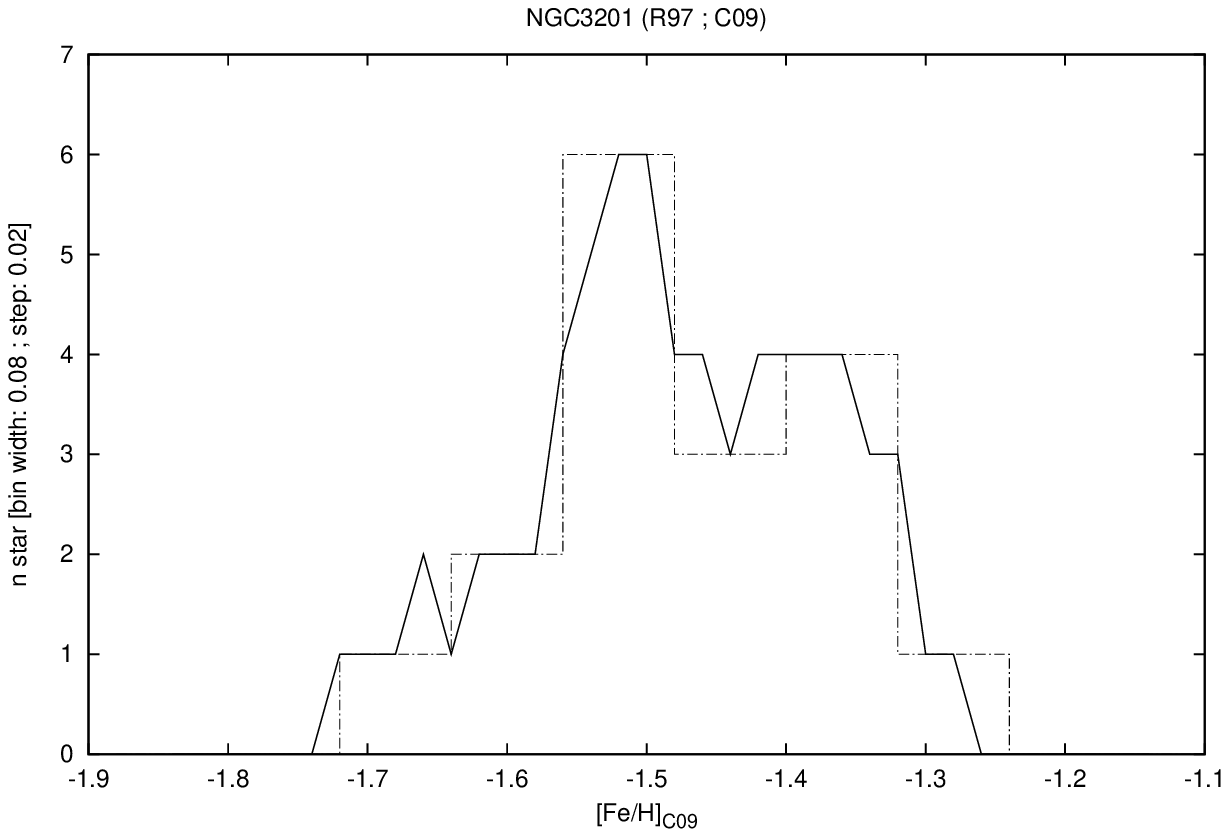}}
\caption{Distribution in metallicity on the C09 scale for NGC\,3201 stars in the S12 (upper plot) and R97 (lower plot) samples.}
\label{fig:3201}
\end{center}\end{figure}

\begin{figure}[ht!]\begin{center}
\resizebox{\hsize}{!}{\includegraphics[]{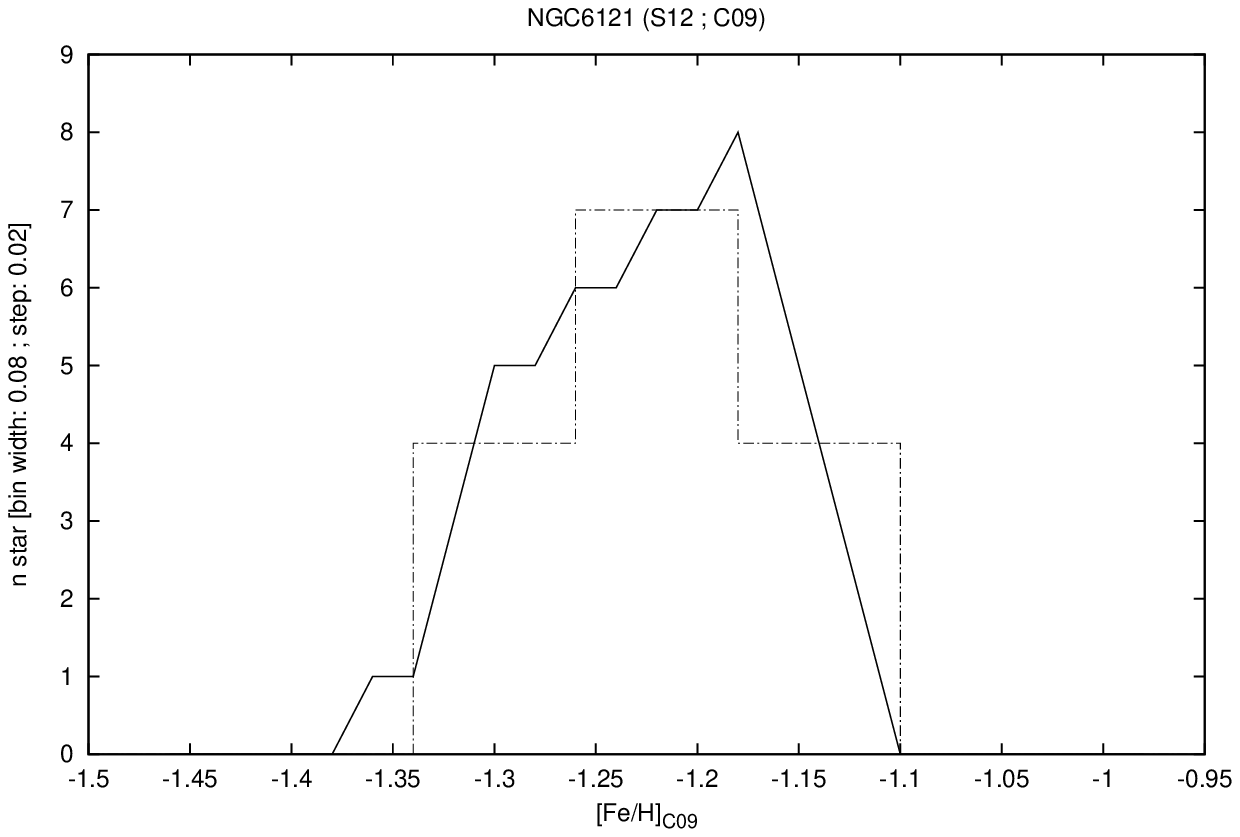}}
\resizebox{\hsize}{!}{\includegraphics[]{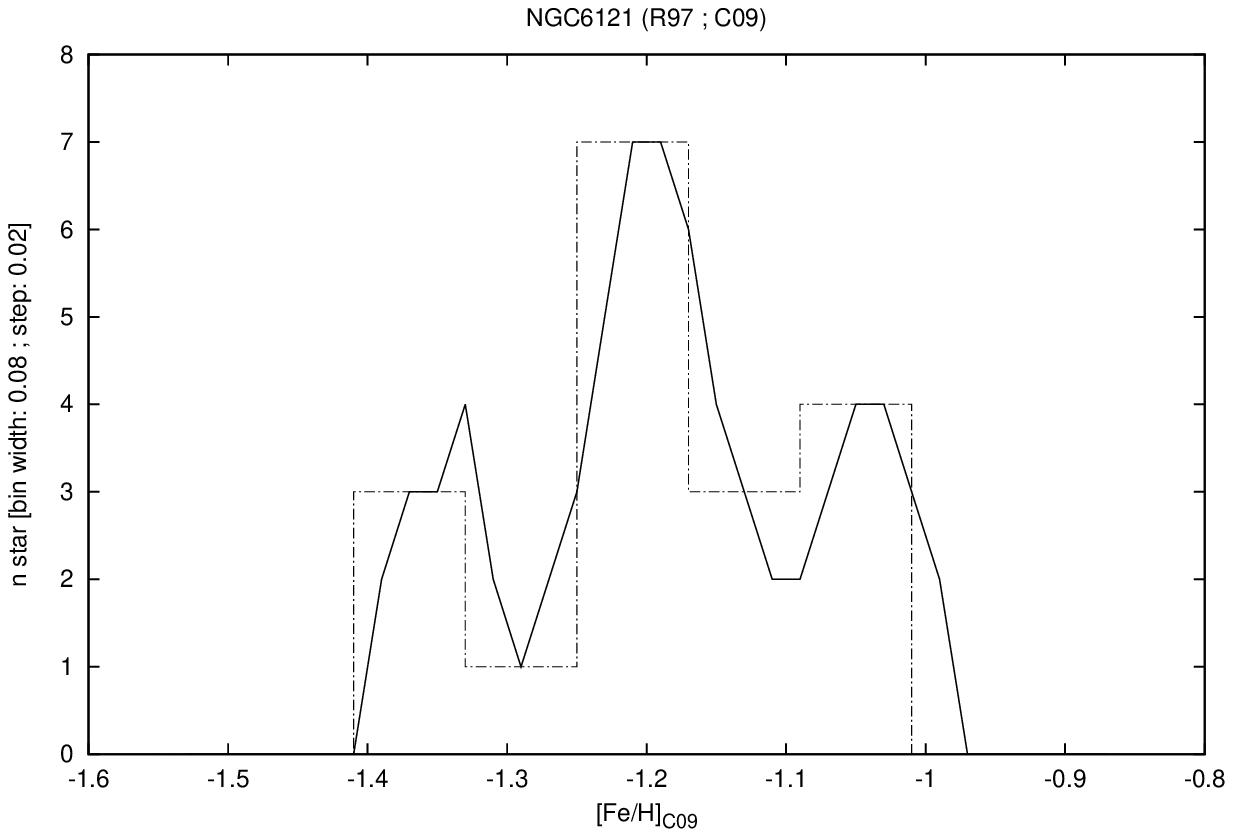}}
\caption{Distribution in metallicity on the C09 scale for NGC\,6121 stars in the S12 (upper plot) and R97 (lower plot) samples.}
\label{fig:6121}
\end{center}\end{figure}

\begin{figure}[ht!]\begin{center}
\resizebox{\hsize}{!}{\includegraphics[]{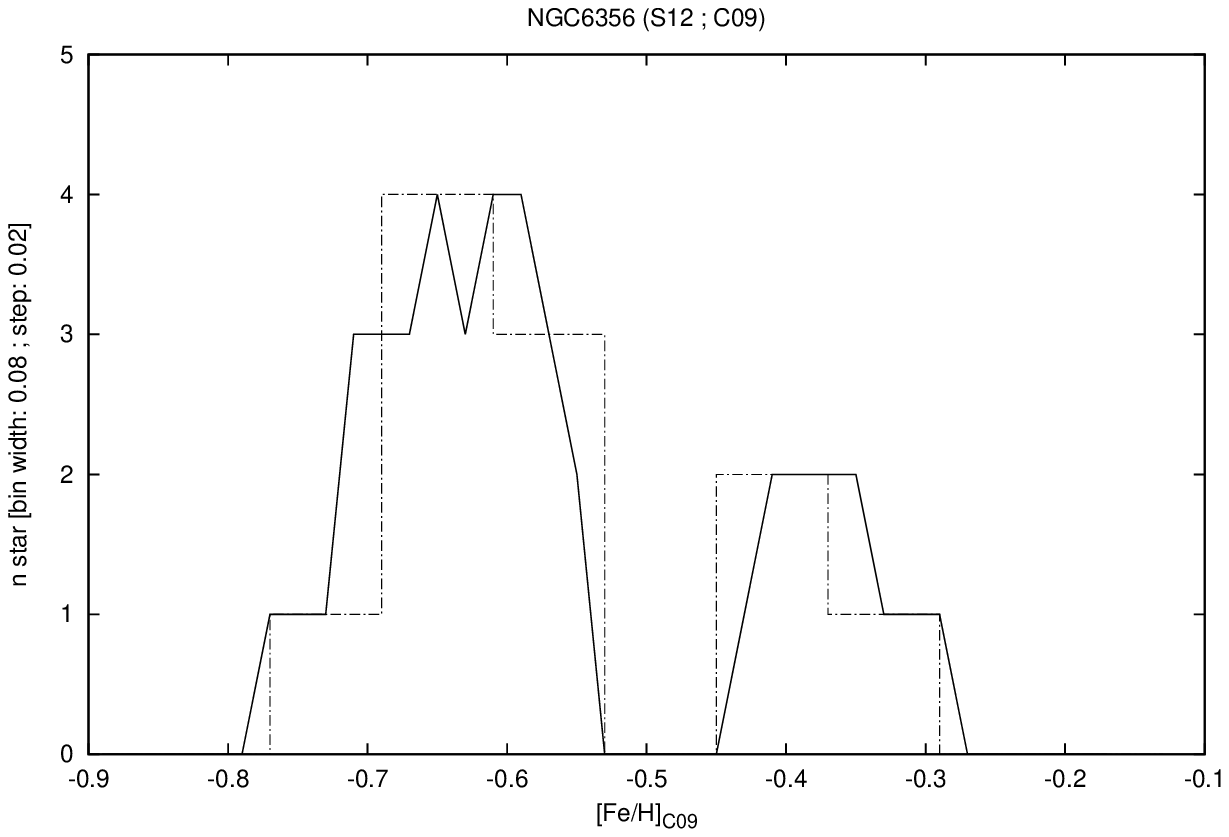}}
\resizebox{\hsize}{!}{\includegraphics[]{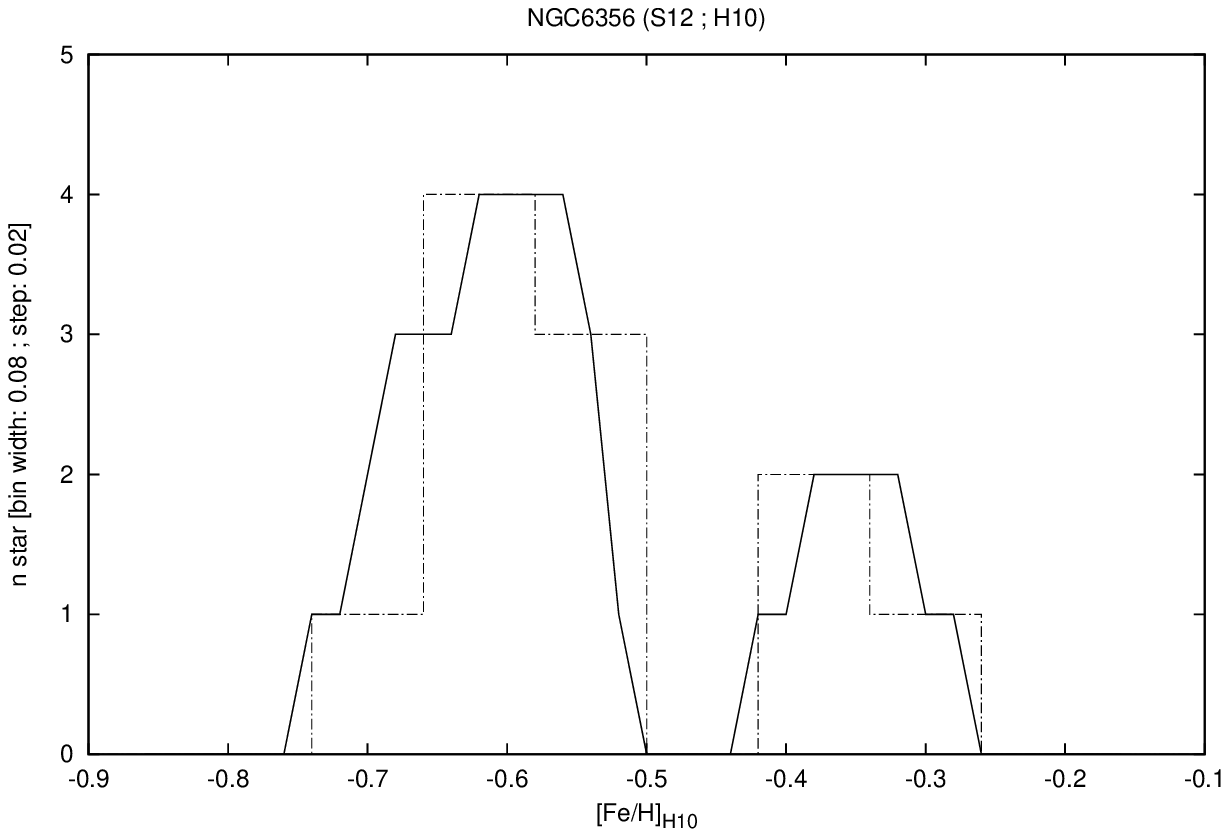}}
\caption{Distribution in metallicity on the C09 (upper plot) and H10 (lower plot) scales for NGC\,6356 stars in the S12 sample.}
\label{fig:6356}
\end{center}\end{figure}

\begin{figure}[ht!]\begin{center}
\resizebox{\hsize}{!}{\includegraphics[]{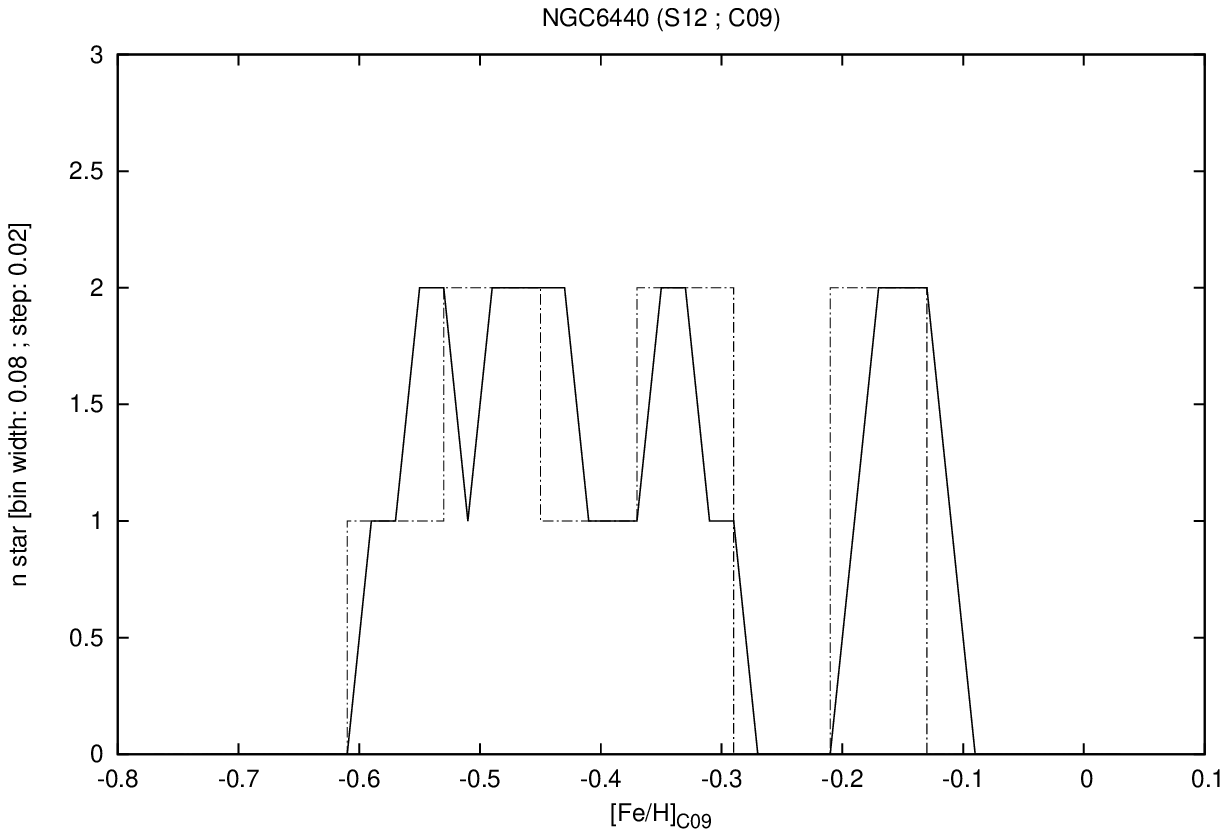}}
\resizebox{\hsize}{!}{\includegraphics[]{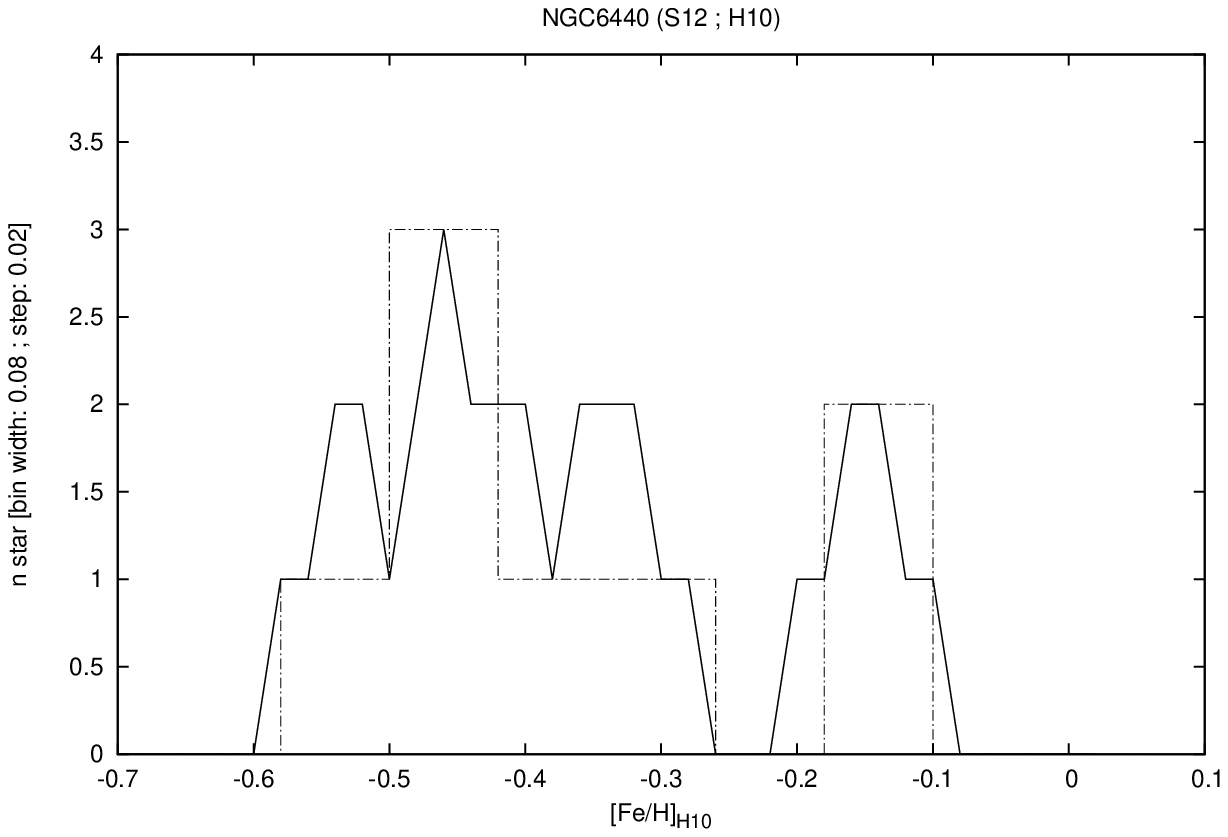}}
\caption{Distribution in metallicity on the C09 (upper plot) and H10 (lower plot) scales for NGC\,6440 stars in the S12 sample.}
\label{fig:6440}
\end{center}\end{figure}

\begin{figure}[ht!]\begin{center}
\resizebox{\hsize}{!}{\includegraphics[]{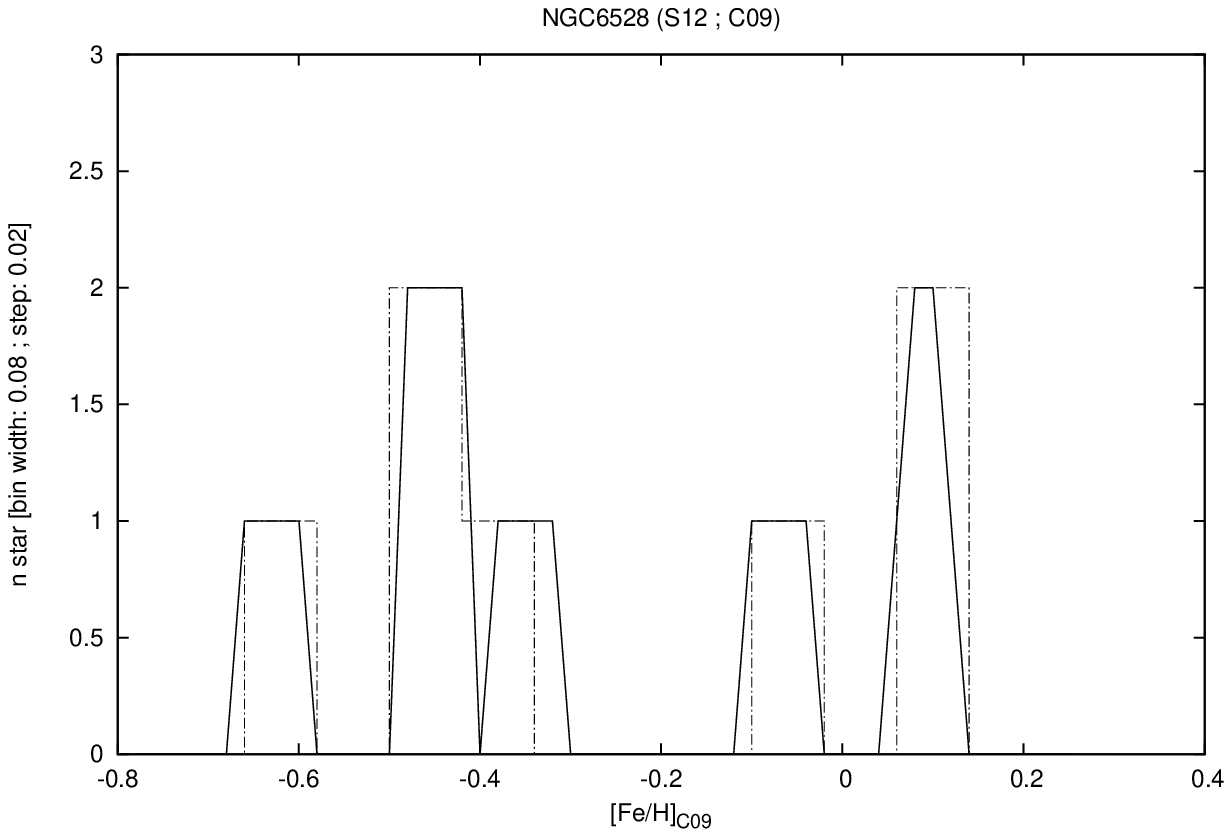}}
\resizebox{\hsize}{!}{\includegraphics[]{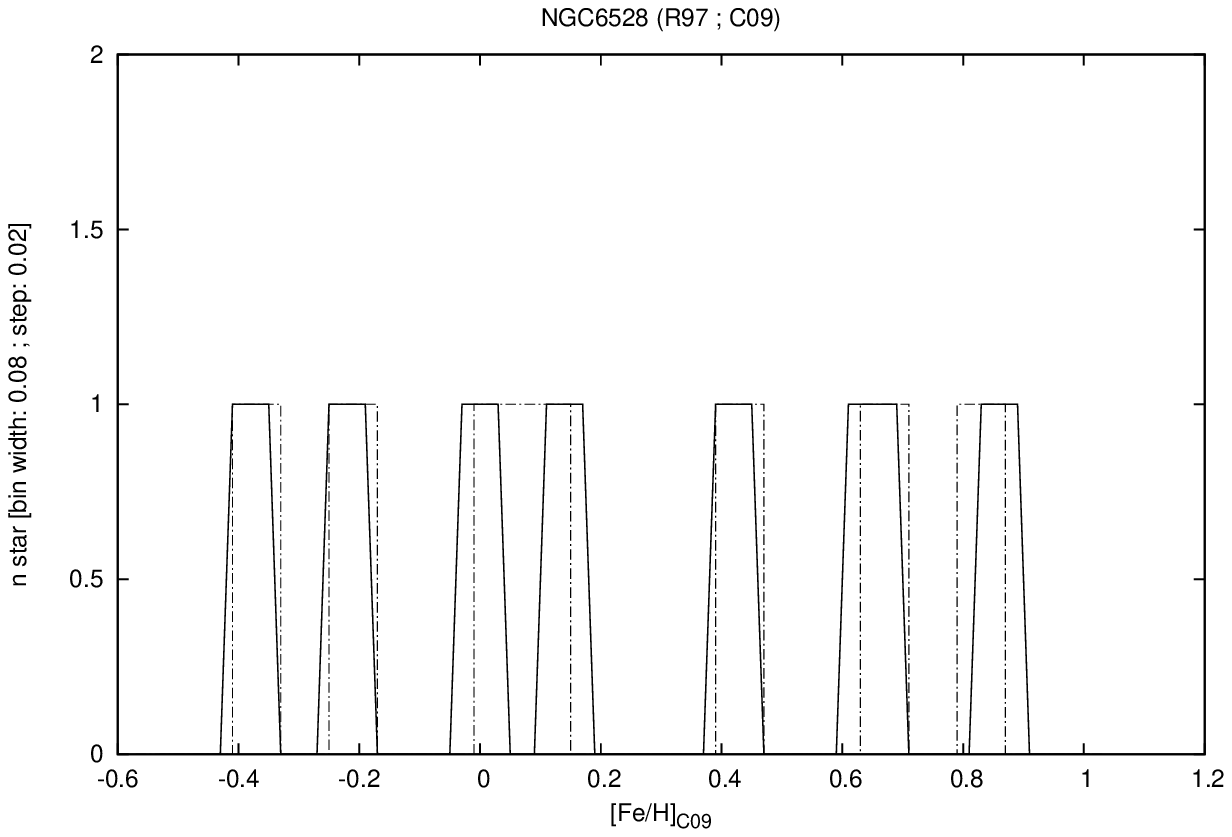}}
\caption{Distribution in metallicity on the C09 scale for NGC\,6528 stars in the S12 (upper plot) and R97 (lower plot) samples.}
\label{fig:6528}
\end{center}\end{figure}

\begin{figure}[ht!]\begin{center}
\resizebox{\hsize}{!}{\includegraphics[]{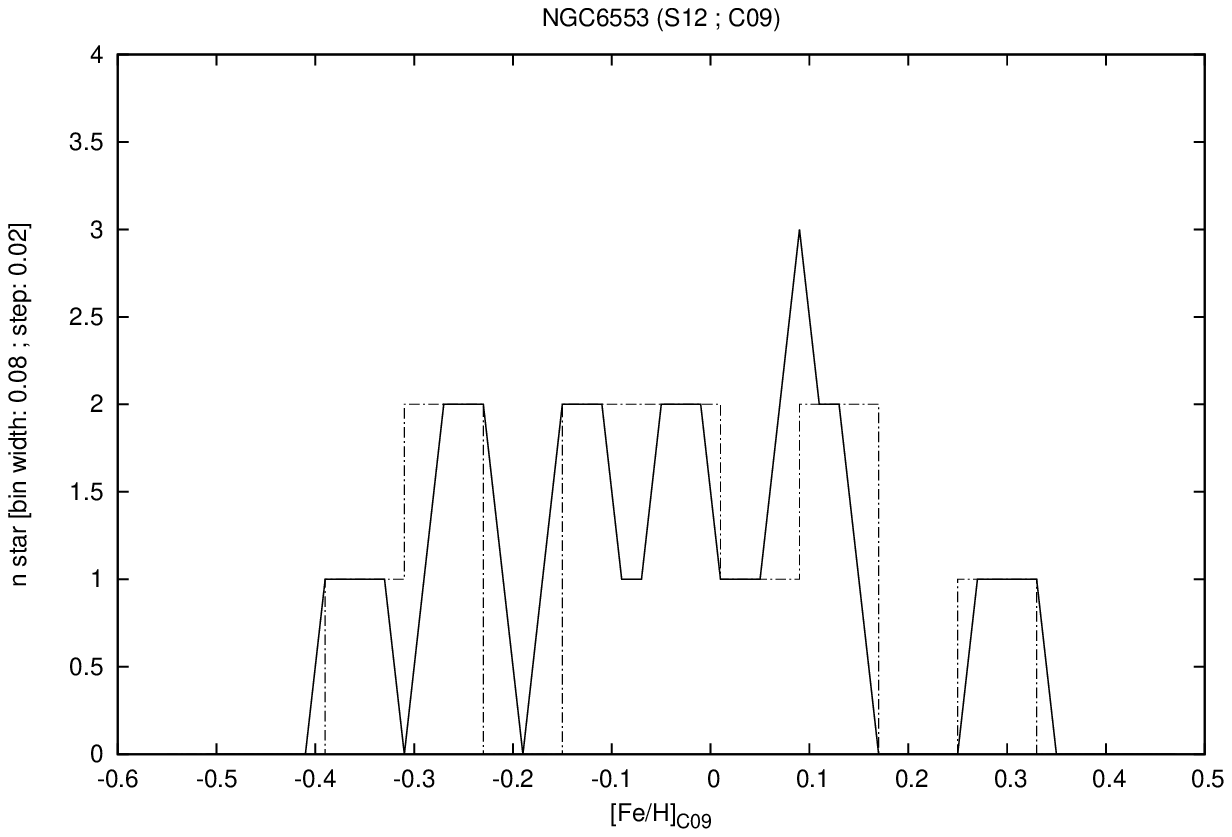}}
\resizebox{\hsize}{!}{\includegraphics[]{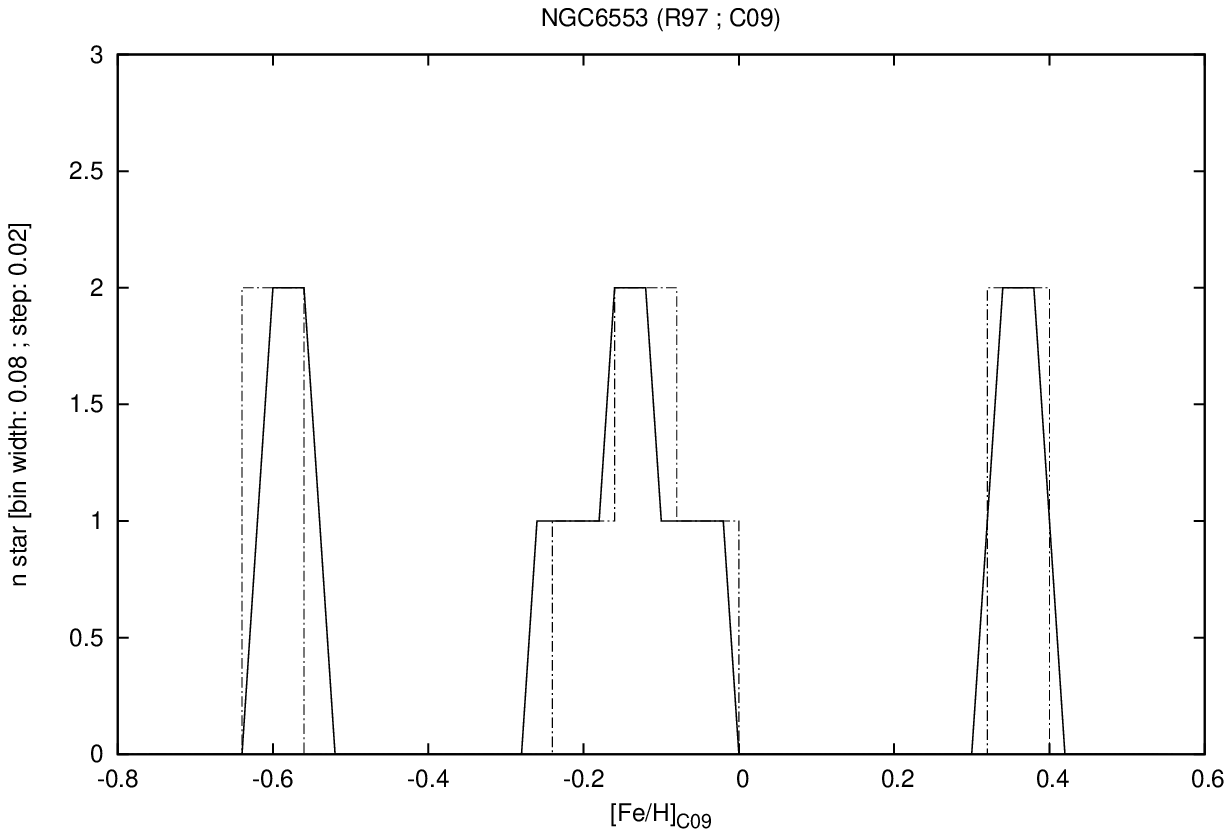}}
\caption{Distribution in metallicity on the C09 scale for NGC\,6553 stars in the S12 (upper plot) and R97 (lower plot) samples.}
\label{fig:6553}
\end{center}\end{figure}

\begin{figure}[ht!]\begin{center}
\resizebox{\hsize}{!}{\includegraphics[]{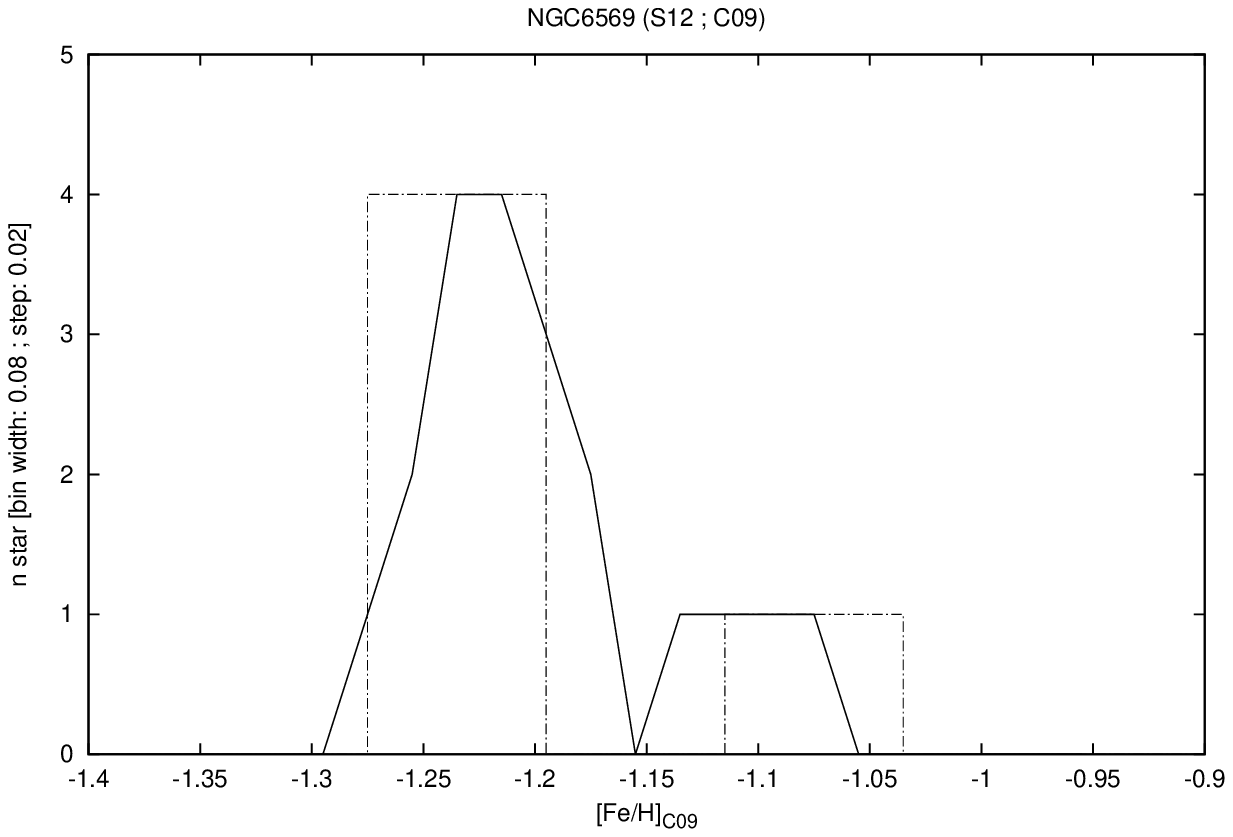}}
\resizebox{\hsize}{!}{\includegraphics[]{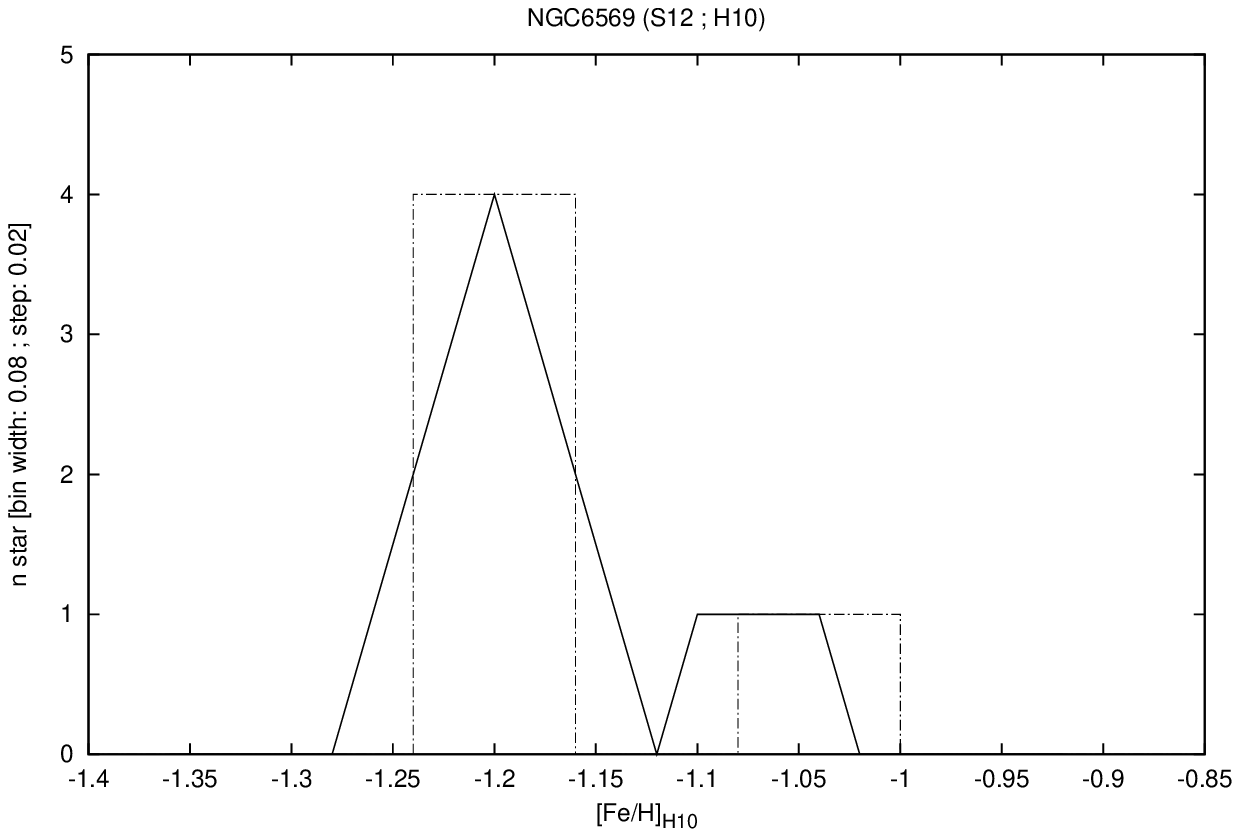}}
\caption{Distribution in metallicity on the C09 (upper plot) and H10 (lower plot) scales for NGC\,6569 stars in the S12 sample.}
\label{fig:6569}
\end{center}\end{figure}

\begin{figure}[ht!]\begin{center}
\resizebox{\hsize}{!}{\includegraphics[]{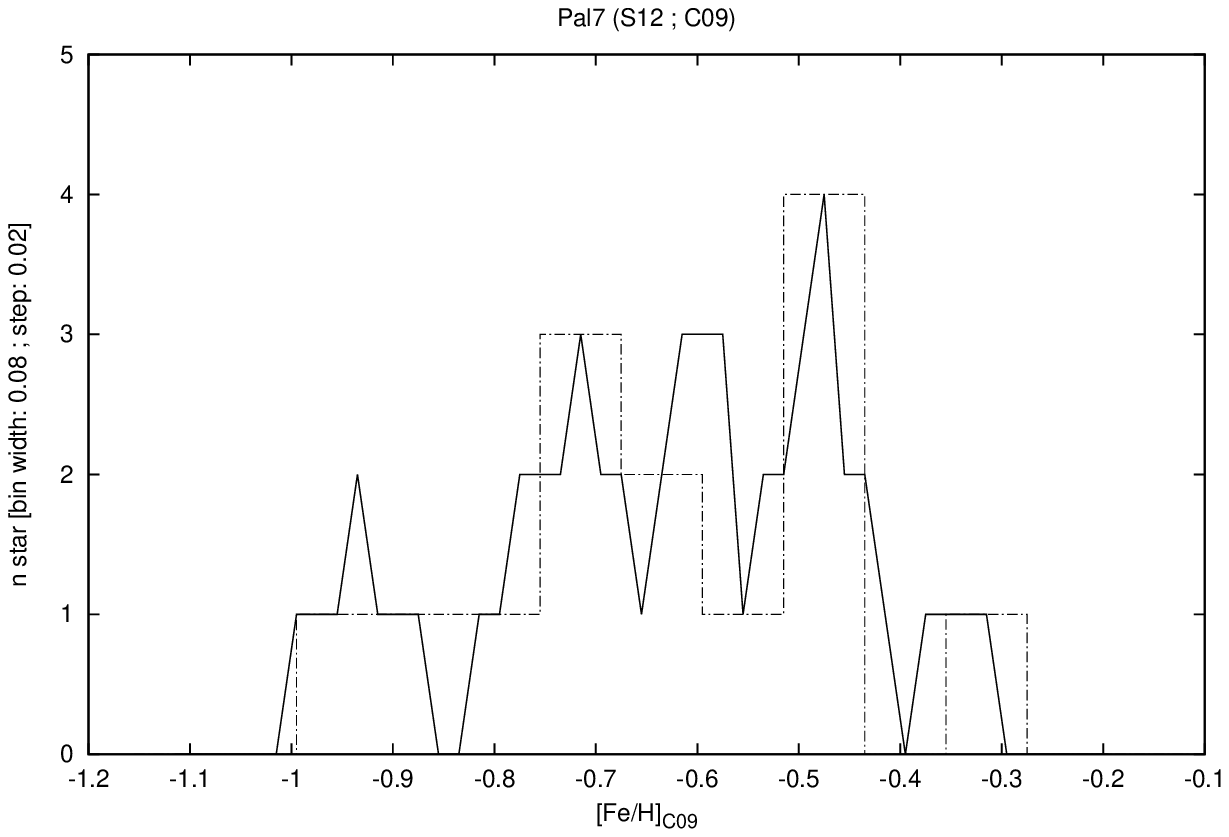}}
\resizebox{\hsize}{!}{\includegraphics[]{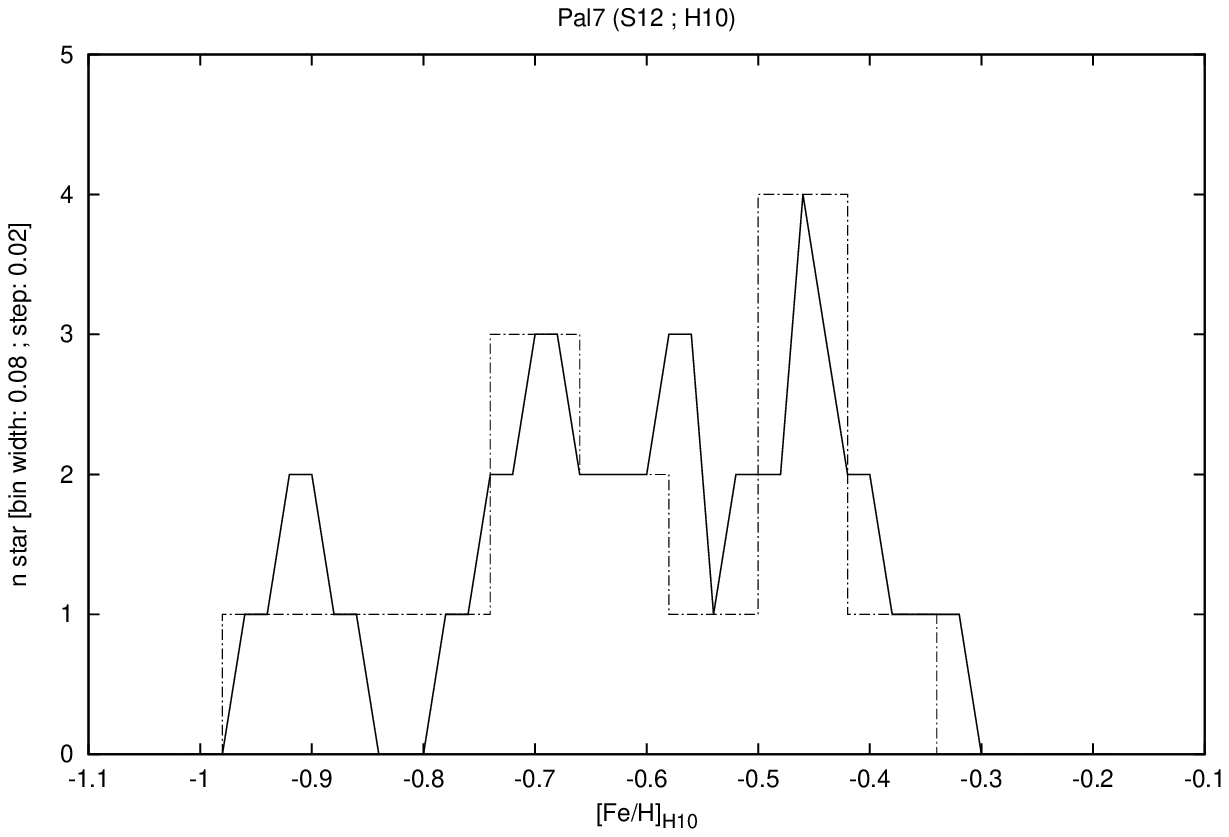}}
\caption{Distribution in metallicity on the C09 (upper plot) and H10 (lower plot) scales for Pal\,7 stars in the S12 sample.}
\label{fig:Pal7}
\end{center}\end{figure}

\begin{figure}[ht!]\begin{center}
\resizebox{\hsize}{!}{\includegraphics[]{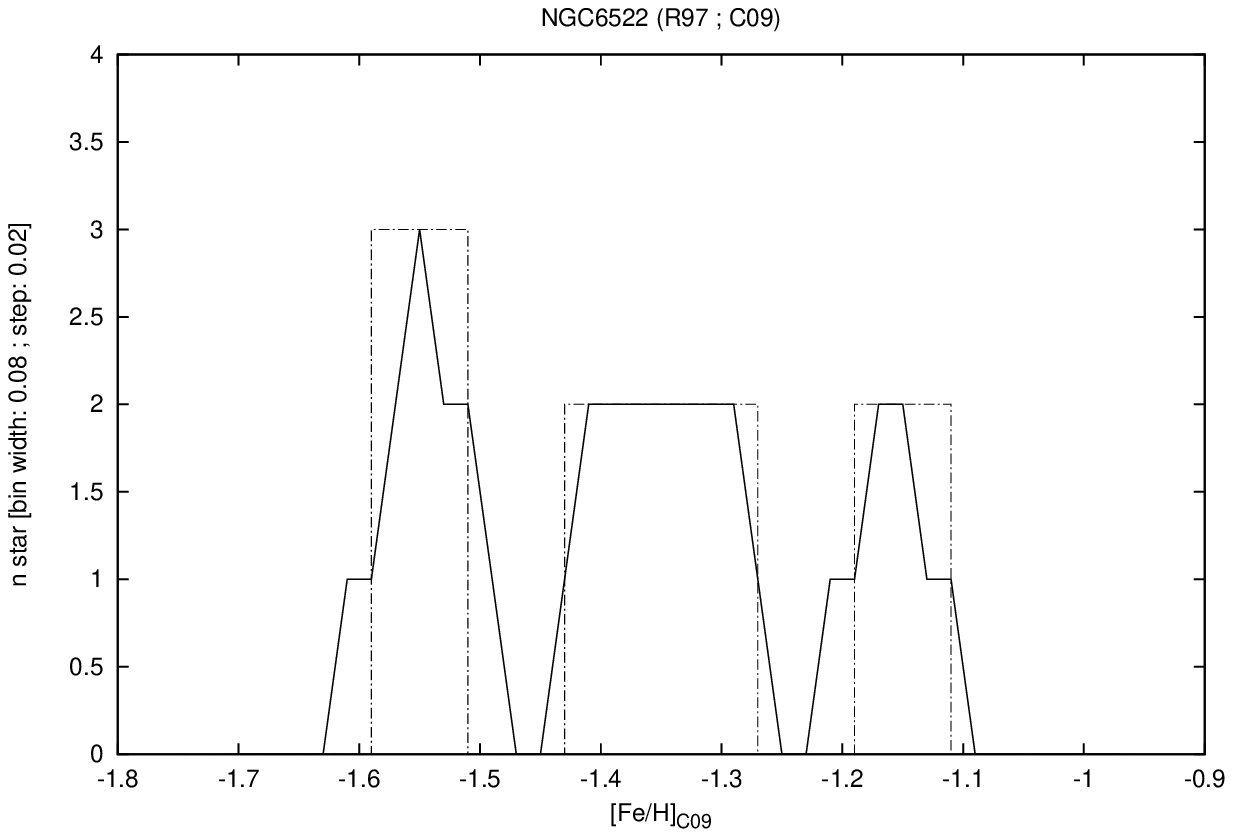}}
\resizebox{\hsize}{!}{\includegraphics[]{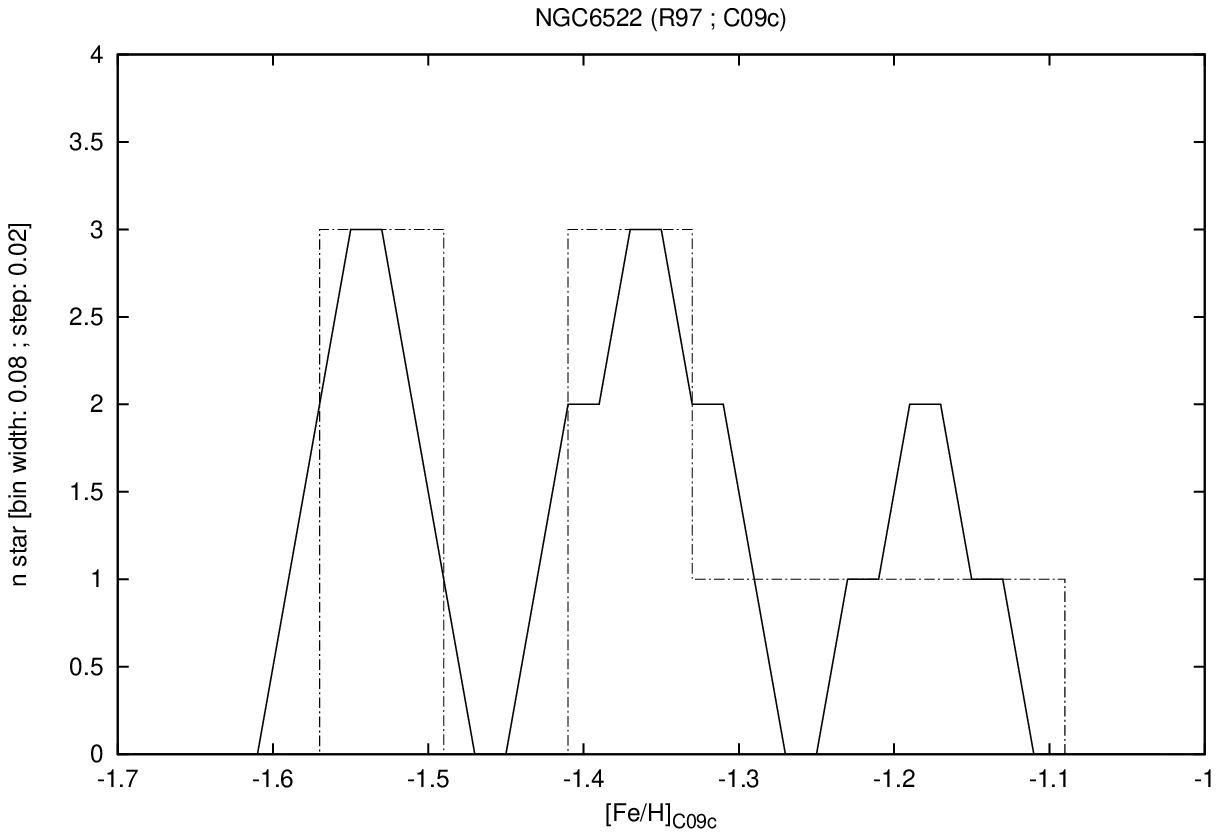}}
\caption{Distribution in metallicity on the C09 (upper plot) and C09c (lower plot) scales for NGC\,6522 stars in the R97 sample.}
\label{fig:6522}
\end{center}\end{figure}

\begin{figure}[ht!]\begin{center}
\resizebox{\hsize}{!}{\includegraphics[]{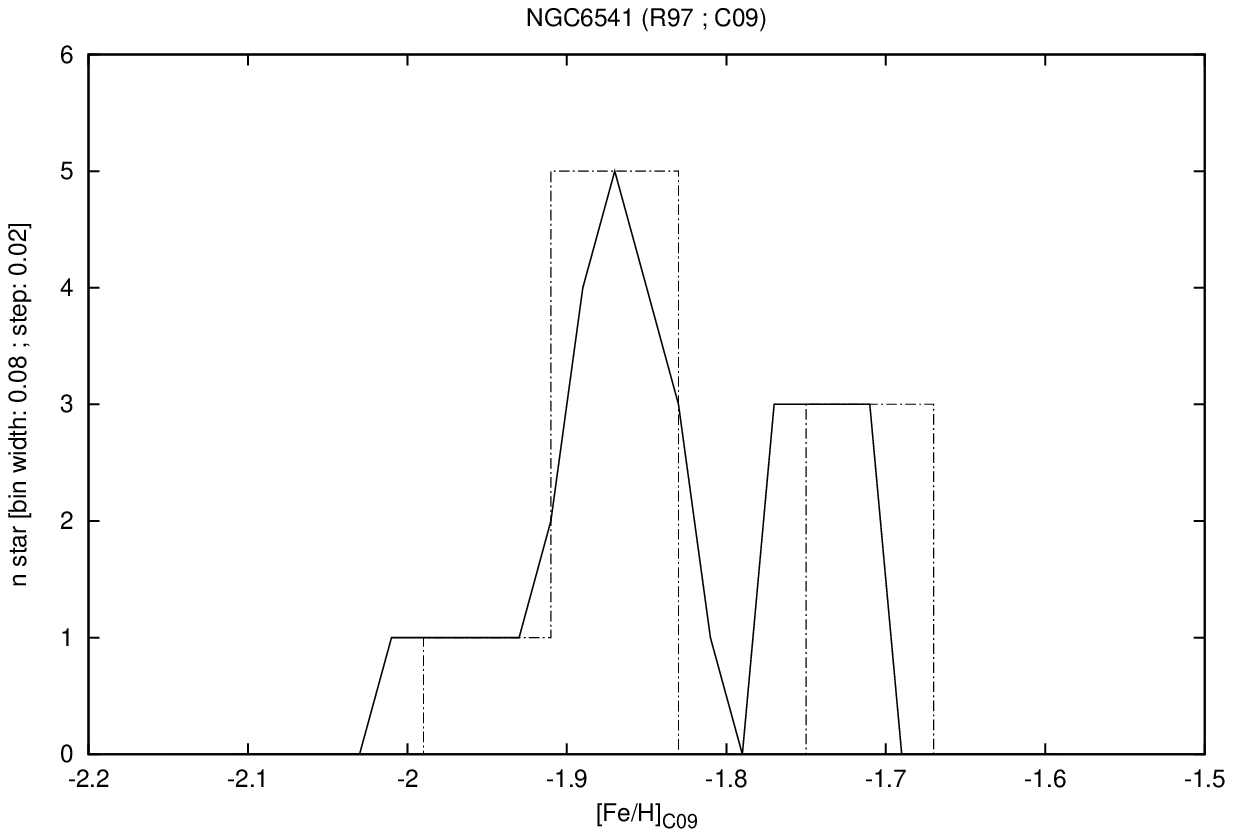}}
\resizebox{\hsize}{!}{\includegraphics[]{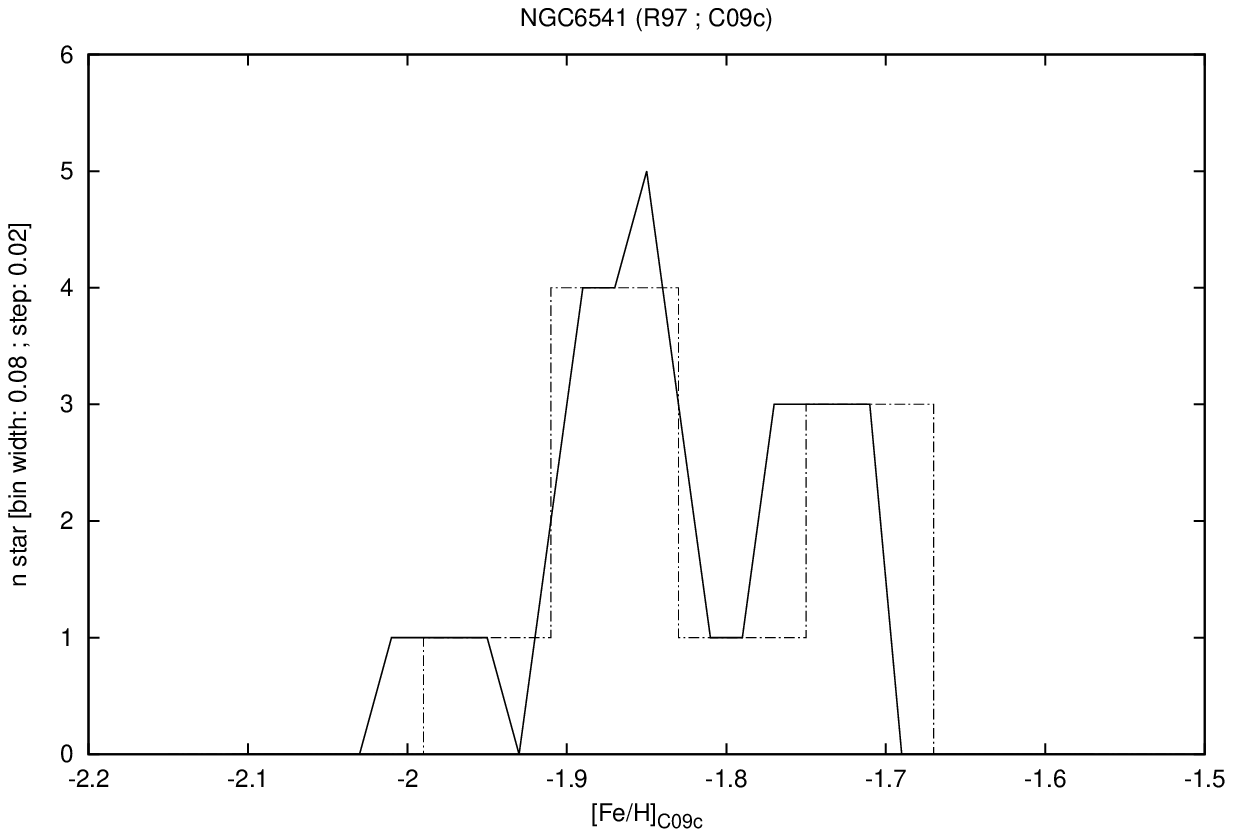}}
\caption{Distribution in metallicity on the C09 (upper plot) and C09c (lower plot) scales for NGC\,6541 stars in the R97 sample.}
\label{fig:6541}
\end{center}\end{figure}

\begin{figure}[ht!]\begin{center}
\resizebox{\hsize}{!}{\includegraphics[]{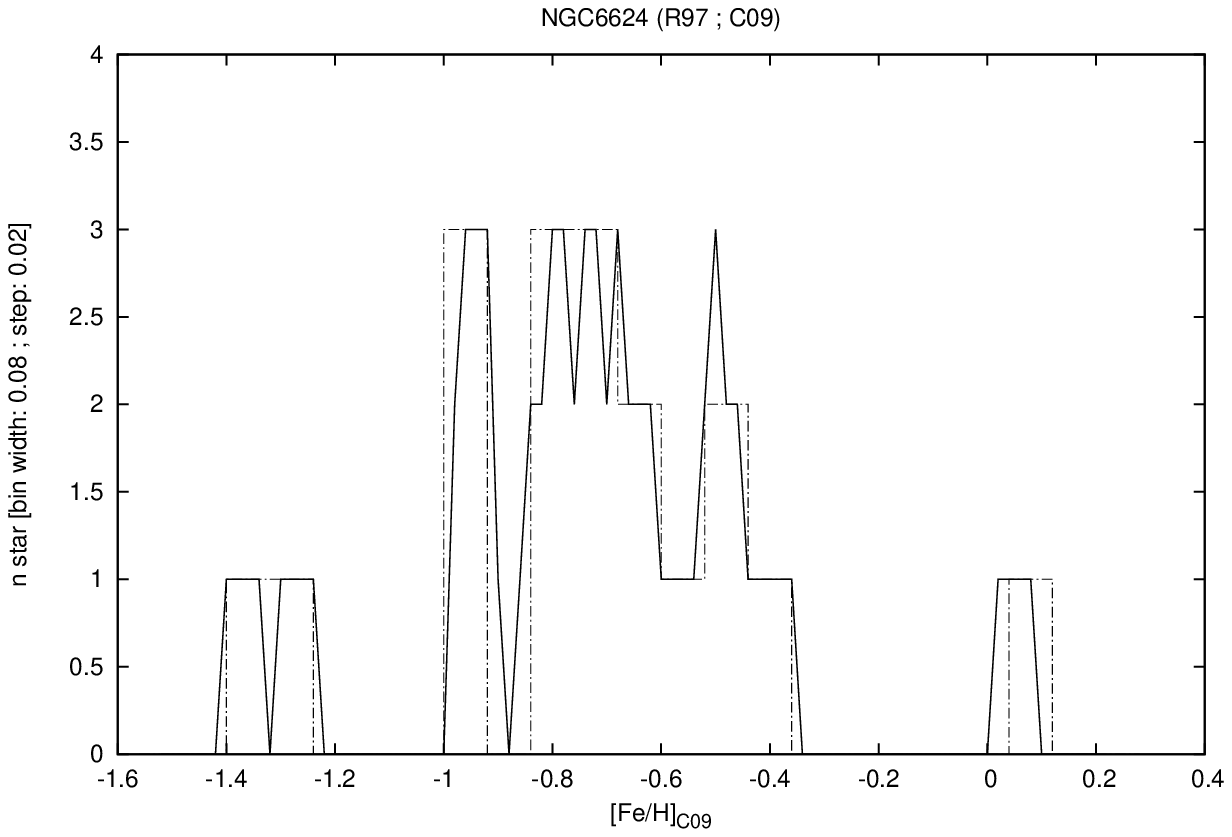}}
\resizebox{\hsize}{!}{\includegraphics[]{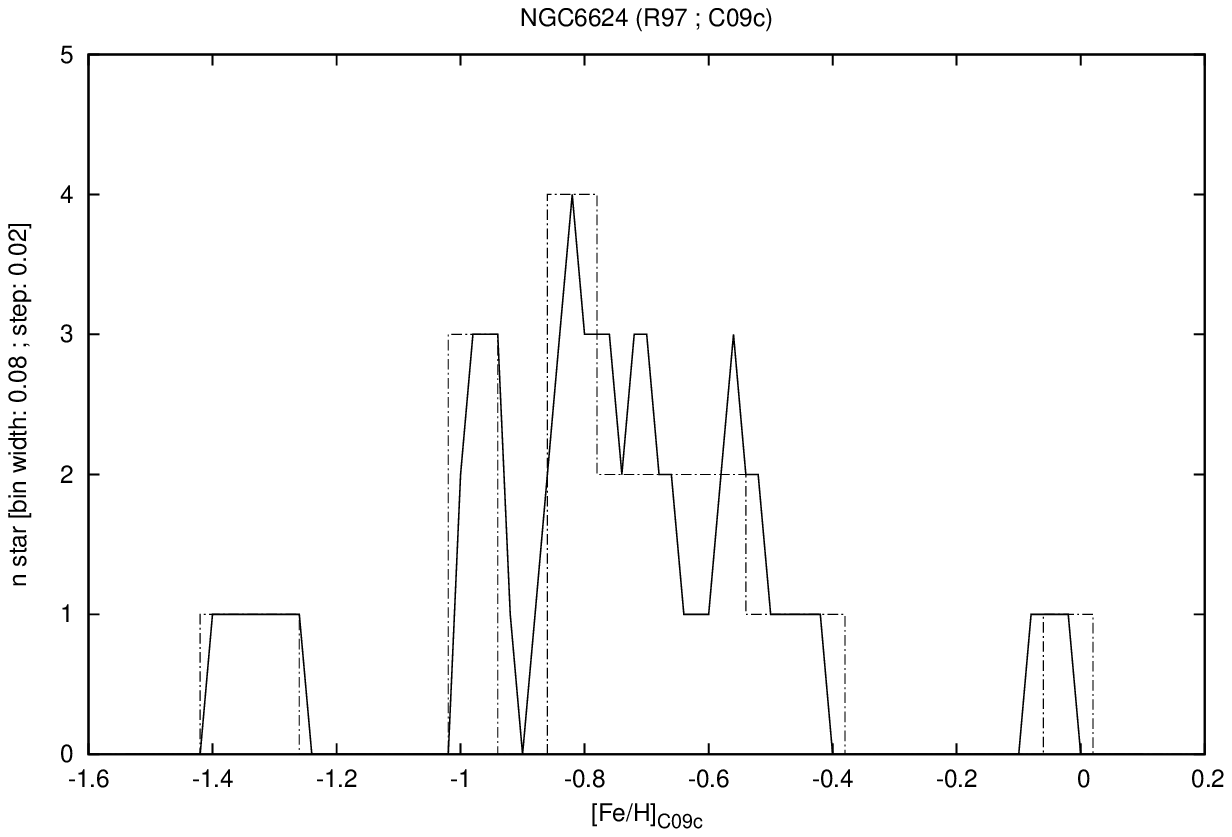}}
\caption{Distribution in metallicity on the C09 (upper plot) and C09c (lower plot) scales for NGC\,6624 stars in the R97 sample.}
\label{fig:6624}
\end{center}\end{figure}

\begin{figure}[ht!]\begin{center}
\resizebox{\hsize}{!}{\includegraphics[]{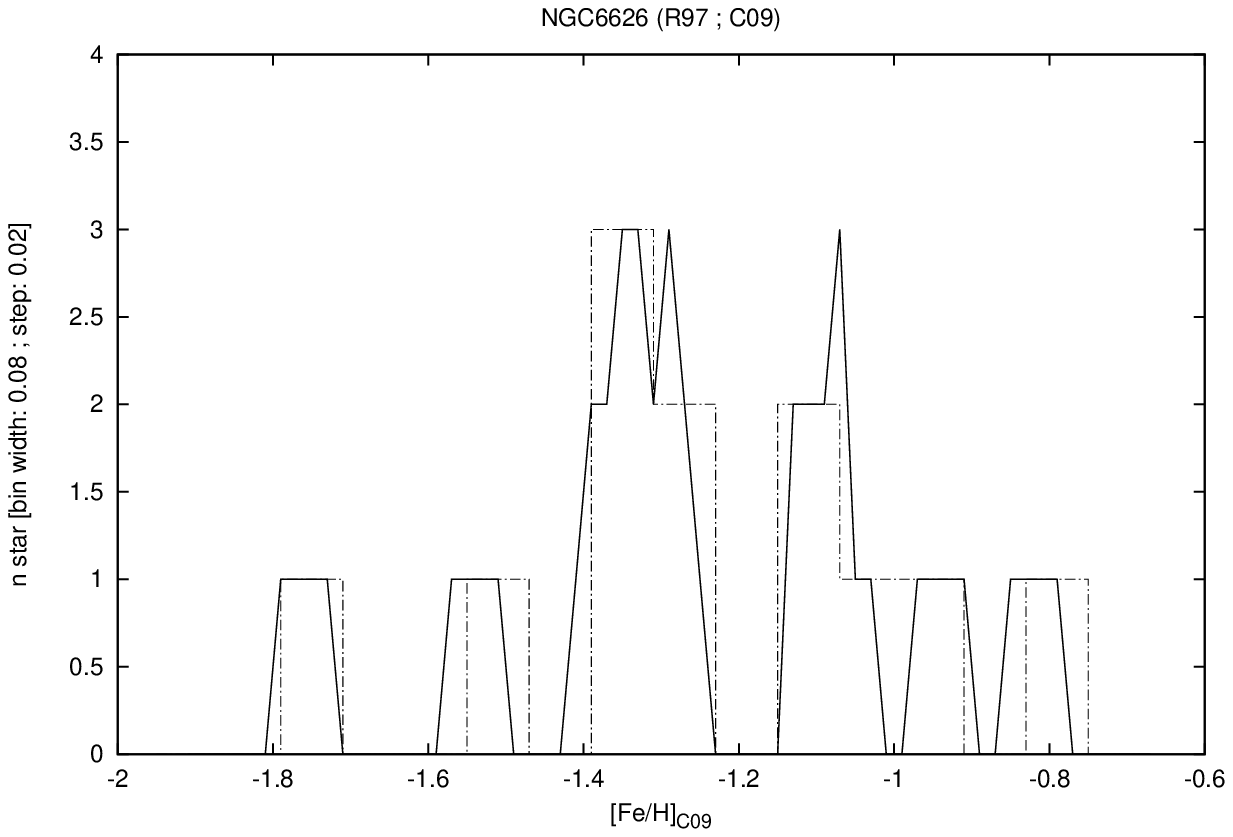}}
\resizebox{\hsize}{!}{\includegraphics[]{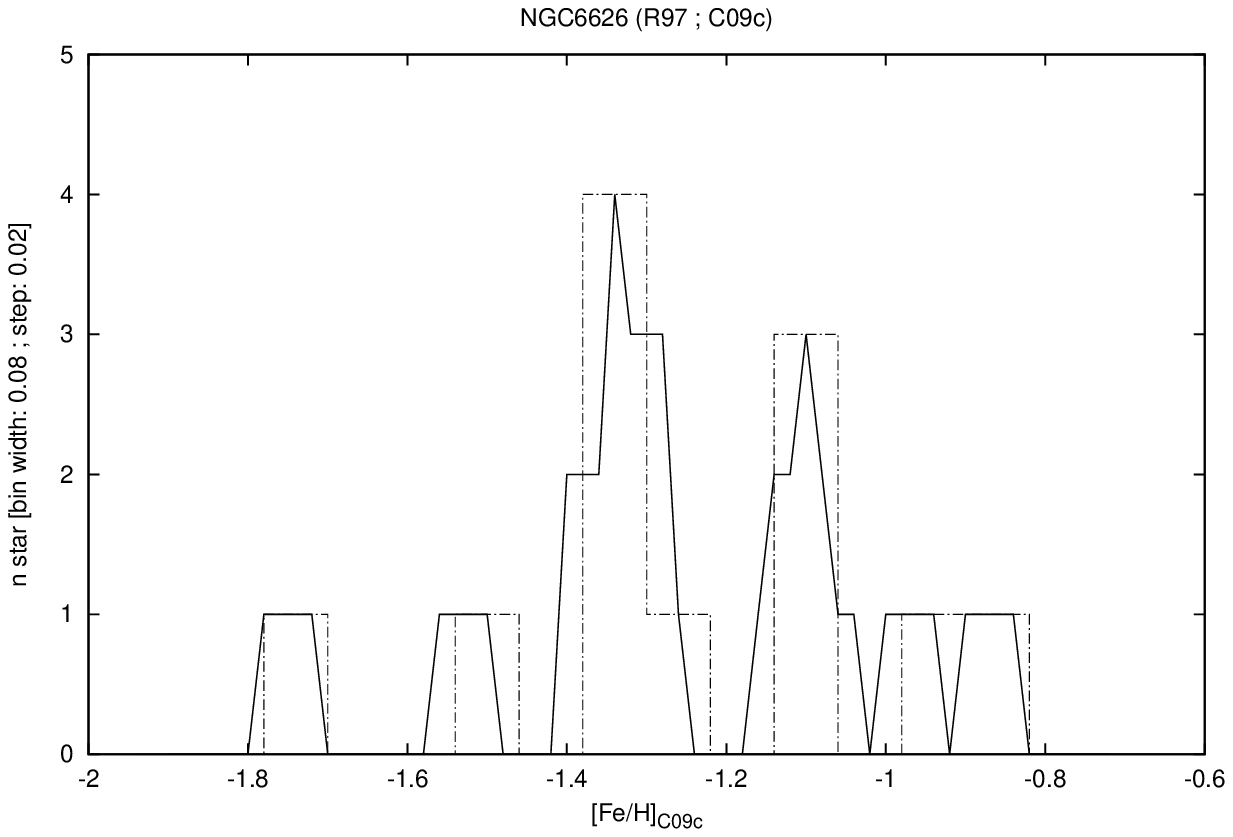}}
\caption{Distribution in metallicity on the C09 (upper plot) and C09c (lower plot) scales for NGC\,6626 stars in the R97 sample.}
\label{fig:6626}
\end{center}\end{figure}

\begin{figure}[ht!]\begin{center}
\resizebox{\hsize}{!}{\includegraphics[]{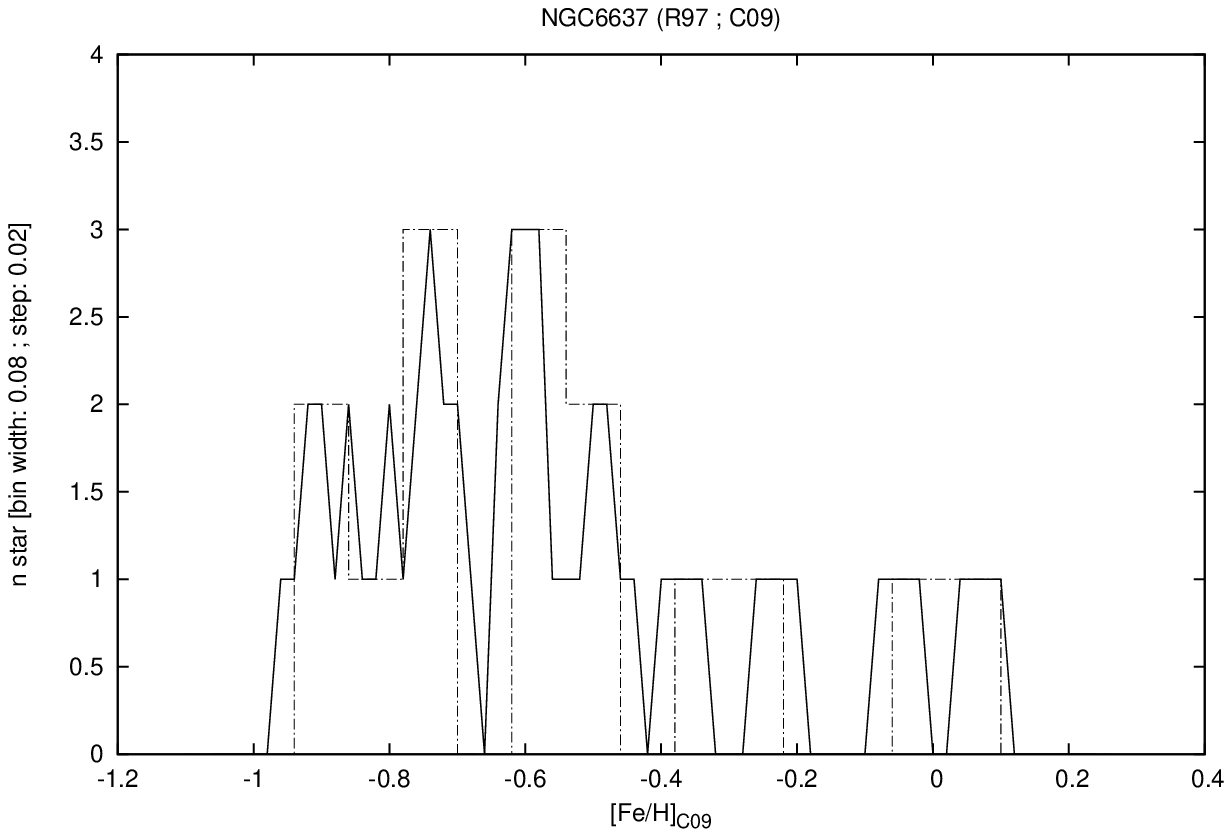}}
\resizebox{\hsize}{!}{\includegraphics[]{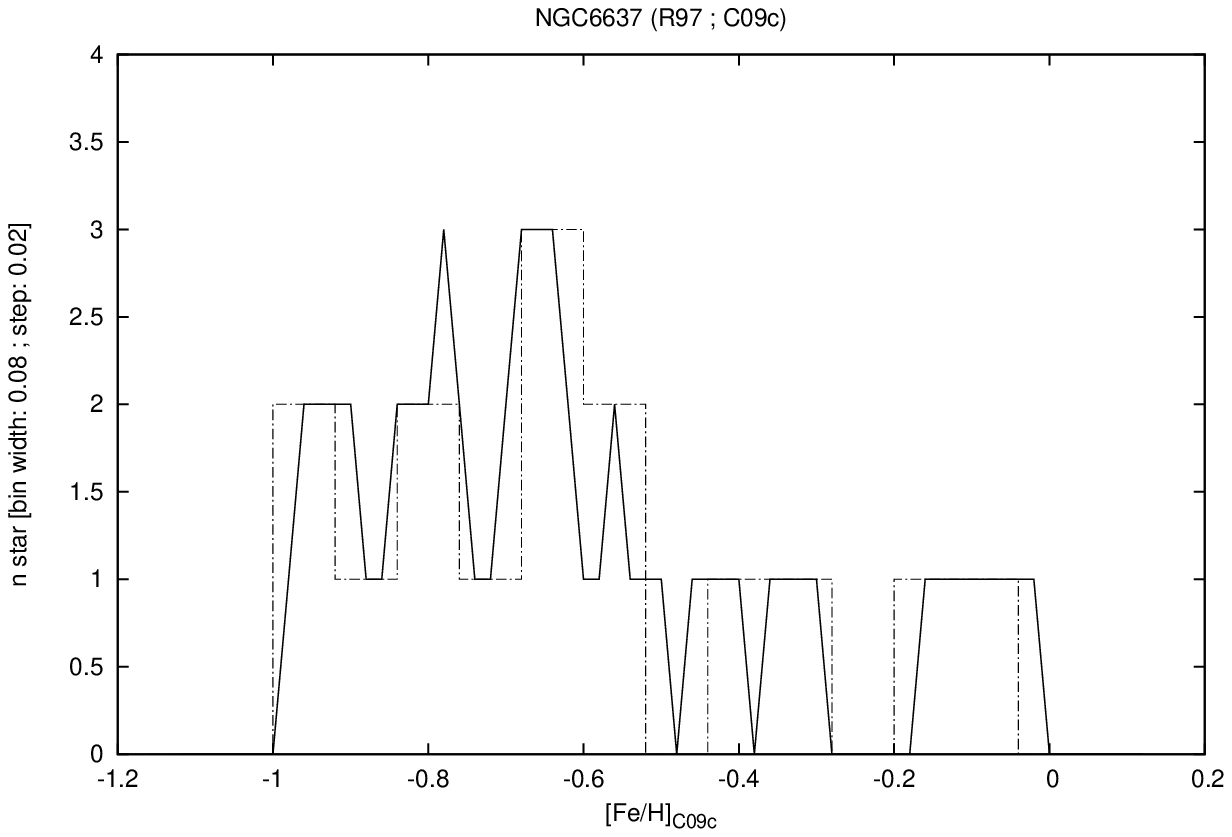}}
\caption{Distribution in metallicity on the C09 (upper plot) and C09c (lower plot) scales for NGC\,6637 stars in the R97 sample.}
\label{fig:6637}
\end{center}\end{figure}

\begin{figure}[ht!]\begin{center}
\resizebox{\hsize}{!}{\includegraphics[]{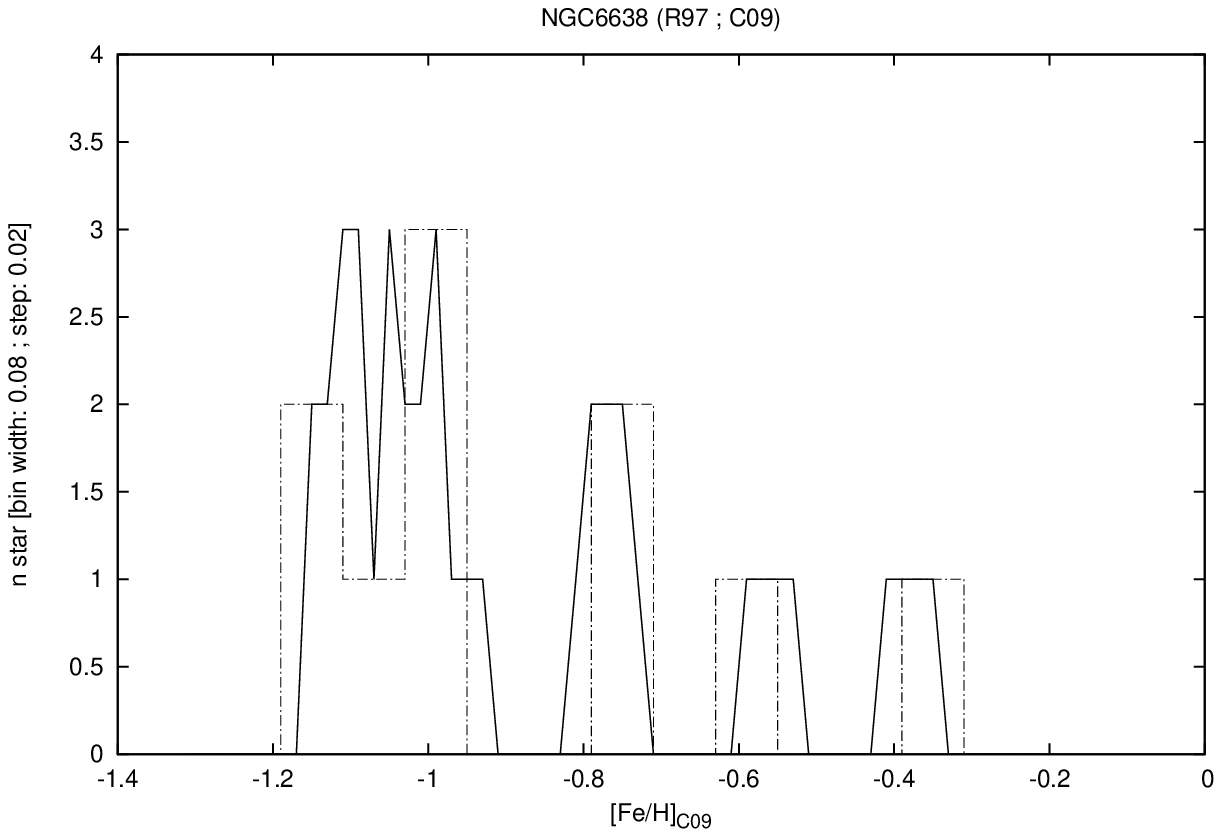}}
\resizebox{\hsize}{!}{\includegraphics[]{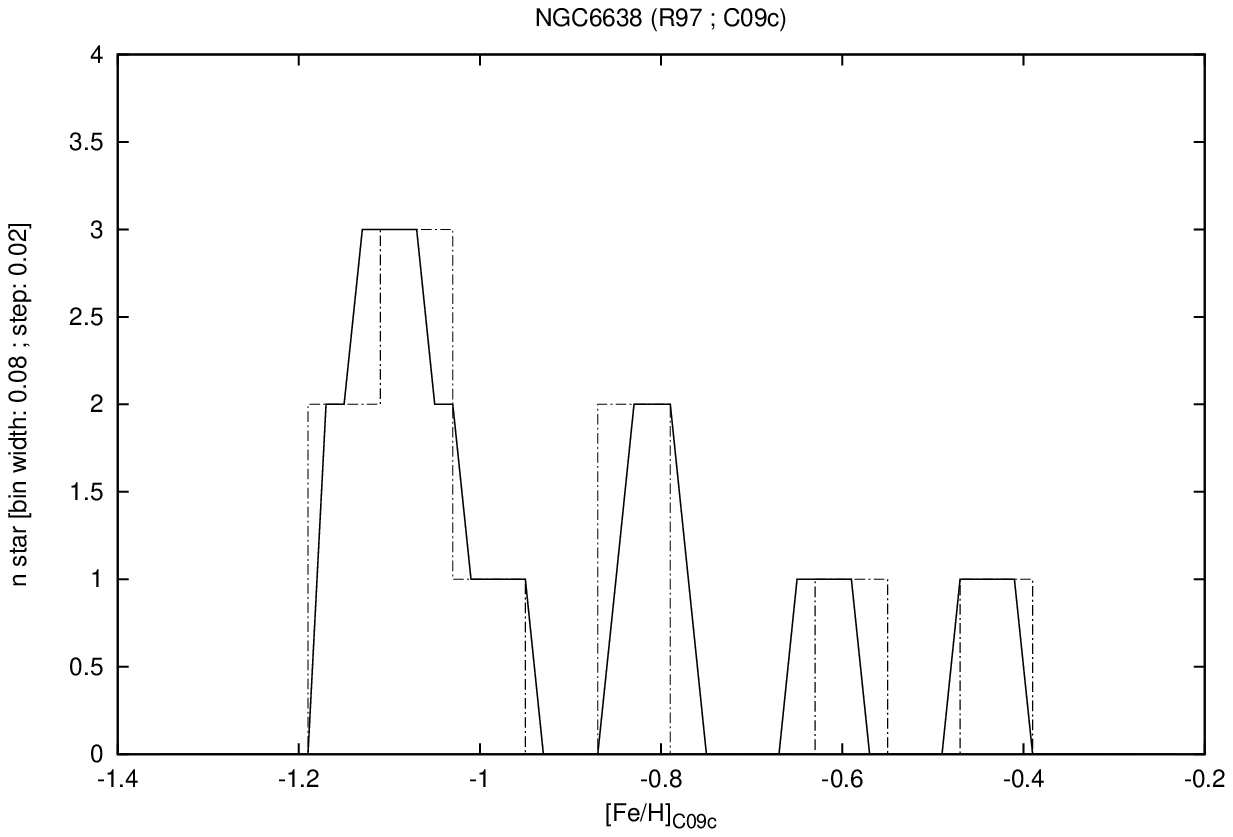}}
\caption{Distribution in metallicity on the C09 (upper plot) and C09c (lower plot) scales for NGC\,6638 stars in the R97 sample.}
\label{fig:6638}
\end{center}\end{figure}

We determined $\Ks{}(HB)$ values for 30 Galactic Globular clusters (GCs).
We then calculated the calibration equations between the metallicity and the reduced equivalent widths (rEW) of the Calcium Triplet (CaT) using nIR photometry.
We considered the GCs in the catalogs of equivalent widths presented in \Sav{} and \Rut{}.
We presented the calibration equations on three metallicity scales: the one presented in \Carr, the C09 with recent high-resolution spectroscopic metallicities (C09c, see Section~\ref{ss:met}), and the values listed in \Har{}.
For the \Sav{} dataset, the cubic relations (Eq. \ref{eq:Fe_W_3SC09}, \ref{eq:Fe_W_3SC09c} and \ref{eq:Fe_W_3SH10}) have the smallest urms, while for the \Rut{} dataset the quadratic relations (Eq. \ref{eq:Fe_W_2RC09}, \ref{eq:Fe_W_2RC09c} and \ref{eq:Fe_W_2RH10}) yield the best fits.
The analysis of these solutions and of the metallicities obtained through them provide the following important points:
\begin{itemize}
\item The comparison between the calibration equations on the three metallicity scales demonstrates that, within $\sim 0.05$~dex, the three scales are equivalent.
This similarity is not surprising since the other two scales are based on \Carr{}, even if they differ in individual values and in the determination.
This result assures the overall validity of the obtained equations, because the differences in the metallicities of some GCs slightly affect  the fit.
The validity of the solutions for the \Sav{} dataset is assured by the small scatter shown by the \Carr{} calibrator GCs, that can be assumed to have a well-determined metallicity (7 out 8 of the calibrators included in the \Sav{} dataset present $|\Delta\FeH{}| \leq 0.05$~dex).

\item The calibration is independent of the passband used to calculate the rEW, as demonstrated by the comparison between the calibrations obtained from visible ($\sim 0.5\;\mu$m) and nIR ($\sim 2\;\mu$m)  photometry.
This result permits one to apply our relations even to $\rEW$ calculated with magnitudes from other passbands.

\item The comparison between the metallicity distributions for NGC\,6656 (M\,22) presented in \citet{DaCosta2009} and the one obtained from our calibrations (see Figures~\ref{fig:6656} and \ref{fig:6656V}) demonstrates that nIR CaT rEW are less affected by internal scatter (most likely produced by differential reddening in this cluster).
The separation between the two main peaks in our distribution is in agreement with the metallicity difference between stars rich and poor in s-process element found by \citet{Marino2009}, while the absolute positions are shift of $\sim 0.08$~dex.
The well-defined peaks demonstrates that the nIR CaT rEW is very promising as a powerful tool to study metallicity distributions of GCs, especially clusters belonging to the Galactic bulge, where differential reddening is much stronger.

\item The analysis of the metallicity distribution of the our sample (Figures~\ref{fig:2808}-\ref{fig:6638}) suggests peculiarities in some of the GCs.
NGC\,6528 and NGC\,6553 (the two most metal-rich ones) show a almost flat distribution that cover approximately 0.5-1~dex.
For NGC\,6553, a similar spread of 0.5~dex was already pointed out by \citet{AlvesBrito2006} for HR spectroscopic values present in the literature.
\item NGC\,3201 presents  a spread in the metallicity distribution that suggests the presence of a complex metallicity population, as already proposed by \citet{Simmerer2013}, where the metallicity of 24 RGB stars were analyzed.
This was also supported by \citet{Munoz2013}.
Our distributions present a similar spread in [Fe/H] of 0.3~dex, but also a hint of a double peak.
7 of the 24 stars are in common with the 17 selected from \Rut{} catalog, while only one is present in the \Sav{} sample.
The stars in common with \Rut{} present a metallicity difference less than $\sim 0.1$~dex with a mean value of $+0.04$~dex, and the same value for the difference with the \Carr{} metallicity; 5 stars present an absolute metallicity difference less than $0.05$~dex with a mean difference of $+0.02$~dex, where \citet{Simmerer2013} declared a total uncertainty in [Fe/H] less than 0.15~dex.
This result confirms the accuracy of the [Fe/H] values obtained through the CaT method with respect to those obtained with HR spectroscopy.
\item We detected the presence of possible complex distributions also for other GCs, namely  NGC\,2808, NGC\,6121 (M\,4), NGC\,6356, NGC\,6440, NGC\,6522, NGC\,6541, NGC\,6569, NGC\,6624, NGC\,6626 (M\,28), NGC\,6637 (M\,69), NGC\,6638, and Pal\,7 (IC\,1276).
NGC\,6440 and NGC\,6569 are also suggested to have a double horizontal branch in \citet{Mauro2012}, while  NGC\,6626 is a metal-poor cluster that presents parameters very similar to NGC\,6656, such as metallicity, absolute magnitude and CMD shape.
For these three GCs, HR spectra were obtained and are under study.
\end{itemize}

{A comparison with previous results of metallicities obtained from CaT method with NIR photometry did not lead to relevant conclusions.}
{Both \citet{WarrenCole2009} and \citet{Lane2010b} calibrated on different scales (\CarGr{} and Harris 1996, respectively) and, except for NGC\,6656, all the other GCs in common are also calibrators of \Carr{}, whose metallicities are in good agreement with our results.}

\begin{acknowledgement}
The authors thank the referee and R. Lane for the useful comments.
FM is thankful for the financial support from FONDECYT for project 3140177.
DG, FM and RC gratefully acknowledge support from the Chilean BASAL   Centro de Excelencia en Astrof\'isica y Tecnolog\'ias Afines (CATA) grant PFB-06/2007.
RC is thankful for the financial support from Fondo GEMINI-CONICYT 32100008.
ANC received support from Comite Mixto ESO-Gobierno de Chile and GEMINI- CONICYT No. 32110005
This publication makes use of data products from the Two Micron All Sky Survey, which is a joint project of the University of Massachusetts and the Infrared Processing and Analysis Center/California Institute of Technology, funded by the National Aeronautics and Space Administration and the National Science Foundation.
\end{acknowledgement}

\bibliographystyle{aa}
\bibliography{bib-CaT,bib-VVV,bib-GC,bib-Harris,bib-SkZpipe,bib-hsw,bib-Met,bib-Surveys,bib-StEvol}

\begin{landscape}

\begin{table}
\begin{center}
\caption{Metallicities for the S12 GCs in C09 scale.
Column (1) lists the cluster ID (the GCs marked with * were not used to calculate the calibration).
Column (2) is the metallicity according the scale.
Column (3) is the calculated rEW.
Column (Xa) is the metallicity based on the calibration using a polynomial of degree X.
Column (Xb) is the difference between the calculated and the scale metallicity.
Column (Xc) is the former difference in r.m.s.}
\label{tab:SC09}
\begin{tabular}{|l|c|c||c|c|c||c|c|c||c|c|c||c|c|c||c|c|c|}
\hline
GC & [Fe/H] &  $\rEW$ &  [Fe/H]$_1$& $\Delta$ & $\Delta/\sigma$ &  [Fe/H]$_2$& $\Delta$ & $\Delta/\sigma$ &  [Fe/H]$_3$& $\Delta$ & $\Delta/\sigma$ &  [Fe/H]$_4$& $\Delta$ & $\Delta/\sigma$ &  [Fe/H]$_5$& $\Delta$ & $\Delta/\sigma$  \\(1) & (2) & (3) & (Ia) &  (Ib) &  (Ic) & (IIa) &  (IIb) &  (IIc) & (IIIa) &  (IIIb) &  (IIIc) & (IVa) &  (IVb) &  (IVc) & (Va) &  (Vb) &  (Vc) \\
\hline
NGC2808  & -1.18 & 4.47   & -1.12 & -0.06 & -0.2  & -1.14 & -0.04 & -0.2  & -1.19 & +0.01 & +0.0  & -1.19 & +0.01 & +0.0  & -1.18 & -0.00 & -0.0 \\
\hline
NGC3201  & -1.51 & 3.91   & -1.40 & -0.11 & -0.4  & -1.47 & -0.04 & -0.2  & -1.47 & -0.04 & -0.2  & -1.47 & -0.04 & -0.2  & -1.47 & -0.04 & -0.1 \\
\hline
NGC6121  & -1.18 & 4.40   & -1.16 & -0.02 & -0.1  & -1.19 & +0.01 & +0.0  & -1.23 & +0.05 & +0.2  & -1.23 & +0.05 & +0.2  & -1.22 & +0.04 & +0.2 \\
\hline
NGC6139* & -1.71 & 3.92   & -1.40 & -0.31 & -1.1  & -1.47 & -0.24 & -1.1  & -1.47 & -0.24 & -1.1  & -1.47 & -0.24 & -1.1  & -1.47 & -0.24 & -1.0 \\
\hline
NGC6254  & -1.57 & 3.70   & -1.51 & -0.06 & -0.2  & -1.59 & +0.02 & +0.1  & -1.57 & -0.00 & -0.0  & -1.57 & -0.00 & -0.0  & -1.58 & +0.01 & +0.0 \\
\hline
NGC6325  & -1.37 & 4.36   & -1.17 & -0.20 & -0.7  & -1.21 & -0.16 & -0.7  & -1.25 & -0.12 & -0.6  & -1.25 & -0.12 & -0.5  & -1.24 & -0.13 & -0.5 \\
\hline
NGC6356  & -0.35 & 5.28   & -0.70 & +0.35 & +1.2  & -0.60 & +0.25 & +1.1  & -0.58 & +0.23 & +1.1  & -0.58 & +0.23 & +1.0  & -0.59 & +0.24 & +1.0 \\
\hline
NGC6380* & -0.40 & 5.11   & -0.79 & +0.39 & +1.3  & -0.72 & +0.32 & +1.4  & -0.73 & +0.33 & +1.5  & -0.73 & +0.33 & +1.5  & -0.74 & +0.34 & +1.4 \\
\hline
NGC6397  & -1.99 & 2.66   & -2.04 & +0.05 & +0.2  & -2.06 & +0.07 & +0.3  & -1.99 & -0.00 & -0.0  & -1.99 & -0.00 & -0.0  & -1.98 & -0.01 & -0.0 \\
\hline
NGC6440  & -0.20 & 5.48   & -0.60 & +0.40 & +1.4  & -0.45 & +0.25 & +1.1  & -0.38 & +0.18 & +0.8  & -0.38 & +0.18 & +0.8  & -0.39 & +0.19 & +0.8 \\
\hline
NGC6441  & -0.44 & 5.20   & -0.74 & +0.30 & +1.0  & -0.65 & +0.21 & +0.9  & -0.65 & +0.21 & +1.0  & -0.65 & +0.21 & +0.9  & -0.66 & +0.22 & +0.9 \\
\hline
NGC6528  & +0.07 & 5.65   & -0.51 & +0.58 & +2.0  & -0.32 & +0.39 & +1.7  & -0.19 & +0.26 & +1.2  & -0.19 & +0.26 & +1.1  & -0.19 & +0.26 & +1.1 \\
\hline
NGC6553  & -0.16 & 5.76   & -0.46 & +0.30 & +1.0  & -0.23 & +0.07 & +0.3  & -0.06 & -0.10 & -0.5  & -0.06 & -0.10 & -0.5  & -0.04 & -0.12 & -0.5 \\
\hline
NGC6558  & -1.37 & 4.62   & -1.04 & -0.33 & -1.1  & -1.05 & -0.32 & -1.4  & -1.09 & -0.28 & -1.3  & -1.09 & -0.28 & -1.2  & -1.08 & -0.29 & -1.2 \\
\hline
NGC6569  & -0.72 & 4.44   & -1.13 & +0.41 & +1.4  & -1.16 & +0.44 & +1.9  & -1.20 & +0.48 & +2.3  & -1.20 & +0.48 & +2.2  & -1.19 & +0.47 & +2.0 \\
\hline
NGC6656  & -1.70 & 3.07   & -1.83 & +0.13 & +0.5  & -1.89 & +0.19 & +0.8  & -1.82 & +0.12 & +0.6  & -1.82 & +0.12 & +0.5  & -1.83 & +0.13 & +0.5 \\
\hline
NGC6838  & -0.82 & 5.07   & -0.81 & -0.01 & -0.0  & -0.75 & -0.07 & -0.3  & -0.77 & -0.05 & -0.2  & -0.77 & -0.05 & -0.2  & -0.77 & -0.05 & -0.2 \\
\hline
NGC7078  & -2.33 & 2.01   & -2.38 & +0.05 & +0.2  & -2.28 & -0.05 & -0.2  & -2.33 & -0.00 & -0.0  & -2.33 & -0.00 & -0.0  & -2.33 & +0.00 & +0.0 \\
\hline
Pal7   * & -0.65 & 5.22   & -0.73 & +0.08 & +0.3  & -0.64 & -0.01 & -0.1  & -0.63 & -0.02 & -0.1  & -0.63 & -0.02 & -0.1  & -0.64 & -0.01 & -0.0 \\
\hline

\end{tabular}
\end{center}
\end{table}

\begin{table}
\begin{center}
\caption{Metallicities for the S12 GCs in C09c scale.
Column (1) lists the cluster ID (the GCs marked with * were not used to calculate the calibration).
Column (2) is the metallicity according the scale.
Column (3) is the calculated rEW.
Column (Xa) is the metallicity based on the calibration using a polynomial of degree X.
Column (Xb) is the difference between the calculated and the scale metallicity.
Column (Xc) is the former difference in r.m.s.}
\label{tab:SC09c}
\begin{tabular}{|l|c|c||c|c|c||c|c|c||c|c|c||c|c|c||c|c|c|}
\hline
GC & [Fe/H] &  $\rEW$ &  [Fe/H]$_1$& $\Delta$ & $\Delta/\sigma$ &  [Fe/H]$_2$& $\Delta$ & $\Delta/\sigma$ &  [Fe/H]$_3$& $\Delta$ & $\Delta/\sigma$ &  [Fe/H]$_4$& $\Delta$ & $\Delta/\sigma$ &  [Fe/H]$_5$& $\Delta$ & $\Delta/\sigma$  \\(1) & (2) & (3) & (Ia) &  (Ib) &  (Ic) & (IIa) &  (IIb) &  (IIc) & (IIIa) &  (IIIb) &  (IIIc) & (IVa) &  (IVb) &  (IVc) & (Va) &  (Vb) &  (Vc) \\
\hline
NGC2808  & -1.18 & 4.47   & -1.11 & -0.07 & -0.3  & -1.13 & -0.05 & -0.3  & -1.16 & -0.02 & -0.1  & -1.16 & -0.02 & -0.1  & -1.13 & -0.05 & -0.2 \\
\hline
NGC3201  & -1.51 & 3.91   & -1.40 & -0.11 & -0.5  & -1.45 & -0.06 & -0.3  & -1.46 & -0.05 & -0.3  & -1.46 & -0.05 & -0.3  & -1.46 & -0.05 & -0.2 \\
\hline
NGC6121  & -1.18 & 4.40   & -1.15 & -0.03 & -0.1  & -1.18 & -0.00 & -0.0  & -1.21 & +0.03 & +0.1  & -1.21 & +0.03 & +0.1  & -1.18 & -0.00 & -0.0 \\
\hline
NGC6139* & -1.71 & 3.92   & -1.39 & -0.32 & -1.4  & -1.45 & -0.26 & -1.4  & -1.45 & -0.26 & -1.5  & -1.46 & -0.25 & -1.4  & -1.46 & -0.25 & -1.2 \\
\hline
NGC6254  & -1.57 & 3.70   & -1.51 & -0.06 & -0.3  & -1.57 & +0.00 & +0.0  & -1.56 & -0.01 & -0.1  & -1.56 & -0.01 & -0.0  & -1.59 & +0.02 & +0.1 \\
\hline
NGC6325  & -1.37 & 4.36   & -1.17 & -0.20 & -0.9  & -1.20 & -0.17 & -0.9  & -1.23 & -0.14 & -0.8  & -1.23 & -0.14 & -0.8  & -1.20 & -0.17 & -0.9 \\
\hline
NGC6356  & -0.35 & 5.28   & -0.70 & +0.35 & +1.5  & -0.60 & +0.25 & +1.4  & -0.59 & +0.24 & +1.4  & -0.59 & +0.24 & +1.3  & -0.62 & +0.27 & +1.4 \\
\hline
NGC6380* & -0.40 & 5.11   & -0.78 & +0.38 & +1.6  & -0.72 & +0.32 & +1.8  & -0.72 & +0.32 & +1.9  & -0.72 & +0.32 & +1.8  & -0.74 & +0.34 & +1.7 \\
\hline
NGC6397  & -1.99 & 2.66   & -2.04 & +0.05 & +0.2  & -2.05 & +0.06 & +0.3  & -2.00 & +0.01 & +0.1  & -2.00 & +0.01 & +0.0  & -1.98 & -0.01 & -0.0 \\
\hline
NGC6440  & -0.20 & 5.48   & -0.59 & +0.39 & +1.7  & -0.46 & +0.26 & +1.4  & -0.41 & +0.21 & +1.2  & -0.41 & +0.21 & +1.2  & -0.45 & +0.25 & +1.2 \\
\hline
NGC6441  & -0.44 & 5.20   & -0.73 & +0.29 & +1.3  & -0.65 & +0.21 & +1.2  & -0.65 & +0.21 & +1.2  & -0.65 & +0.21 & +1.1  & -0.68 & +0.24 & +1.2 \\
\hline
NGC6528  & -0.17 & 5.65   & -0.50 & +0.33 & +1.4  & -0.33 & +0.16 & +0.9  & -0.24 & +0.07 & +0.4  & -0.25 & +0.08 & +0.5  & -0.25 & +0.08 & +0.4 \\
\hline
NGC6553  & -0.16 & 5.76   & -0.45 & +0.29 & +1.2  & -0.25 & +0.09 & +0.5  & -0.13 & -0.03 & -0.2  & -0.15 & -0.01 & -0.0  & -0.11 & -0.05 & -0.3 \\
\hline
NGC6558  & -0.97 & 4.62   & -1.03 & +0.06 & +0.3  & -1.04 & +0.07 & +0.4  & -1.07 & +0.10 & +0.6  & -1.06 & +0.09 & +0.5  & -1.04 & +0.07 & +0.4 \\
\hline
NGC6569  & -0.79 & 4.44   & -1.12 & +0.33 & +1.4  & -1.15 & +0.36 & +2.0  & -1.18 & +0.39 & +2.2  & -1.18 & +0.39 & +2.2  & -1.15 & +0.36 & +1.8 \\
\hline
NGC6656  & -1.70 & 3.07   & -1.83 & +0.13 & +0.5  & -1.87 & +0.17 & +1.0  & -1.83 & +0.13 & +0.7  & -1.82 & +0.12 & +0.7  & -1.86 & +0.16 & +0.8 \\
\hline
NGC6838  & -0.82 & 5.07   & -0.80 & -0.02 & -0.1  & -0.75 & -0.07 & -0.4  & -0.76 & -0.06 & -0.3  & -0.75 & -0.07 & -0.4  & -0.77 & -0.05 & -0.2 \\
\hline
NGC7078  & -2.33 & 2.01   & -2.37 & +0.04 & +0.2  & -2.29 & -0.04 & -0.2  & -2.32 & -0.01 & -0.0  & -2.33 & -0.00 & -0.0  & -2.33 & +0.00 & +0.0 \\
\hline
Pal7   * & -0.65 & 5.22   & -0.72 & +0.07 & +0.3  & -0.64 & -0.01 & -0.0  & -0.64 & -0.01 & -0.1  & -0.63 & -0.02 & -0.1  & -0.66 & +0.01 & +0.1 \\
\hline

\end{tabular}
\end{center}
\end{table}

\begin{table}
\begin{center}
\caption{Metallicities for the S12 GCs in H10 scale.
Column (1) lists the cluster ID (the GCs marked with * were not used to calculate the calibration).
Column (2) is the metallicity according the scale.
Column (3) is the calculated rEW.
Column (Xa) is the metallicity based on the calibration using a polynomial of degree X.
Column (Xb) is the difference between the calculated and the scale metallicity.
Column (Xc) is the former difference in r.m.s.}
\label{tab:SH10}
\begin{tabular}{|l|c|c||c|c|c||c|c|c||c|c|c||c|c|c||c|c|c|}
\hline
GC & [Fe/H] &  $\rEW$ &  [Fe/H]$_1$& $\Delta$ & $\Delta/\sigma$ &  [Fe/H]$_2$& $\Delta$ & $\Delta/\sigma$ &  [Fe/H]$_3$& $\Delta$ & $\Delta/\sigma$ &  [Fe/H]$_4$& $\Delta$ & $\Delta/\sigma$ &  [Fe/H]$_5$& $\Delta$ & $\Delta/\sigma$  \\(1) & (2) & (3) & (Ia) &  (Ib) &  (Ic) & (IIa) &  (IIb) &  (IIc) & (IIIa) &  (IIIb) &  (IIIc) & (IVa) &  (IVb) &  (IVc) & (Va) &  (Vb) &  (Vc) \\
\hline
NGC2808  & -1.14 & 4.47   & -1.05 & -0.09 & -0.5  & -1.13 & -0.01 & -0.1  & -1.15 & +0.01 & +0.1  & -1.17 & +0.03 & +0.1  & -1.16 & +0.02 & +0.1 \\
\hline
NGC3201  & -1.59 & 3.91   & -1.38 & -0.21 & -1.1  & -1.48 & -0.11 & -0.7  & -1.46 & -0.13 & -0.8  & -1.49 & -0.10 & -0.6  & -1.49 & -0.10 & -0.6 \\
\hline
NGC6121  & -1.16 & 4.40   & -1.10 & -0.06 & -0.3  & -1.18 & +0.02 & +0.1  & -1.20 & +0.04 & +0.2  & -1.22 & +0.06 & +0.3  & -1.21 & +0.05 & +0.3 \\
\hline
NGC6139  & -1.65 & 3.92   & -1.38 & -0.27 & -1.4  & -1.48 & -0.17 & -1.1  & -1.46 & -0.19 & -1.2  & -1.48 & -0.17 & -1.0  & -1.49 & -0.16 & -0.9 \\
\hline
NGC6254  & -1.56 & 3.70   & -1.51 & -0.05 & -0.3  & -1.60 & +0.04 & +0.2  & -1.56 & +0.00 & +0.0  & -1.58 & +0.02 & +0.1  & -1.59 & +0.03 & +0.2 \\
\hline
NGC6325  & -1.25 & 4.36   & -1.12 & -0.13 & -0.7  & -1.21 & -0.04 & -0.3  & -1.22 & -0.03 & -0.2  & -1.24 & -0.01 & -0.1  & -1.24 & -0.01 & -0.1 \\
\hline
NGC6356  & -0.40 & 5.28   & -0.57 & +0.17 & +0.9  & -0.54 & +0.14 & +0.8  & -0.55 & +0.15 & +0.9  & -0.53 & +0.13 & +0.8  & -0.53 & +0.13 & +0.7 \\
\hline
NGC6380  & -0.75 & 5.11   & -0.67 & -0.08 & -0.4  & -0.67 & -0.08 & -0.5  & -0.69 & -0.06 & -0.3  & -0.67 & -0.08 & -0.5  & -0.67 & -0.08 & -0.5 \\
\hline
NGC6397  & -2.02 & 2.66   & -2.13 & +0.11 & +0.6  & -2.07 & +0.05 & +0.3  & -2.02 & +0.00 & +0.0  & -1.96 & -0.06 & -0.3  & -1.96 & -0.06 & -0.3 \\
\hline
NGC6440  & -0.36 & 5.48   & -0.45 & +0.09 & +0.5  & -0.37 & +0.01 & +0.1  & -0.36 & +0.00 & +0.0  & -0.35 & -0.01 & -0.0  & -0.36 & -0.00 & -0.0 \\
\hline
NGC6441  & -0.46 & 5.20   & -0.61 & +0.15 & +0.8  & -0.60 & +0.14 & +0.8  & -0.62 & +0.16 & +1.0  & -0.59 & +0.13 & +0.8  & -0.59 & +0.13 & +0.8 \\
\hline
NGC6528  & -0.11 & 5.65   & -0.35 & +0.24 & +1.3  & -0.22 & +0.11 & +0.7  & -0.18 & +0.07 & +0.5  & -0.20 & +0.09 & +0.6  & -0.20 & +0.09 & +0.5 \\
\hline
NGC6553  & -0.18 & 5.76   & -0.28 & +0.10 & +0.5  & -0.13 & -0.05 & -0.3  & -0.07 & -0.11 & -0.7  & -0.11 & -0.07 & -0.4  & -0.11 & -0.07 & -0.4 \\
\hline
NGC6558  & -1.32 & 4.62   & -0.96 & -0.36 & -1.9  & -1.03 & -0.29 & -1.8  & -1.06 & -0.26 & -1.6  & -1.06 & -0.26 & -1.6  & -1.06 & -0.26 & -1.5 \\
\hline
NGC6569  & -0.76 & 4.44   & -1.07 & +0.31 & +1.6  & -1.15 & +0.39 & +2.4  & -1.17 & +0.41 & +2.5  & -1.18 & +0.42 & +2.5  & -1.18 & +0.42 & +2.4 \\
\hline
NGC6656  & -1.70 & 3.07   & -1.88 & +0.18 & +1.0  & -1.90 & +0.20 & +1.2  & -1.84 & +0.14 & +0.9  & -1.81 & +0.11 & +0.7  & -1.81 & +0.11 & +0.6 \\
\hline
NGC6838  & -0.78 & 5.07   & -0.70 & -0.08 & -0.4  & -0.70 & -0.08 & -0.5  & -0.73 & -0.05 & -0.3  & -0.71 & -0.07 & -0.4  & -0.71 & -0.07 & -0.4 \\
\hline
NGC7078  & -2.37 & 2.01   & -2.52 & +0.15 & +0.8  & -2.27 & -0.10 & -0.6  & -2.34 & -0.03 & -0.2  & -2.38 & +0.01 & +0.0  & -2.38 & +0.01 & +0.0 \\
\hline
Pal7     & -0.75 & 5.22   & -0.60 & -0.15 & -0.8  & -0.58 & -0.17 & -1.0  & -0.60 & -0.15 & -0.9  & -0.58 & -0.17 & -1.0  & -0.58 & -0.17 & -1.0 \\
\hline

\end{tabular}
\end{center}
\end{table}

\begin{table}
\begin{center}
\caption{Metallicities for the R97 GCs in C09 scale.
Column (1) lists the cluster ID (the GCs marked with * were not used to calculate the calibration).
Column (2) is the metallicity according the scale.
Column (3) is the calculated rEW.
Column (Xa) is the metallicity based on the calibration using a polynomial of degree X.
Column (Xb) is the difference between the calculated and the scale metallicity.
Column (Xc) is the former difference in r.m.s.}
\label{tab:RC09}
\begin{tabular}{|l|c|c||c|c|c||c|c|c||c|c|c||c|c|c||c|c|c|}
\hline
GC & [Fe/H] &  $\rEW$ &  [Fe/H]$_1$& $\Delta$ & $\Delta/\sigma$ &  [Fe/H]$_2$& $\Delta$ & $\Delta/\sigma$ &  [Fe/H]$_3$& $\Delta$ & $\Delta/\sigma$  \\(1) & (2) & (3) & (Ia) &  (Ib) &  (Ic) & (IIa) &  (IIb) &  (IIc) & (IIIa) &  (IIIb) &  (IIIc) \\
\hline
NGC2808  & -1.18 & 3.80   & -1.26 & +0.08 & +0.4  & -1.31 & +0.13 & +1.3  & -1.32 & +0.14 & +1.3 \\
\hline
NGC3201  & -1.51 & 3.59   & -1.39 & -0.12 & -0.6  & -1.47 & -0.04 & -0.4  & -1.47 & -0.04 & -0.3 \\
\hline
NGC4372  & -2.19 & 2.07   & -2.25 & +0.06 & +0.3  & -2.21 & +0.02 & +0.2  & -2.20 & +0.01 & +0.1 \\
\hline
NGC4590  & -2.27 & 1.98   & -2.30 & +0.03 & +0.1  & -2.23 & -0.04 & -0.4  & -2.23 & -0.04 & -0.3 \\
\hline
NGC6121  & -1.18 & 3.98   & -1.16 & -0.02 & -0.1  & -1.18 & -0.00 & -0.0  & -1.19 & +0.01 & +0.1 \\
\hline
NGC6397  & -1.99 & 2.29   & -2.13 & +0.14 & +0.6  & -2.14 & +0.15 & +1.4  & -2.12 & +0.13 & +1.2 \\
\hline
NGC6522  & -1.45 & 3.72   & -1.31 & -0.14 & -0.7  & -1.37 & -0.08 & -0.8  & -1.38 & -0.07 & -0.6 \\
\hline
NGC6528  & +0.07 & 5.35   & -0.38 & +0.45 & +2.1  & +0.15 & -0.08 & -0.8  & +0.22 & -0.15 & -1.3 \\
\hline
NGC6541  & -1.82 & 2.98   & -1.73 & -0.09 & -0.4  & -1.84 & +0.02 & +0.2  & -1.82 & +0.00 & +0.0 \\
\hline
NGC6544  & -1.47 & 3.54   & -1.41 & -0.06 & -0.3  & -1.50 & +0.03 & +0.3  & -1.50 & +0.03 & +0.3 \\
\hline
NGC6553  & -0.16 & 5.06   & -0.54 & +0.38 & +1.8  & -0.17 & +0.01 & +0.1  & -0.14 & -0.02 & -0.2 \\
\hline
NGC6624  & -0.42 & 4.56   & -0.83 & +0.41 & +1.9  & -0.67 & +0.25 & +2.5  & -0.69 & +0.27 & +2.3 \\
\hline
NGC6626* & -1.46 & 3.81   & -1.26 & -0.20 & -0.9  & -1.31 & -0.15 & -1.5  & -1.32 & -0.14 & -1.3 \\
\hline
NGC6637  & -0.59 & 4.64   & -0.79 & +0.20 & +0.9  & -0.60 & +0.01 & +0.1  & -0.61 & +0.02 & +0.2 \\
\hline
NGC6638  & -0.99 & 4.29   & -0.98 & -0.01 & -0.0  & -0.92 & -0.07 & -0.7  & -0.94 & -0.05 & -0.5 \\
\hline
NGC6809  & -1.93 & 2.93   & -1.76 & -0.17 & -0.8  & -1.86 & -0.07 & -0.7  & -1.84 & -0.09 & -0.8 \\
\hline
NGC7099  & -2.33 & 1.86   & -2.37 & +0.04 & +0.2  & -2.26 & -0.07 & -0.7  & -2.28 & -0.05 & -0.5 \\
\hline

\end{tabular}
\end{center}
\end{table}

\begin{table}
\begin{center}
\caption{Metallicities for the R97 GCs in C09c scale.
Column (1) lists the cluster ID (the GCs marked with * were not used to calculate the calibration).
Column (2) is the metallicity according the scale.
Column (3) is the calculated rEW.
Column (Xa) is the metallicity based on the calibration using a polynomial of degree X.
Column (Xb) is the difference between the calculated and the scale metallicity.
Column (Xc) is the former difference in r.m.s.}
\label{tab:RC09c}
\begin{tabular}{|l|c|c||c|c|c||c|c|c||c|c|c||c|c|c||c|c|c|}
\hline
GC & [Fe/H] &  $\rEW$ &  [Fe/H]$_1$& $\Delta$ & $\Delta/\sigma$ &  [Fe/H]$_2$& $\Delta$ & $\Delta/\sigma$ &  [Fe/H]$_3$& $\Delta$ & $\Delta/\sigma$  \\(1) & (2) & (3) & (Ia) &  (Ib) &  (Ic) & (IIa) &  (IIb) &  (IIc) & (IIIa) &  (IIIb) &  (IIIc) \\
\hline
NGC2808  & -1.18 & 3.80   & -1.27 & +0.09 & +0.5  & -1.32 & +0.14 & +1.5  & -1.32 & +0.14 & +1.4 \\
\hline
NGC3201  & -1.51 & 3.59   & -1.39 & -0.12 & -0.7  & -1.47 & -0.04 & -0.5  & -1.47 & -0.04 & -0.4 \\
\hline
NGC4372  & -2.19 & 2.07   & -2.25 & +0.06 & +0.4  & -2.21 & +0.02 & +0.2  & -2.21 & +0.02 & +0.2 \\
\hline
NGC4590  & -2.27 & 1.98   & -2.30 & +0.03 & +0.2  & -2.23 & -0.04 & -0.4  & -2.24 & -0.03 & -0.3 \\
\hline
NGC6121  & -1.18 & 3.98   & -1.17 & -0.01 & -0.1  & -1.19 & +0.01 & +0.1  & -1.20 & +0.02 & +0.2 \\
\hline
NGC6397  & -1.99 & 2.29   & -2.13 & +0.14 & +0.8  & -2.13 & +0.14 & +1.5  & -2.13 & +0.14 & +1.3 \\
\hline
NGC6522  & -1.45 & 3.72   & -1.31 & -0.14 & -0.8  & -1.38 & -0.07 & -0.8  & -1.38 & -0.07 & -0.7 \\
\hline
NGC6528  & -0.17 & 5.35   & -0.39 & +0.22 & +1.4  & +0.04 & -0.21 & -2.2  & +0.07 & -0.24 & -2.3 \\
\hline
NGC6541  & -1.82 & 2.98   & -1.73 & -0.09 & -0.5  & -1.83 & +0.01 & +0.1  & -1.82 & -0.00 & -0.0 \\
\hline
NGC6544  & -1.47 & 3.54   & -1.42 & -0.05 & -0.3  & -1.50 & +0.03 & +0.3  & -1.50 & +0.03 & +0.3 \\
\hline
NGC6553  & -0.16 & 5.06   & -0.56 & +0.40 & +2.4  & -0.25 & +0.09 & +1.0  & -0.24 & +0.08 & +0.8 \\
\hline
NGC6624  & -0.69 & 4.56   & -0.84 & +0.15 & +0.9  & -0.72 & +0.03 & +0.3  & -0.72 & +0.03 & +0.3 \\
\hline
NGC6626* & -1.28 & 3.81   & -1.26 & -0.02 & -0.1  & -1.31 & +0.03 & +0.3  & -1.32 & +0.04 & +0.4 \\
\hline
NGC6637  & -0.59 & 4.64   & -0.80 & +0.21 & +1.3  & -0.66 & +0.07 & +0.7  & -0.66 & +0.07 & +0.6 \\
\hline
NGC6638  & -0.99 & 4.29   & -0.99 & +0.00 & +0.0  & -0.95 & -0.04 & -0.4  & -0.95 & -0.04 & -0.3 \\
\hline
NGC6809  & -1.93 & 2.93   & -1.76 & -0.17 & -1.0  & -1.85 & -0.08 & -0.8  & -1.84 & -0.09 & -0.8 \\
\hline
NGC7099  & -2.33 & 1.86   & -2.37 & +0.04 & +0.2  & -2.27 & -0.06 & -0.6  & -2.28 & -0.05 & -0.5 \\
\hline

\end{tabular}
\end{center}
\end{table}

\begin{table}
\begin{center}
\caption{Metallicities for the R97 GCs in H10 scale.
Column (1) lists the cluster ID (the GCs marked with * were not used to calculate the calibration).
Column (2) is the metallicity according the scale.
Column (3) is the calculated rEW.
Column (Xa) is the metallicity based on the calibration using a polynomial of degree X.
Column (Xb) is the difference between the calculated and the scale metallicity.
Column (Xc) is the former difference in r.m.s.}
\label{tab:RH10}
\begin{tabular}{|l|c|c||c|c|c||c|c|c||c|c|c||c|c|c||c|c|c|}
\hline
GC & [Fe/H] &  $\rEW$ &  [Fe/H]$_1$& $\Delta$ & $\Delta/\sigma$ &  [Fe/H]$_2$& $\Delta$ & $\Delta/\sigma$ &  [Fe/H]$_3$& $\Delta$ & $\Delta/\sigma$  \\(1) & (2) & (3) & (Ia) &  (Ib) &  (Ic) & (IIa) &  (IIb) &  (IIc) & (IIIa) &  (IIIb) &  (IIIc) \\
\hline
NGC2808  & -1.14 & 3.80   & -1.18 & +0.04 & +0.2  & -1.27 & +0.13 & +1.2  & -1.26 & +0.12 & +1.2 \\
\hline
NGC3201  & -1.59 & 3.59   & -1.32 & -0.27 & -1.8  & -1.42 & -0.17 & -1.5  & -1.43 & -0.16 & -1.5 \\
\hline
NGC4372  & -2.17 & 2.07   & -2.28 & +0.11 & +0.7  & -2.19 & +0.02 & +0.2  & -2.19 & +0.02 & +0.2 \\
\hline
NGC4590  & -2.23 & 1.98   & -2.34 & +0.11 & +0.7  & -2.23 & -0.00 & -0.0  & -2.20 & -0.03 & -0.3 \\
\hline
NGC6121  & -1.16 & 3.98   & -1.07 & -0.09 & -0.6  & -1.15 & -0.01 & -0.1  & -1.12 & -0.04 & -0.4 \\
\hline
NGC6397  & -2.02 & 2.29   & -2.14 & +0.12 & +0.8  & -2.11 & +0.09 & +0.8  & -2.15 & +0.13 & +1.3 \\
\hline
NGC6522  & -1.34 & 3.72   & -1.23 & -0.11 & -0.7  & -1.33 & -0.01 & -0.1  & -1.33 & -0.01 & -0.1 \\
\hline
NGC6528  & -0.11 & 5.35   & -0.19 & +0.08 & +0.5  & +0.04 & -0.15 & -1.3  & -0.04 & -0.07 & -0.7 \\
\hline
NGC6541  & -1.81 & 2.98   & -1.70 & -0.11 & -0.7  & -1.78 & -0.03 & -0.2  & -1.85 & +0.04 & +0.4 \\
\hline
NGC6544  & -1.40 & 3.54   & -1.34 & -0.06 & -0.4  & -1.45 & +0.05 & +0.5  & -1.47 & +0.07 & +0.6 \\
\hline
NGC6553  & -0.18 & 5.06   & -0.38 & +0.20 & +1.3  & -0.24 & +0.06 & +0.6  & -0.24 & +0.06 & +0.6 \\
\hline
NGC6624  & -0.44 & 4.56   & -0.70 & +0.26 & +1.7  & -0.69 & +0.25 & +2.3  & -0.64 & +0.20 & +1.9 \\
\hline
NGC6626  & -1.32 & 3.81   & -1.17 & -0.15 & -1.0  & -1.27 & -0.05 & -0.5  & -1.26 & -0.06 & -0.6 \\
\hline
NGC6637  & -0.64 & 4.64   & -0.65 & +0.01 & +0.1  & -0.63 & -0.01 & -0.1  & -0.58 & -0.06 & -0.6 \\
\hline
NGC6638  & -0.95 & 4.29   & -0.87 & -0.08 & -0.5  & -0.91 & -0.04 & -0.4  & -0.86 & -0.09 & -0.9 \\
\hline
NGC6809  & -1.94 & 2.93   & -1.73 & -0.21 & -1.3  & -1.81 & -0.13 & -1.2  & -1.88 & -0.06 & -0.6 \\
\hline
NGC7099  & -2.27 & 1.86   & -2.42 & +0.15 & +1.0  & -2.27 & -0.00 & -0.0  & -2.21 & -0.06 & -0.6 \\
\hline

\end{tabular}
\end{center}
\end{table}

\end{landscape}

\begin{landscape}

\begin{longtable}{|l|c|c|c|c|c|c|c|c|c|c|c|}
\caption[Data]{Data}\label{tab:data}\\
 \hline
Cluster &  ra(J2000) & dec(J2000) & $J$ & $\sigma_J$ & $H$ & $\sigma_H$ & $\Ks{}$ & $\sigma_{Ks}$ &  EW & $\sigma_{EW}$ & source \\
\hline
\endfirsthead
\caption{continued.} \\
\hline
Cluster &  ra(J2000) & dec(J2000) & $J$ & $\sigma_J$ & $H$ & $\sigma_H$ & $\Ks{}$ & $\sigma_{Ks}$ &  EW & $\sigma_{EW}$ & source \\
\hline
\endhead

 \hline

\endfoot

\hline \hline

\endlastfoot

\hline
NGC2808 & 137.944550 & -64.870630 & 11.6660 & 0.0260 & 10.9670 & 0.0300 & 10.8120 & 0.0210 & 5.660 & 0.233 & S12 \\
NGC2808 & 138.069940 & -64.867500 & 10.2290 & 0.0280 & 9.3530 & 0.0240 & 9.1010 & 0.0230 & 6.700 & 0.282 & S12 \\
NGC2808 & 137.940140 & -64.858480 & 12.8840 & 0.0320 & 12.3110 & 0.0410 & 12.1220 & 0.0270 & 4.710 & 0.288 & S12 \\
NGC2808 & 138.076100 & -64.852680 & 11.9450 & 0.0330 & 11.1770 & 0.0320 & 10.9950 & 0.0240 & 5.930 & 0.228 & S12 \\
NGC2808 & 138.070550 & -64.847130 & 12.8550 & 0.0620 & 12.2170 & 0.0630 & 12.0500 & 0.0540 & 4.960 & 0.189 & S12 \\
NGC2808 & 138.090370 & -64.842770 & 13.1930 & 0.0440 & 12.5020 & 0.0390 & 12.3290 & 0.0300 & 5.340 & 0.230 & S12 \\
NGC2808 & 138.008410 & -64.834180 & 13.2110 & 0.0250 & 12.5690 & 0.0220 & 12.3560 & 0.0360 & 5.070 & 0.233 & S12 \\
NGC2808 & 138.031650 & -64.826760 & 12.4970 & 0.0370 & 11.7930 & 0.0350 & 11.6670 & 0.0320 & 5.240 & 0.214 & S12 \\
\hline

\end{longtable}

\end{landscape}

\end{document}